\documentclass[reqno]{amsart}

\usepackage{amsmath}
\usepackage{amscd}
\usepackage{amssymb}
\usepackage{enumerate}
\usepackage{amsfonts}
\usepackage{graphicx}

\numberwithin{equation}{section}

\newcommand{\rar}[1]{\stackrel{#1}{\longrightarrow}}

\newcommand{\al}{\alpha}
\newcommand{\be}{\beta}
\newcommand{\ga}{\gamma}

\newcommand{\de}{\delta}

\newcommand{\la}{\lambda}
\newcommand{\La}{\Lambda}
\newcommand{\ze}{\zeta}
\newcommand{\ka}{\kappa}
\newcommand{\eps}{\epsilon}
\newcommand{\sg}{\sigma}
\newcommand{\te}{\theta}
\newcommand{\om}{\omega}

\newcommand{\bC}{{\mathbb C}}

\newcommand{\bR}{{\mathbb R}}
\newcommand{\bZ}{{\mathbb Z}}

\newcommand{\fP}{{\mathfrak P}}
\newcommand{\fX}{{\mathfrak X}}
\newcommand{\fY}{{\mathfrak Y}}
\newcommand{\fZ}{{\mathfrak Z}}

\newcommand{\eq}[1]{(\ref{#1})}

\newcommand{\abs}[1]{\vert #1\vert}
\newcommand{\norm}[1]{\vert\vert #1\vert\vert}

\newtheorem{thm}{Theorem}[section]
\newtheorem{cor}[thm]{Corollary}
\newtheorem{lem}[thm]{Lemma}
\newtheorem{prop}[thm]{Proposition}

\theoremstyle{remark}
\newtheorem{rem}[thm]{Remark}

\newcommand{\Res}[1]{\underset{#1}{\operatorname{Res}}\,}
\newcommand{\matr}[4]{\left(\begin{array}{cc} \! #1 & \! #2 \! \\ \! #3 & \! #4 \! \end{array}\right)}
\newcommand{\diag}[2]{\matr{#1}{0}{0}{#2}}
\newcommand{\tr}{\mathrm{Tr}\,}

\newcommand{\mx}{m_\fX}

\newcommand{\zp}{\bZ_{\geq 0}}
\newcommand{\zpk}{\bZ_{\geq k}}
\newcommand{\hypergeom}[5]{\, _{#1}\!F_{#2}\!
\left(\begin{array}{c|}#3\\#4\end{array}\,\,#5\right)}
\newcommand{\qhypergeom}[5]{\, _{#1}\!\phi_{#2}\!
\left(\begin{array}{c|}#3\\#4\end{array}\,\,q;#5\right)}

\newcommand{\yb}{\bar{y}}
\newcommand{\zb}{\bar{z}}
\newcommand{\wb}{\bar{w}}
\newcommand{\fb}{\bar{f}}
\newcommand{\gb}{\bar{g}}

\newcommand{\tht}{\eq}

\title[Distribution of the first particle]{Distribution of the first particle in discrete orthogonal polynomial ensembles}

\author[A.~Borodin \and D.~Boyarchenko]{Alexei Borodin\address{Alexei Borodin: School of Mathematics, Insitute for Advanced Study, Princeton, NJ 08540, e-mail: borodine@math.upenn.edu} \and Dmitriy Boyarchenko\address{Dmitriy Boyarchenko: Department of Mathematics, University of Pennsylvania, Philadelphia, PA 19104-6395, e-mail: dmitriy@math.upenn.edu}}

\begin{document}

\begin{abstract}
We show that the distribution function of the first particle in a
discrete orthogonal polynomial ensemble can be obtained through a
certain recurrence procedure, if the (difference or $q$-)
log-derivative of the weight function is rational. In a number of
classical special cases the recurrence procedure is equivalent to
the difference and q-Painlev\'e equations of \cite{JS},
\cite{Sak}.

Our approach is based on the formalism of discrete integrable
operators and discrete Riemann--Hilbert problems developed in
\cite{IMRN}, \cite{B}.
\end{abstract}

\maketitle

\setcounter{tocdepth}{1}

\tableofcontents

\section{Introduction}\label{s:intro}

\subsection{} The basic problem considered in this paper is the following. Let
$\fX$ be a locally finite subset of $\bR$ and $w:\fX\to\bR_{>0}$
be a positive--valued function on $\fX$ with finite moments:
\[
\sum_{x\in\fX} \abs{x}^n w(x)<\infty,\qquad n=0,1,\dotsc\,.
\]

Fix a positive integer $k$ (the number of particles) and consider
the probability measure on all $k$-point subsets of $\fX$ given by
\begin{equation}\label{1}
\operatorname{Prob}\{x_1,\dotsc,x_k\}=\operatorname{const}\cdot
\prod_{1\le i<j\le k}(x_i-x_j)^2\prod_{i=1}^k w(x_i).
\end{equation}
We are interested in the distribution of $\max\{x_1,\dotsc,x_k\}$
with respect to this measure.

The problem is motivated by random matrix theory on one side, and
by combinatorial and representation theoretic models on the other
one.

In random matrix theory, probability measures of the form
\[
\operatorname{const}\cdot \prod_{1\le i<j\le
k}(x_i-x_j)^2\prod_{i=1}^k w(x_i)dx_i
\]
on $k$-point subsets of $\bR$, with $w(x)$ being a smooth function
on a subinterval of $\bR$, play a prominent role. Most
computations for such models are conveniently done by means of the
orthogonal polynomials associated with the weight function $w(x)$.
On this ground, these measures are often called {\it orthogonal
polynomial ensembles}. See \cite{Me}, \cite{NW} and references
therein for a further discussion.

The problem of describing the distribution of the $\max\{x_i\}$ in
the continuous setting for the classical weights has been solved
in the following sense: the distribution function was explicitly
written in terms of a specific solution of one of the six (2nd
order nonlinear ordinary differential) Painlev\'e equations. It
was done in \cite{TW2} for the Hermite weight $w(x)=\exp(-x^2)$,
$x\in\bR$, and for the Laguerre weight $x^a\exp(-x)$, $x>0$; in
\cite{TW2}, \cite{HS} for the Jacobi weight $(1-x)^a(1+x)^b$,
$x\in(-1,1)$; and in \cite{WF}, \cite{BD} for the quasi-Jacobi
weight $(1-ix)^s(1+ix)^{\bar{s}}$, $x\in\bR$.\footnote{Continuous problems of this type have been extensively studied. We refer to the introduction of \cite{BD} for a brief review and references.}
 Thus, it is natural
to ask what would be an analog of these results when we take $w$
to be a classical {\it discrete} weight function.

On the other hand, in recent years the random variable
$\max\{x_i\}$ with $x_i$'s distributed according to \tht{1} with
certain specific weights, came up as the main quantity of interest
in a number of problems originating in combinatorics,
first-passage percolation, representation theory, and growth
processes, see e.g. \cite{BOl}, \cite{J1}, \cite{J2}, \cite{Bai} and
references therein.

\subsection{}\label{assumps} In order to state our first result we need to introduce more
notation. Let us denote the points of $\fX$ by $\pi_s$,
$s=0,1,\dotsc,N$, with $\pi_0<\pi_1<\dotsb$ or $\pi_0>\pi_1>\dotsb$\,. Here
$N=\abs{\fX}-1$ may be finite or infinite. We use the following
two basic assumptions:
\begin{itemize}
\item There exists an affine transformation $\sigma:\bR\to\bR$
such that $\sigma \pi_{s+1}=\pi_s$ for all $s$, $0\le s< N$.
\item There exist polynomials $P(x)$ and $Q(x)$ such that
\[
\frac{w(\pi_{s-1})}{w(\pi_s)}=\frac{P(\pi_s)}{Q(\pi_s)},\qquad
1\le s\le N,
\]
and $P(\pi_0)=0$.
\end{itemize}
The orthogonality data for a number (but not all) hypergeometric
polynomials of the Askey scheme satisfy these assumptions, see \S\ref{s:Lax-Askey}
below for details.\footnote{For some classical families of polynomials both assumptions are satisfied but the orthogonality set $\fX$ is not locally finite. We extend our results to those cases, see \S\ref{s:Fredholm} below.}

We prove that under the two conditions above, there exists a
certain recurrence procedure which computes the {\it gap probability}
\[
D_s=\operatorname{Prob}\left\{x_i\notin\{\pi_s,\pi_{s+1},\dots\}\text{  for all }i\right\}=\begin{cases}
\operatorname{Prob}\{\max\{x_i\}<\pi_s\},&\text{if  } \pi_0<\pi_1<\dotsb\,,\\  \operatorname{Prob}\{\min\{x_i\}>\pi_s\},&
\text{if  } \pi_0>\pi_1>\dotsb\,,
\end{cases}
\]
with $x_i$'s distributed according to \tht{1}. In fact, the
recurrence procedure produces a sequence $(A_s,M_s(\ze))$, where
$A_s$ is a nilpotent 2 by 2 matrix and $M_s(\ze)$ is a matrix
polynomial
\[
M_s(\ze)=M_s^{(l)}\ze^l+\dots +M_s^{(0)}, \qquad
M_s^{(i)}\in\operatorname{Mat}(2,\bC),
\]
of degree $l=\max\{\deg P,\deg Q\}$. The elementary step of the
recurrence is provided by the equality
\begin{equation}\label{2}
\left(I+\frac{A_s}{\sigma \ze-\pi_s}\right)M_s(\ze)
=M_{s+1}(\ze)\left(I+\frac{A_{s+1}}{\ze-\pi_{s+1}}\right).
\end{equation}
It is not hard to see that if $\det M_s(\pi_{s+1})\ne 0$ (which is
always the case in our setting) then \tht{2} defines
$(A_{s+1},M_{s+1})$ uniquely provided that we know $(A_s,M_s)$.
However, the existence of $(A_{s+1},M_{s+1})$ is not obvious and
needs to be proved.\footnote{In fact, for 2 by 2 matrices one can
easily see that $(A_{s+1}, M_{s+1}^{(l)},\dots,M_{s+1}^{(0)})$ are
rational functions of $(A_s, M_s^{(l)},\dots,M_s^{(0)})$.} Again,
in our setting it always holds.

We then show that the ratio
\[
\left(\frac{D_{s+3}}{D_{s+2}}-\frac{D_{s+2}}{D_{s+1}}\right)
\left(\frac{D_{s+2}}{D_{s+1}}-\frac{D_{s+1}}{D_{s}}\right)^{-1}
\]
is an explicit rational function of $(A_s,
M_s^{(l)},\dots,M_s^{(0)})$ and $(A_{s+1},
M_{s+1}^{(l)},\dots,M_{s+1}^{(0)})$.

Since $D_s=\operatorname{Prob}\{\max\{x_i\}<\pi_s\}$ is nonzero
only if $s\ge k$ (recall that $k$ is the number of $x_i$'s in
\tht{1}), it is enough to provide the initial conditions
$D_k,D_{k+1},D_{k+2}$, $A_k$, $M_k(\ze)$ in order to be able to
compute $D_s$ for arbitrary $s$. These initial conditions are
readily expressed in terms of $\{\pi_s\}$ and $\{w(\pi_s)\}$, see
\S\ref{s:initial} below.

For certain classical weights $w$ the recurrence relation \tht{2}
can be substantially simplified. To illustrate the situation, let
us consider $\fX=\bZ_{\ge 0}$ and $w(x)=a^x/x!$, where $a>0$ is a
parameter. This weight function corresponds to the Charlier
orthogonal polynomials.

In this case, $A_s$ and $M_s(\ze)$ can be parameterized by three
scalar sequences $a_s$, $b_s$, $c_s$ as follows:
\begin{equation}\label{3}
M_s(\ze)=\bmatrix 1&0\\0&0\endbmatrix \ze+\bmatrix
b_s&b_sc_s\\a/c_s&a\endbmatrix,\qquad A_s=(k+b_s)\bmatrix
-1&-a_sc_s\\1/(a_sc_s)&1\endbmatrix.
\end{equation}
Then the equality \tht{2} leads to the following recurrence
relations:
\begin{gather}
\label{4}
a_{s+1}=\frac{(b_s+aa_s)(k+b_s+aa_s)}{aa_s(s+1+b_s+aa_s)}\,, \\
\label{5}
b_{s+1}=\frac{s+1}{1-a_{s+1}}-(s+1+k+b_s+aa_s), \\
\label{6} c_{s+1}=\frac{aa_s}{k+b_s+aa_s}\,c_s.
\end{gather}
The connection of these sequences and the distribution $D_s$ is
given by
\[
\left(\frac{D_{s+3}}{D_{s+2}}-\frac{D_{s+2}}{D_{s+1}}\right)
\left(\frac{D_{s+2}}{D_{s+1}}-\frac{D_{s+1}}{D_{s}}\right)^{-1}
=\frac{c_s (b_s+a a_s)(b_{s+1}+a a_{s+1})}{a(s+2)a_{s+1}^2
c_{s+1}}\,.
\]
The corresponding initial conditions can be found in subsection \ref{ss:charlier}.

Under the change of variables
\[
f_s=a_s^{-1}, \qquad g_s=aa_s+b_s+s+1,
\]
\tht{4}-\tht{5} turn into
\begin{eqnarray*}
f_s f_{s+1}=\frac{a g_s}{(g_s-s-1)(g_s+k-s-1)},\\
g_s+g_{s+1}=\frac{a}{f_{s+1}}-\frac{s+1}{1-f_{s+1}}-k+2s+3.
\end{eqnarray*}
This recurrence is immediately identified with the difference
Painlev\'e IV equation (dPIV) of \cite{Sak}.

\subsection{} It turns out that the situation for the Charlier weight described
above is rather typical. We are also able to reduce \tht{2} to
scalar rational recurrence relations for the weight functions
corresponding to Meixner, Krawtchouk, q-Charlier, alternative q-Charlier, little
q-Laguerre/Wall, little q-Jacobi, and q-Krawtchouk orthogonal
polynomials. In the appropriate variables,  Meixner and Krawtchouk
cases lead to dPV of \cite{Sak}, little q-Jacobi and q-Krawtchouk
lead to q-PVI of \cite{JS}, \cite{Sak}, and q-Charlier and little
q-Laguerre yield a certain degeneration of q-PVI.

It is remarkable that in almost all the cases we can solve explicitly, 
we end up
with one of the equations of Sakai's hierarchy which was
constructed out of purely algebraic geometric considerations, see
\cite{Sak}. (We were not able to see such a reduction in the alternative q-Charlier case, but we do not claim that there is none.) So far we have not found a conceptual explanation for
this fact.

One can notice, however, that recurrence relations originating
from \tht{2} must have some kind of singularity confinement
property. (This property was the starting point of Sakai's work.)
For example, the parameterization \tht{3} does not make much sense
if, say, $A_s$ has a zero (2,1) element. Then the values of $a_s$
and $c_s$ are not well--defined. In terms of the recurrence
relations, this is reflected by vanishing of one of the
denominators in  \tht{4}-\tht{6}. However, the matrix sequence
$\{(A_s,M_s)\}$ does not feel this singularity, which means that
the sequences $\{a_s\}$, $\{b_s\}$, $\{c_s\}$ can be ``continued
through'' their singular values. Of course, all Sakai's equations
have this kind of singularity confinement by construction.

Let us also point out that it is not clear at this point whether
the weights of higher hypergeometric polynomials of the Askey
scheme will also lead to one of Sakai's equations. All the
cases that we were able to solve explicitly have {\it linear} matrices
$M_s(\ze)$ in \tht{2}, while, say, for the Hahn weight $M_s(\ze)$
is quadratic. Handling such cases seems to be a problem of the
next level of difficulty. It remains an interesting open problem
to derive explicit rational recurrence relations for $\deg M_s=2$.

For Charlier and Meixner weights, $D_s$ can also be written as
Toeplitz determinants with symbols
\begin{equation}\label{7}
(1+z)^k\exp(az^{-1})\quad\text{  and  }\quad (1+z)^k(1+bz^{-1})^c
\end{equation}
respectively, see \cite{TW3}, \cite{J2}, \cite{BOk}. Here $a,b,c$
are parameters. Among previous results on the subject let us
mention
\begin{itemize}
\item the derivation of dPII for Toeplitz determinants with the
symbol $\exp(\theta(z+z^{-1}))$, see \cite{Bai}, \cite{B},
\cite{AvM} (note also the derivation of the same equation for the
quantity closely related to these Toeplitz determinants in
\cite{PS}, \cite{TW3});
\item derivation of dPV for Toeplitz determinants
with symbol $(1+z)^k(1+bz^{-1})^c$ and $k$ being not necessarily
integral in \cite{B};
\item derivation of rational recurrence
relations for Toeplitz determinants with symbols of the form
\begin{equation}\label{8}
\exp(P_1(z)+P_2(z^{-1}))\,z^\gamma(1-d_1z)^{\gamma_1'}
(1-d_2z)^{\gamma_2'}(1-d_1^{-1}z^{-1})^{\gamma_1''}(1-d_2^{-1}z^{-1})^{\gamma_2''}
\end{equation}
where $P_1$ and $P_2$ are polynomials with $|\deg P_1-\deg P_2|\le
1$, $\gamma_1',\gamma_2',\gamma_1'',\gamma_2'',d_1,d_2$ are
constants, see \cite{AvM}. Interestingly enough, for the symbols
\tht{7}, which are special cases of \tht{8}, the relations of
\cite{AvM} do not seem to have much in common with those of
\cite{B} and the present paper.
\end{itemize}

\subsection{} The methods used in this paper are based on the formalism of
discrete integrable operators and discrete Riemann--Hilbert
problem (DRHP) developed in \cite{IMRN}, \cite{B}. The first step
is to represent $D_s$ as a Fredholm determinant of an integrable
operator: $D_s=\det(1-K_s)$, where $K_s$ is an operator in
$\ell^2(\{\pi_s,\pi_{s+1},\dots\})$ with the kernel
\[
K_s(x,y)=\frac 1{\Vert p_{k-1}\Vert^2_{\ell^2(\fX,w)}}\,
\frac{p_{k-1}(x)p_{k}(y)-p_{k}(x)p_{k-1}(y)}{x-y}\,\sqrt{w(x)w(y)}\,.
\]
Here $p_k$ and $p_{k-1}$ are monic $k$th and $(k-1)$st orthogonal
polynomials on $\fX$ with respect to the weight function $w$.
Using the results of \cite{IMRN}, \cite{B}, the computation of
such a Fredholm determinant can be reduced to solving a DRHP on
$\{\pi_0,\dots,\pi_{s-1}\}$ with a jump matrix easily expressible
in terms of $w$. Our assumptions on $\fX$ and $w$, see above, then
allow us to obtain a {\it Lax pair} for the solution $m_s(\ze)$ of
this DRHP, which has the form
\begin{equation}\label{9}
m_{s+1}(\ze)=\left(I+\frac{A_s}{\ze-\pi_s}\right)m_s(\ze),\qquad
m_s(\sigma\zeta)=M_s(\ze)m_{s+1}(\ze)D^{-1}(\ze).
\end{equation}
Here $M_s(\ze)$ is a matrix polynomial and $D(\ze)$ is a fixed
diagonal matrix polynomial. Compatibility condition for this pair
of equations is exactly \tht{2}.

The paper is organized as follows. In \S\ref{s:DRHP} we reduce the
problem of computing discrete orthogonal polynomials with a given
weight to a DRHP. In \S\ref{s:Lax} we derive the Lax pair \tht{9}.
In \S\ref{s:Fredholm} we show how to express the Fredholm
determinants $D_s$ in terms of $(A_s,M_s)$. In
\S\ref{s:compatibility} we prove that the compatibility condition
\tht{2} always has a unique solution. In \S\ref{s:initial} we
derive the initial conditions $A_k,M_k,D_k,D_{k+1}$. In
\S\ref{s:Lax-Askey} we write down explicitly the Lax pairs for 14
families of discrete hypergeometric orthogonal polynomials of the
Askey scheme. In \S\ref{s:general} we solve \tht{2} in terms of rational recurrences for the matrix elements of $A_s$ and $M_s$ for $\deg M_s=1$.
In \S\ref{s:dP} we show how to reduce \tht{2} to
dPIV and dPV equations in the case when $\deg M_s=1$ and
$\sigma\zeta=\zeta-1$. In \S\ref{s:Jimbo-Sakai} we reduce \tht{2}
to $q$-PVI or its degeneration in the case when $\deg M_s=1$ and
$\sigma\zeta=q^{\pm1}\ze$. Finally, in \S\ref{s:Charlier} we solve
\tht{2} in terms of difference and $q$-Painlev\'e equations for 7
families of classical orthogonal polynomials. At the end of the
paper we also provide a few plots of the ``density function''
(difference or $q$-derivative of $D_s$) obtained using the formulas of \S\ref{s:Charlier}.

This research was partially conducted during the period the first
author (A.~B.) served as a Clay Mathematics Institute Long-Term
Prize Fellow. He was also partially supported by the NSF grant
DMS-9729992.

\subsection{} The following notation is used throughout our paper. For an integer $k$, we write
$\zpk=\{k,k+1,k+2,\dotsc\}$. If $a,q\in\bC$ and $k\in\zp$, one
defines the {\it Pochhammer symbol} and its $q$-analogue (often
also called the {\it $q$-shifted factorial}) by
\[
(a)_0:=1,\quad (a)_k:=a(a+1)(a+2)\dotsm(a+k-1)\ {\rm if}\ k\geq 1
\]
and
\[
(a;q)_0:=1,\quad (a;q)_k:=(1-a)(1-aq)(1-aq^2)\dotsm(1-aq^{k-1})\
{\rm if}\ k\geq 1,
\]
respectively. One usually writes
\[
(a_1,\dotsc,a_r)_k=\prod_{j=1}^r (a_j)_k  \quad\text{and}\quad
(a_1,\dotsc,a_r;q)_k=\prod_{j=1}^r (a_j;q)_k.
\]
If $r,s\in\zp$ and $a_1,\dotsc,a_r,b_1,\dotsc,b_s,z,q\in\bC$, the
{\it hypergeometric series} and the {\it basic hypergeometric
series} are defined by
\[
\hypergeom{r}{s}{a_1,\dotsc,a_r}{b_1,\dotsc,b_s}{z}:=\sum_{k=0}^\infty
\frac{(a_1,\dotsc,a_r)_k}{(b_1,\dotsc,b_s)_k}\frac{z^k}{k!}
\]
and
\[
\qhypergeom{r}{s}{a_1,\dotsc,a_r}{b_1,\dotsc,b_s}{z}:=\sum_{k=0}^\infty
\frac{(a_1,\dotsc,a_r;q)_k}{(b_1,\dotsc,b_s;q)_k}
(-1)^{(1+s-r)k}q^{(1+s-r){k\choose 2}} \frac{z^k}{(q;q)_k},
\]
respectively.

\section{Discrete Riemann-Hilbert Problems and Orthogonal
Polynomials}\label{s:DRHP}

\subsection{} In this section we explain how solutions of discrete
Riemann-Hilbert problems (DRHP) for jump matrices of a special
type can be expressed in terms of the corresponding orthogonal
polynomials. Let $\fX$ be a discrete locally finite subset of
$\bC$, and let $w:\fX\rightarrow Mat(2,\bC)$ be a function. As in
\cite{IMRN}, \cite{B}, we say that an analytic function
\[
m:\bC\setminus\fX\longrightarrow Mat(2,\bC)
\]
solves the DRHP $(\fX,w)$ if $m$ has simple poles at the points of
$\fX$ and its residues at these points are given by the {\it jump} (or {\it residue})
{\it condition}
\begin{equation}\label{res}
\Res{\ze=x} m(\ze) = \lim\limits_{\ze\to x} (m(\ze)w(x)),\quad
x\in\fX.
\end{equation}
\begin{lem}\label{determinants}
If $m(\ze)$ is a solution of the DRHP $(\fX,w)$ and the matrix
$w(x)$ is nilpotent for all $x\in\fX$, then the function $\det
m(\ze)$ is entire. If, in addition, $\det m(\ze)\to 1$ as
$\ze\to\infty$, then $\det m(\ze)\equiv 1$.
\end{lem}
\begin{proof}
For each $x\in\fX$, the jump condition \eq{res} implies that the
function $m(\ze)\cdot\bigl(I-(\ze-x)^{-1}w(x)\bigr)$ is analytic
in a neighborhood of $x$. Since $w(x)$ is nilpotent, this product
has the same determinant as $m(\ze)$, which shows that $\det
m(\ze)$ has no pole at $x$. The second statement of the lemma
follows from Liouville's theorem.
\end{proof}

\subsection{} We now assume that the matrix $w(x)$ has the following form:
\begin{equation}\label{jump}
w(x)=\matr{0}{\om(x)}{0}{0},
\end{equation}
where $\om:\fX\rightarrow\bC$ is a function. Recall that a
collection $\{P_n(\zeta)\}_{n=0}^\infty$ of complex polynomials is
called the collection of {\em orthogonal polynomials associated to
the weight function} $\om$ if
\begin{itemize}
\item $P_n$ is a polynomial of degree $n$ for all
$n=1,2,\dotsc$, and $P_0\equiv const$;
\item if $m\neq n$, then
\[
\sum_{x\in\fX} P_m(x) P_n(x) \om(x) = 0.
\]
\end{itemize}

We will always take $P_n$ to be {\it monic\/}: $P_n(x)=x^n+\text{ lower terms }$.

In order for the definition to make sense, we assume that all
moments of the weight function $\om$ are finite, i.e.,
\begin{equation}\label{moments}
\text{the series } \sum_{x\in\fX} \abs{\om(x) x^j} \ {\rm
converges\ for\ all\ } j\geq 0.
\end{equation}
Under this condition, one can consider the following inner product
on the space $\bC[\ze]$ of all complex polynomials:
\[
(f(\ze),g(\ze))_\om:=\sum_{x\in\fX} f(x) g(x) \om(x).
\]
It is clear that there exists a collection of orthogonal
polynomials $\{P_n(\ze)\}$ associated to $\om$ such that
$(P_n,P_n)_\om\neq 0$ for all $n$ if and only if the restriction
of $(\cdot,\cdot)_\om$ to the space $\bC[\ze]^{\leq d}$ of
polynomials of degree at most $d$ is nondegenerate for all $d\geq
0$. If this condition holds, we say that the
weight function $\om$ is {\it nondegenerate}. In this case, it is also
clear that the collection $\{P_n\}$ is unique.

\begin{rem}\label{r:finite}
If the set $\fX$ is finite, one has to modify the definitions
above. Indeed, if $\fX$ consists of $N+1$ points ($N\in\zp$), the
inner product $(\cdot,\cdot)_\om$ is necessarily degenerate on
$\bC[\ze]^{\leq d}$ for all $d>N$. So, instead, we require that
$(\cdot,\cdot)_\om$ be nondegenerate on $\bC[\ze]^{\leq d}$ for
$0\leq d\leq N$, and we are only interested in a collection
$\{P_n(\ze)\}_{n=0}^N$ of orthogonal polynomials of degrees up to
$N$. On the other hand, the condition \eq{moments} is empty in
this case.
\end{rem}

\begin{rem}\label{r:positive}
If the values of the weight function $\om$ are real and strictly
positive and the orthogonality set $\fX$ is contained in $\bR$,
then $\om$ is automatically nondegenerate.
\end{rem}

\subsection{} The first basic result of the paper is
\begin{thm}[Solution of DRHP]\label{t:DRHP}
Let $\fX$ be a discrete locally finite subset of $\bC$ and
$\om:\fX\rightarrow\bC$ be a nondegenerate weight function satisfying
\eq{moments}. Let $\{P_n(\ze)\}_{n=0}^N$ be the collection of monic
orthogonal polynomials associated to $\om$, where
$N=\mathrm{card}(\fX)-1\in\zp\cup\{\infty\}$. Assume that the jump
matrix $w(x)$ is given by \eq{jump}. Then for any
$k=1,2,\dotsc,N$, the DRHP $(\fX,w)$ has a unique solutions
$\mx(\ze)$ satisfying the asymptotic condition
\begin{equation}\label{asymp}
\mx(\ze)\cdot\matr{\ze^{-k}}{0}{0}{\ze^k} = I +
O\biggl(\frac{1}{\ze}\biggr) \quad {\rm as}\ \ze\to\infty,
\end{equation}
where $I$ is the identity matrix. If we write
\[
\mx(\ze)=\matr{\mx^{11}(\ze)}{\mx^{12}(\ze)}{\mx^{21}(\ze)}{\mx^{22}(\ze)},
\]
then $\mx^{11}(\ze)=P_k(\ze)$ and
$\mx^{21}(\ze)=(P_{k-1},P_{k-1})_\om^{-1}\cdot P_{k-1}(\ze)$.
\end{thm}

\begin{rem}\label{r:precise}
The asymptotic condition \eq{asymp} needs to be made more precise.
Indeed, if $\fX$ is infinite, then the LHS of \eq{asymp} has poles
accumulating at infinity. In this case, we require that the
asymptotics be uniform on a sequence of expanding contours (e.g.,
circles whose radii tend to $\infty$) whose distance from $\fX$
remains bounded away from zero. A similar remark applies to all
asymptotic formulas below.
\end{rem}

\begin{rem}\label{r:continuous}
A result for continuous weight functions similar to Theorem
\ref{t:DRHP} was proved in \cite{FIK}.
\end{rem}

\begin{proof}
Fix a natural number $k\leq N$ and define a matrix-valued function
\[
m=\matr{m^{11}}{m^{12}}{m^{21}}{m^{22}} : \bC \rar{} Mat(2,\bC)
\]
by
\begin{eqnarray*}
m^{11}(\ze)&=&P_k(\ze), \\
m^{12}(\ze)&=&\sum_{x\in\fX}\frac{P_k(x)\om(x)}{\ze-x}, \\
m^{21}(\ze)&=&c\cdot P_{k-1}(\ze), \\
m^{22}(\ze)&=&c\cdot\sum_{x\in\fX}\frac{P_{k-1}(x)\om(x)}{\ze-x},
\end{eqnarray*}
where $c$ is the unique constant for which
$m^{22}(\ze)=\ze^{-k}+O(\ze^{-k-1})$ as $\ze\to\infty$ (we will
show below that such a $c$ exists, and in fact
$c=(P_{k-1},P_{k-1})_\om^{-1}$). We first observe that
\eq{moments} implies that the series for $m^{12}(\ze)$ converges
uniformly and absolutely on compact subsets of $\bC\setminus\fX$,
and hence $m^{12}(\ze)$ is analytic on the complement of $\fX$.
Moreover, since for each $x\in\fX$, the series
\[
\sum_{y\in\fX\setminus\{x\}} \frac{P_k(y)\om(y)}{\ze-y}
\]
converges uniformly and absolutely on compact subsets of
$(\bC\setminus\fX)\cup\{x\}$, we see that $m^{12}(\ze)$ has a
simple pole at $x$, with the residue given by
\[
\Res{\ze=x} m^{12}(\ze) = P_k(x)\om(x).
\]
Similarly, $m^{22}(\ze)$ is analytic away from $\fX$, with a
simple pole at each $x\in\fX$ satisfying
\[
\Res{\ze=x} m^{22}(\ze) = c\cdot P_{k-1}(x)\om(x).
\]
This shows that our matrix $m(\ze)$ satisfies the jump condition
\eq{res}. To verify the asymptotic condition (and find the
constant $c$), note that
\[
m^{11}(\ze)\cdot\ze^{-k}=1+O(1/\ze),\quad
m^{21}(\ze)\cdot\ze^{-k}=O(1/\ze),\quad {\rm as}\ \ze\to\infty.
\]
Next, we write
\begin{equation}\label{star}
\frac{1}{\ze-x}=\frac{1}{\ze}\cdot\left(1-\frac{x}{\ze}\right)^{-1}=
\frac{1}{\ze}+\frac{x}{\ze^2}+\dotsb+\frac{x^{k-1}}{\ze^k}+O(\ze^{-k-1}).
\end{equation}
By the definition of orthogonal polynomials, we have
\[
\sum_{x\in\fX} P_k(x)\om(x)x^i=0,\quad 0\leq i\leq k-1,
\]
and hence substituting \eq{star} into the definition of
$m^{12}(\ze)$ yields
\[
m^{12}(\ze)=O(\ze^{-k-1})\quad {\rm as}\ \ze\to\infty.
\]
Also, \eq{star} gives
\[
m^{22}(\ze)=c\cdot\sum_{x\in\fX} P_{k-1}(x) \om(x)
x^{k-1}\cdot\ze^{-k} + O(\ze^{-k-1})\quad {\rm as}\ \ze\to\infty.
\]
Hence, if we set
\[
c=\left(\sum_{x\in\fX} P_{k-1}(x) \om(x)
x^{k-1}\right)^{-1}=(P_{k-1},P_{k-1})_\om^{-1},
\]
then the matrix $m(\ze)$ satisfies all the conditions of the
theorem.

To prove uniqueness, let $m_\fX$ be an arbitrary solution of the
DRHP $(\fX,w)$ satisfying the asymptotic condition \eq{asymp}.
Since the functions $m$ and $m_\fX$ have the same (simple) poles
and satisfy the same residue conditions at these poles, it is
clear that the function $f(\ze)=m(\ze)^{-1}\cdot\mx(\ze)$ is
entire (note that $m(\ze)$ is invertible by Lemma
\ref{determinants}). The asymptotic conditions on the matrices $m$
and $\mx$ imply that $f(\ze)\to I$ as $\ze\to\infty$. By
Liouville's theorem, $f\equiv I$, which means that $m=\mx$.
\end{proof}

\section{Lax pairs for solutions of DRHP}\label{s:Lax}

\subsection{} Let $\fX$ be a discrete locally finite subset of $\bC$, let
$w:\fX\rightarrow Mat(2,\bC)$ be a function, and fix a natural
number $k<\mathrm{card}(\fX)$. If $w$ arises from a nondegenerate
weight function $\om$ on $\fX$ with finite moments, as in
\S\ref{s:DRHP}, then from Theorem \ref{t:DRHP} we know that the
DRHP $(\fX,w)$ admits a unique solution $\mx(\ze)$ with the
asymptotics of $\mathrm{diag}(\ze^k,\ze^{-k})$ at infinity.

\smallbreak

\noindent
{\sc Convention.} From now on, we assume that the weight function
under consideration is everywhere real and strictly positive, and
the orthogonality set is contained in $\bR$.

\smallbreak

If $\fZ\subseteq\fX$ is any subset of cardinality $>k$, it follows
from Remark \ref{r:positive} and our convention that the
restriction of $\om$ to $\fZ$ is also nondegenerate, whence by
Theorem \ref{t:DRHP}, the DRHP $(\fZ,w\big\vert_\fZ)$ has a unique
solution $m_\fZ(\ze)$ such that
$m_\fZ(\ze)\cdot\Bigl(\begin{smallmatrix} \ze^{-k} & 0 \\ 0 &
\ze^k \end{smallmatrix} \Bigr)\to I$ as $\ze\to\infty$.

\medbreak

Let us now assume that the set $\fX$ is parameterized as
$\fX=\{\pi_x\}_{x=0}^N$, where $N\in\zp\cup\{\infty\}$. For every
$s\geq 0$, we consider the subset $\fZ_s:=\{\pi_x\}_{x\leq
s-1}\subseteq\fX$. If $s>k$, then $\mathrm{card}(\fZ_s)>k$, so by
the previous paragraph, we have the corresponding solution
$m_s(\ze):=m_{\fZ_s}(\ze)$. It can also be shown that even though
$\mathrm{card}(\fZ_k)=k$, the DRHP $(\fZ_k,w\big\vert_{\fZ_k})$
still has a unique solution $m_k(\ze)$. Indeed, uniqueness can be
proved by the same argument as in Theorem \ref{t:DRHP}, and in
Proposition \ref{mk-init} below we give an explicit formula for
$m_k(\ze)$. Note also that $m_s(\ze)$ is defined for all $s\geq
k$; in particular, if $N$ is finite, we have
$m_s(\ze)=m_{\fX}(\ze)$ for all $s>N$.

\medbreak

Our next basic assumption is:
\begin{equation}\label{e:sigma}
\text{there exists an affine transformation } \sg:\bC\rar{}\bC
\text{ such that } \sg\pi_{x+1}=\pi_x \text{ for all } 0\leq x<N.
\end{equation}
The {\it Lax pair} in our setting will consist of two equations,
one of which relates $m_{s+1}(\ze)$ with $m_s(\ze)$ and the other
one relates $m_s(\sg\ze)$ with $m_{s+1}(\ze)$. We will denote the
derivative of $\sg$ (which is a constant) by $\eta$, so that
\begin{equation}\label{e:eta}
\sg(\ze_1)-\sg(\ze_2)=\eta\cdot(\ze_1-\ze_2)\quad \text{for all }
\ze_1,\ze_2\in\bC.
\end{equation}
The cases of special interest are those when $\fX$ is the
orthogonality set for one of the families of discrete
hypergeometric orthogonal polynomials of the Askey scheme. In this
situation, $\fX$ is a subset of either a one-dimensional lattice
or a one-dimensional $q$-lattice in $\bC$, and $\sg$ is given by
either $\sg\ze=\ze-1$ or $\sg\ze=q^{\pm 1}\ze$. These cases will
be treated in greater detail in \S\ref{s:Lax-Askey}; for now we
concentrate on the general theory.

\medbreak

\subsection{} The main result of this section is
\begin{thm}[Lax pair]\label{t:Lax}
For each $s=k,k+1,k+2,\dotsc$, let $\fZ_s=\{\pi_x\in\fX\big\vert
x\leq s-1\}$, and let $m_s(\ze)$ be the unique solution of the
DRHP $(\fZ_s,w\big\vert_{\fZ_s})$ such that
$m_s(\ze)\cdot\Bigl(\begin{smallmatrix} \ze^{-k} & 0 \\ 0 & \ze^k
\end{smallmatrix} \Bigr)\to I$ as $\ze\to\infty$, where $w$ is
given by\footnote{To simplify notation, we assume that the weight
function $\om(x)$ is always defined for $x\in\zp$, $x\leq N$.}
\[
w(\pi_x)=\matr{0}{\om(x)}{0}{0}.
\]
\begin{enumerate}[(a)]
\item For each $s\in\zpk$, $s\leq N$, there exists a constant nilpotent
matrix $A_s$ such that
\begin{equation}\label{e:Lax1}
m_{s+1}(\ze)=\left(I+\frac{A_s}{\ze-\pi_s}\right) m_s(\ze).
\end{equation}
\item Assume that there exist entire functions $d_1(\ze)$,
$d_2(\ze)$ such that
\[
\frac{\om(x-1)}{\om(x)}=\eta\cdot\frac{d_1(\pi_x)}{d_2(\pi_x)}\quad\mathrm{for\
all}\ 1\leq x\leq N,
\]
\[
d_1(\pi_0)=0,\quad\mathrm{and}\quad d_2(\sg^{-1}\pi_N)=0\
\mathrm{if}\ N\ \mathrm{is\ finite}.
\]
Let
\[
D(\ze)=\matr{d_1(\ze)}{0}{0}{d_2(\ze)}.
\]
Then for every $s\in\zpk$, we have
\begin{equation}\label{e:Lax2}
m_s(\sg\ze) = M_s(\ze) m_{s+1}(\ze) D^{-1}(\ze),
\end{equation}
where $M_s(\ze)$ is entire\footnote{Again, note that $M_s(\ze)$ is
defined for \textit{all} $s\in\zpk$, in particular for $s>N$, even
if $N$ is finite.}.
\item With the assumptions of part (b), suppose that the functions
$d_1(\ze)$ and $d_2(\ze)$ are polynomials of degree at most $n$ in
$\ze$, and write $d_1(\ze)=\la_1\ze^n+(\mathrm{lower\ terms})$,
$d_2(\ze)=\la_2\ze^n+(\mathrm{lower\ terms})$. Set
$\ka_1=\eta^k\la_1$, $\ka_2=\eta^{-k}\la_2$. Then the matrix
$M_s(\ze)$ is polynomial of degree at most $n$ in $\ze$, with the
coefficient of $\ze^n$ equal to $\mathrm{diag}(\ka_1,\ka_2)$.
\end{enumerate}
\end{thm}
Equations \eq{e:Lax1} and \eq{e:Lax2} constitute the Lax pair.
\begin{rem}\label{r:d2Nfinite}
As follows from the proof below, the condition
$d_2(\sg^{-1}\pi_N)=0$ in part (b) of the theorem is only required
to assert that $M_s(\ze)$ is entire for $s>N$. If
$d_2(\sg^{-1}\pi_N)\neq 0$, then (b) and (c) still hold for
all $s,\, k\leq s\leq N$.
\end{rem}
\begin{proof}
(a) Fix $s\in\zp$, $s\leq N$, and consider the matrix-valued
function $N(\ze):=m_{s+1}(\ze)m_s^{-1}(\ze)$ (recall that
$m_s(\ze)$ is invertible by Lemma \ref{determinants}). It is clear
that $N(\ze)$ has only one simple pole, at $\ze=\pi_s$. Hence the
function $N(\ze)-(\ze-\pi_s)^{-1} A_s$ is entire, where
$A_s=\mathrm{Res}_{\ze=\pi_s}N(\ze)$. But $m_s(\ze)$ and
$m_{s+1}(\ze)$ have the same asymptotics at infinity, so
$N(\ze)-(\ze-\pi_s)^{-1} A_s\to I$ as $\ze\to\infty$. By
Liouville's theorem, $N(\ze)-(\ze-\pi_s)^{-1} A_s\equiv I$, which
gives \eq{e:Lax1}. Taking determinants of both sides of
\eq{e:Lax1} and using Lemma \ref{determinants} gives
$\det\bigl(I+(\ze-\pi_s)^{-1} A_s\bigr)=1$ for all $\ze$, which
forces the matrix $A_s$ to be nilpotent.

\noindent
(b) We have $M_s(\ze)=m_s(\sg\ze)D(\ze)m_{s+1}^{-1}(\ze)$.
Therefore it is clear that $M_s(\ze)$ is analytic away from
$\{\pi_0,\pi_1,\dotsc,\pi_s\}$. Let $1\leq x\leq s$. Then for
$\ze$ near $\pi_x$, we can write
\[
m_s(\sg\ze)=H_1(\ze)\left(I+\frac{w(x-1)}{\sg\ze-\pi_{x-1}}\right)\quad
{\rm and}\quad
m_{s+1}^{-1}(\ze)=\left(I-\frac{w(x)}{\ze-\pi_x}\right)H_2(\ze),
\]
where $H_1$ and $H_2$ are analytic and invertible matrix-valued
functions defined in a neighborhood of $\pi_x$. So $M_s(\ze)$ is
analytic near $\pi_x$ if and only if so is the product
\[
\left(I+\frac{w(x-1)}{\sg\ze-\pi_{x-1}}\right) D(\ze)
\left(I-\frac{w(x)}{\ze-\pi_x}\right) =
\matr{d_1(\ze)}{\dfrac{\eta^{-1}d_2(\ze)\om(x-1)-d_1(\ze)\om(x)}{\ze-\pi_x}}{0}{d_2(\ze)},
\]
i.e., if and only if
\[
\frac{\om(x-1)}{\om(x)}=\eta\cdot\frac{d_1(\pi_x)}{d_2(\pi_x)}.
\]
Similarly, we note that $m_s(\sg\ze)$ is analytic near $\pi_0$,
whereas
\[
m_{s+1}^{-1}(\ze)=\left(I-\frac{w(0)}{\ze-\pi_0}\right)H(\ze),
\]
with $H$ analytic and invertible near $\pi_0$, which implies that
$M_s(\ze)$ is analytic near $\pi_0$ if and only if $d_1(\pi_0)=0$.
Finally, if $N$ is finite, we have to make sure that $M_s(\ze)$
has no pole at $\sg^{-1}\pi_N$ for $s\geq N+1$. A necessary and
sufficient condition for that is $d_2(\sg^{-1}\pi_N)=0$.

\noindent (c) Using the asymptotic conditions on the matrices $m_s(\sg\ze)$ and
$m_{s+1}(\ze)$, we obtain, as $\ze\to\infty$,
\begin{eqnarray*}
M_s(\ze)&=&m_s(\sg\ze) D(\ze) m_{s+1}^{-1}(\ze) \\
&=&\biggl(I+O\Bigl(\frac{1}{\ze}\Bigr)\biggr)
\matr{\eta^k\ze^k}{0}{0}{\eta^{-k}\ze^{-k}}
\matr{\la_1\ze^n+O(\ze^{n-1})}{0}{0}{\la_2\ze^n+O(\ze^{n-1})}
\matr{\ze^{-k}}{0}{0}{\ze^k}\biggl(I+O\Bigl(\frac{1}{\ze}\Bigl)\biggl) \\
&=& \diag{\ka_1}{\ka_2}\cdot\ze^n + O(\ze^{n-1}).
\end{eqnarray*}
Since $M_s(\ze)$ is entire by part (b), this completes the proof.
\end{proof}

\subsection{} To conclude this section, we remark that the method by which we
have obtained the second equation of the Lax pair can be applied
to derive a (second-order) difference equation for the orthogonal
polynomials corresponding to the weight function $\om$. More
precisely, one proves, by the same argument as Theorem
\ref{t:Lax}, the following
\begin{prop}\label{p:Lax}
Under the assumptions of Theorem \ref{t:Lax}(b), the function
$M(\ze)=\mx(\sg\ze)D(\ze)\mx^{-1}(\ze)$ is entire. The analogue of
Theorem \ref{t:Lax}(c) also holds for the matrix $M(\ze)$.
\end{prop}
Now we have $\mx(\sg\ze)=M(\ze)\mx(\ze)D(\ze)^{-1}$. Considering
the $(1,1)$ and $(2,1)$ elements of both sides and using the
explicit formula for $\mx(\ze)$ given in Theorem \ref{t:DRHP}, we
obtain a system of equations of the form
\begin{eqnarray}
\label{eq1} P_k(\sg\ze) &=& \bigl(M^{11}(\ze) P_k(\ze) + c
M^{12}(\ze)
P_{k-1}(\ze)\bigr) d_1(\ze)^{-1}, \\
\label{eq2} c\cdot P_{k-1}(\sg\ze) &=& \bigl(M^{21}(\ze) P_k(\ze)
+ c M^{22}(\ze) P_{k-1}(\ze)\bigr) d_1(\ze)^{-1},
\end{eqnarray}
where $c=(P_{k-1},P_{k-1})_\om^{-1}$ and the $M^{ij}(\ze)$ are the
elements of the matrix $M(\ze)$. If we find $c\cdot P_{k-1}(\ze)$
from the first equation and substitute it into the second one, we
will get a relation between $P_k(\ze)$, $P_k(\sg\ze)$ and
$P_k(\sg^2\ze)$. Even though the functions $M^{ij}(\ze)$ may
involve unknown parameters, one might hope that the equation we
obtain in the end will only involve known parameters. We will work
out an explicit example in subsection \ref{ss:charlier-diff}, for
the weight function corresponding to the Charlier polynomials. We
will see that the relation we get is exactly the standard
difference equation for Charlier polynomials.

\section{Recurrence relation for Fredholm
determinants}\label{s:Fredholm}

\subsection{}\label{ss:fr-1} We remain in the general setting of \S\ref{s:Lax}.
Thus we consider a discrete locally finite subset
$\fX=\{\pi_x\}_{x=0}^N\subseteq\bR$, where
$N\in\zp\cup\{\infty\}$, and we are also given a strictly positive
weight function $\om:\{x\in\zp\big\vert x\leq N\}\rightarrow\bC$
whose moments are finite. For all $s\in\zp$, $s\leq N$, we let
\[
\fZ_s=\{\pi_x\}_{x=0}^{s-1},\quad \fY_s=\fX\setminus\fZ_s.
\]
Finally, we fix a natural number $k\leq N$ and let
\[
\mx(\ze)=\matr{\mx^{11}(\ze)}{\mx^{12}(\ze)}{\mx^{21}(\ze)}{\mx^{22}(\ze)}
\]
be the unique solution of the discrete Riemann-Hilbert problem
$(\fX,w)$ with the asymptotics of $\mathrm{diag}(\ze^k,\ze^{-k})$
at infinity provided by Theorem \ref{t:DRHP}.

\smallbreak

Let $\al,\be:\{x\in\zp\big\vert x\leq N\}\rightarrow\bC$ be two
functions such that $\al(x)\be(x)=\om(x)$ for all $x$. 
Consider the following kernel on $\fX\times\fX$:
\begin{equation}\label{eq:kernel}
K(\pi_x,\pi_y)=\left\{
\begin{array}{ll}
\al(x)\be(y)
\dfrac{\phi(\pi_x)\psi(\pi_y)-\psi(\pi_x)\phi(\pi_y)}{\pi_x-\pi_y},
& x\neq y, \\
\al(x)\be(x)
\bigl(\phi'(\pi_x)\psi(\pi_x)-\psi'(\pi_x)\phi(\pi_x)\bigr), &
x=y,
\end{array}\right.
\end{equation}
where $\phi(\ze)=\mx^{11}(\ze)=P_k(\ze)$ and
$\psi(\ze)=\mx^{21}(\ze)=(P_{k-1},P_{k-1})_\om^{-1}\cdot
P_{k-1}(\ze)$. Up to the factor of $\al(x)\be(y)$, this is
precisely the Christoffel-Darboux kernel for the family of
orthogonal polynomials corresponding to the weight function $\om$.
We will not need the specific form of the functions $\al$ and
$\be$ in our computations. Note also that changing $\al$ and $\be$
while keeping their product fixed results in conjugation of the
kernel $K$ and hence has no effect on the Fredholm determinants
studied below.

\subsection{}\label{ss:fr-2} For each $s\in\zpk$, $s\leq N$, we
let $K_s$ be the restriction of $K$ to $\fY_s\times\fY_s$. We
also denote by $K$ and $K_s$ the
operators on $l^2(\fX)$ and $l^2(\fY_s)$ defined by the kernels
$K$ and $K_s$, respectively. The main goal of our paper is to
derive a recurrence relation for the Fredholm determinants
\[
D_s=\det(1-K_s),\quad s\in\zpk,\ s\leq N.
\]
The resulting equation will be in terms of the elements of the
matrices $A_s$ and $M_s(\ze)$ from the Lax pair \eq{e:Lax1},
\eq{e:Lax2}. Note that the Fredholm determinants are always well
defined because $K$ is a finite rank operator, as follows from the
Christoffel-Darboux formula (see, e.g., \cite{S}):
\[
\frac{K(\pi_x,\pi_y)}{\al(x)\be(y)}=\sum_{m=0}^{k-1}
\frac{P_m(\pi_x)P_m(\pi_y)}{(P_m,P_m)_\om}.
\]
For a probabilistic interpretation of these determinants, see subsection
\ref{ss:fr-5} below.
\begin{lem}\label{l:invertible}
For each $s\in\zpk$, $s\leq N$, the operator $1-K_s$ is
invertible, and $D_s=\det(1-K_s)\ne 0$.
\end{lem}
\begin{proof}
The operator $K$ is the orthogonal projection onto the subspace of
$l^2(\fX)$ spanned by the polynomials $P_0,\dotsc,P_{k-1}$. In
particular, it has finite rank, its only eigenvalues are $0$ and
$1$, and the eigenvectors corresponding to the eigenvalue $1$ are
linear combinations of $P_0,\dotsc,P_{k-1}$, i.e., the polynomials
of degree $\leq k-1$. It follows that the operator $K_s$ also has
finite rank, hence $1-K_s$ is not invertible if and only if $1$ is
an eigenvalue of $K_s$. Suppose that $k\leq s\leq N$ and there
exists $f\in l^2(\fY_s)$ such that $K_s f = f$. We extend $f$ by
zero to $\fX$, and we denote the extension by $\tilde{f}\in
l^2(\fX)$. Now $K\tilde{f}\big\vert_{\fY_s}=K_s
f=f=\tilde{f}\big\vert_{\fY_s}$ and
$\tilde{f}\big\vert_{\fZ_s}=0$. This implies that
$K\tilde{f}\big\vert_{\fZ_s}=0$ because otherwise we would have
$\norm{K\tilde{f}}_\om>\norm{\tilde{f}}_\om$, which is impossible
because $K$ is a projection operator. This shows that
$K\tilde{f}=\tilde{f}$, whence by the remarks above, $\tilde{f}$
is a polynomial of degree $\leq k-1$. But it vanishes on the set
$\fZ_s$ of cardinality $\geq k$, which implies that $\tilde{f}$ is identically equal to zero. Hence, $1-K_s$ is an invertible operator, and $\det(1-K_s)\ne 0$.
\end{proof}

\subsection{}\label{ss:fr-3} For each $s\in\zpk$, $s\leq N$, we have the
unique solution $m_s(\ze)$ of the DRHP $(\fZ_s,w\big\vert_{\fZ_s})$ having the
same asymptotics at infinity as $\mx(\ze)$. As in \S\ref{s:Lax},
we assume that there exist entire functions $d_1(\ze)$, $d_2(\ze)$
such that
\begin{equation}\label{d1d2}
d_1(\pi_0)=0 \quad \text{and} \quad
\frac{\om(x-1)}{\om(x)}=\eta\cdot\frac{d_1(\pi_x)}{d_2(\pi_x)}\
\mathrm{for\ all}\ 1\leq x\leq N.
\end{equation}
The assumption that $d_2(\sg^{-1}\pi_N)=0$ if $N$ is finite is not
essential for us now since it was only used in \S\ref{s:Lax} to
show that the function $M_s(\ze)$ is entire for $s>N$, whereas
here we are only interested in the case $k\leq s\leq N$ (cf.
Remark \ref{r:d2Nfinite}). Note that $d_1(\pi_x) d_2(\pi_x)\neq 0$
for all $1\leq x\leq N$. We let
\[
D(\ze)=\diag{d_1(\ze)}{d_2(\ze)}\quad\mathrm{and}\quad
M_s(\ze)=m_s(\sg\ze) D(\ze) m_{s+1}^{-1}(\ze).
\]
By part (b) of Theorem \ref{t:Lax}, the function $M_s(\ze)$ is
entire for all $k\leq s\leq N$. Using part (a) of this theorem, we
obtain the following general form of the Lax pair for the
solutions $m_s(\ze)$:
\begin{equation}\label{lax1}
m_{s+1}(\ze)=\left(I+\frac{A_s}{\ze-\pi_s}\right) m_s(\ze), \quad
A_s=\matr{p_s}{q_s}{r_s}{-p_s},\ p_s^2=-q_s r_s;
\end{equation}
\begin{equation}\label{lax2}
m_s(\sg\ze)=M_s(\ze) m_{s+1}(\ze) D(\ze)^{-1}.
\end{equation}
In \S\ref{s:compatibility}, we will explain how one can obtain a
recurrence relation for the matrix elements of $M_s(\ze)$ and
$A_s$. In the present section, we assume that the function
$M_s(\ze)$ is known, and derive a recurrence relation for the
Fredholm determinants using this function and the parameters
$p_s$, $q_s$.

For all $s,\,k\leq s\leq N$, we put
\begin{equation}\label{matrix-elements}
m_s(\pi_s)=\matr{m_s^{11}}{m_s^{12}}{m_s^{21}}{m_s^{22}},\quad
M_s(\pi_{s+1})=\matr{\mu_s^{11}}{\mu_s^{12}}{\mu_s^{21}}{\mu_s^{22}},\quad
M'_s(\pi_{s+1})=\matr{\nu_s^{11}}{\nu_s^{12}}{\nu_s^{21}}{\nu_s^{22}},
\end{equation}
where $M_s'(\ze)=\frac{d}{d\ze} M_s(\ze)$. Then we have the
following recurrence relations for the matrix elements $m_s^{11}$
and for the Fredholm determinants $D_s$ in terms of the parameters
$p_s$, $q_s$, $\mu_s^{ij}$, $\nu_s^{ij}$.
\begin{thm}[Recurrence relation for Fredholm determinants]\label{t:Fredholm}
Assume that $p_s\neq 0$ for all $s\geq k$, and hence also
$q_s,r_s\neq 0$ for $s\geq k$.
\begin{enumerate}[(a)]
\item For each $s\in\zpk$, $s\leq N$, we have
\begin{equation}\label{ms11}
m_{s+1}^{11}=d_2(\pi_{s+1})^{-1}\left(\mu_s^{22}+\frac{p_s}{q_s}\mu_s^{12}\right)m_s^{11}.
\end{equation}
\item For each $s\in\zpk$, $s\leq N-2$, we have
\begin{equation}\label{fr}
\frac{D_{s+2}}{D_{s+1}}-\frac{D_{s+1}}{D_s}= \frac{\om(s)\cdot
u_s\cdot (m_s^{11})^2}{\eta\cdot d_1(\pi_{s+1})\cdot
d_2(\pi_{s+1})},
\end{equation}
where
\[
u_s=(\nu_s^{21}\mu_s^{22}-\nu_s^{22}\mu_s^{21}) +
\frac{p_s}{q_s}\cdot
(\nu_s^{21}\mu_s^{12}-\nu_s^{12}\mu_s^{21}+\nu_s^{11}\mu_s^{22}-\nu_s^{22}\mu_s^{11})
+ \frac{p_s^2}{q_s^2}\cdot
(\nu_s^{11}\mu_s^{12}-\nu_s^{12}\mu_s^{11})
\]
for all $s$.
\end{enumerate}
\end{thm}
\begin{rem}\label{r:psnotzero}
We will show later (see Proposition \ref{p:ps-nonzero} and Remark
\ref{r:pk-nonzero}) that under the assumptions of subsection \ref{assumps}, $p_s$ is nonzero for all $s\geq
k$.
\end{rem}
\begin{proof}
(a) First we take the residues of both sides of \eq{lax1} at
$\ze=\pi_s$ and use the jump condition on the LHS:
\[
\lim\limits_{\ze\to\pi_s} m_{s+1}(\ze) w(\pi_s) = A_s m_s(\pi_s).
\]
Since the first column of the matrix $w(\pi_s)$ is zero, we deduce
that the first column of the matrix $A_s m_s(\pi_s)$ is zero,
whence $m_s^{21}=-(p_s/q_s) m_s^{11}$. Next we substitute
$\ze=\pi_{s+1}$ in \eq{lax2} and rewrite it as follows:
\[
m_{s+1}(\pi_{s+1})=M_s^{-1}(\pi_{s+1}) m_s(\pi_s) D(\pi_{s+1}).
\]
Since $\det M_s(\ze)=\det D(\ze)$ by Lemma \ref{determinants}, the
last equation can be written explicitly as
\[
\matr{m_{s+1}^{11}}{m_{s+1}^{12}}{m_{s+1}^{21}}{m_{s+1}^{22}}=
\frac{1}{d_1(\pi_{s+1}) d_2(\pi_{s+1})}
\matr{\mu_s^{22}}{-\mu_s^{12}}{-\mu_s^{21}}{\mu_s^{11}}
\matr{m_s^{11}}{m_s^{12}}{m_s^{21}}{m_s^{22}}
\diag{d_1(\pi_{s+1})}{d_2(\pi_{s+1})}.
\]
Equating the $(1,1)$ and $(2,1)$ elements of both sides yields
\[
m_{s+1}^{11}=d_2(\pi_{s+1})^{-1}(\mu_s^{22}m_s^{11}-\mu_s^{12}m_s^{21})
=d_2(\pi_{s+1})^{-1}\left(\mu_s^{22}+\frac{p_s}{q_s}\mu_s^{12}\right)m_s^{11}
\]
and
\[
m_{s+1}^{21}=d_2(\pi_{s+1})^{-1}(-\mu_s^{21}m_s^{11}+\mu_s^{11}m_s^{21})
=-d_2(\pi_{s+1})^{-1}\left(\mu_s^{21}+\frac{p_s}{q_s}\mu_s^{11}\right)m_s^{11}.
\]

\smallbreak

\noindent
(b) For all $s\in\zpk$, $s\leq N$, we put $R_s=K_s(1-K_s)^{-1}$
(note that this is well defined by Lemma \ref{l:invertible}), so
that $R_s+1=(1-K_s)^{-1}$, and hence
$(R_s+1)(s,s)=\det(1-K_{s+1})/\det(1-K_s)$, i.e.,
\begin{equation}\label{ds}
R_s(s,s)=\frac{D_{s+1}}{D_s}-1.
\end{equation}
Now Theorem 2.3(ii) in \cite{B} gives (Situation 2.2 ibid.
explains why this theorem is applicable here)
\[
R_s(s,s)=g^t(s) m_s^{-1}(\pi_s) m_s'(\pi_s) f(s),
\]
where $f(s)=(\al(s),0)^t$ and $g(s)=(0,-\be(s))^t$. For the
purpose of our calculation, we rewrite this formula as follows:
\begin{equation}\label{rs}
R_s(s,s)=-\om(s)\cdot e_2^t m_s^{-1}(\pi_s) m_s'(\pi_s) e_1,
\end{equation}
where $e_1=(1,0)^t$ and $e_2=(0,1)^t$. Similarly, we get
\begin{equation}\label{rs1}
R_{s+1}(s+1,s+1)=-\om(s+1)\cdot e_2^t m_{s+1}^{-1}(\pi_{s+1})
m_{s+1}'(\pi_{s+1}) e_1.
\end{equation}
We substitute $\ze=\pi_{s+1}$ into \eq{lax2} again:
\begin{equation}\label{ms}
m_s(\pi_s)=M_s(\pi_{s+1}) m_{s+1}(\pi_{s+1}) D(\pi_{s+1})^{-1},
\end{equation}
and we differentiate \eq{lax2} at $\ze=\pi_{s+1}$ to obtain
\begin{equation}\label{msp}
m_s'(\pi_s)=\eta^{-1} \frac{d}{d\ze}\Big\vert_{\ze=\pi_{s+1}}
\left( M_s(\ze) m_{s+1}(\ze) D(\ze)^{-1} \right).
\end{equation}
If the derivative in \eq{msp} falls onto the third factor, the
contribution of the corresponding term to the RHS of \eq{rs} is
(by \eq{ms})
\[
-\eta^{-1} \om(s) \cdot e_2^t D(\pi_{s+1})
\frac{d}{d\ze}\Big\vert_{\ze=\pi_{s+1}} \left(D(\ze)^{-1}\right)
e_1 = const\cdot(0, 1)\diag{\ast}{\ast} \left(
\begin{array}{c}
1 \\
0
\end{array}\right)=0.
\]
If the derivative in \eq{msp} falls onto the second factor, the
contribution of the corresponding term to the RHS of \eq{rs} is,
by \eq{ms}, \eq{d1d2} and \eq{rs1},
\begin{eqnarray*}
&& -\eta^{-1}\om(s)\cdot e_2^t D(\pi_{s+1})
m_{s+1}^{-1}(\pi_{s+1}) m_{s+1}'(\pi_{s+1}) D(\pi_{s+1})^{-1} e_1
\\ &=& -\eta^{-1} \om(s)
\frac{d_2(\pi_{s+1})}{d_1(\pi_{s+1})} \cdot e_2^t
m_{s+1}^{-1}(\pi_{s+1}) m_{s+1}'(\pi_{s+1}) e_1 \\
&=& -\om(s+1)\cdot e_2^t m_{s+1}^{-1}(\pi_{s+1})
m_{s+1}'(\pi_{s+1}) e_1 \\
&=& R_{s+1}(s+1,s+1).
\end{eqnarray*}
Therefore, if we subtract $R_{s+1}(s+1,s+1)$ from both sides of
\eq{rs}, we get the following equation:
\begin{equation}\label{rs-recurrence}
R_{s+1}(s+1,s+1)-R_s(s,s) = \eta^{-1}\om(s)\cdot e_2^t
m_s^{-1}(\pi_s) M_s'(\pi_{s+1}) m_{s+1}(\pi_{s+1})
D(\pi_{s+1})^{-1} e_1.
\end{equation}
Since $\det m_s\equiv 1$ for all $s$ by Lemma \ref{determinants},
the RHS of \eq{rs-recurrence} equals
\begin{eqnarray*}
&& \eta^{-1}\om(s)\cdot (0, 1)
\matr{m_s^{22}}{-m_s^{12}}{-m_s^{21}}{m_s^{11}}
\matr{\nu_s^{11}}{\nu_s^{12}}{\nu_s^{21}}{\nu_s^{22}}
\matr{m_{s+1}^{11}}{m_{s+1}^{12}}{m_{s+1}^{21}}{m_{s+1}^{22}}
\diag{d_1(\pi_{s+1})^{-1}}{d_2(\pi_{s+1})^{-1}} \left(
\begin{array}{c}
1 \\
0
\end{array}\right) \\
&=& \eta^{-1}\om(s) d_1(\pi_{s+1})^{-1}
(\nu_s^{21}m_s^{11}m_{s+1}^{11}+\nu_s^{22}m_s^{11}m_{s+1}^{21}-
\nu_s^{11}m_s^{21}m_{s+1}^{11}-\nu_s^{12}m_s^{21}m_{s+1}^{21}).
\end{eqnarray*}
Substituting the formulas for $m_s^{21}$, $m_{s+1}^{11}$,
$m_{s+1}^{21}$ derived in part (a) into the last expression and
using \eq{ds}, we find that \eq{rs-recurrence} is equivalent to
\eq{fr}.
\end{proof}

\begin{cor}\label{c:Fredholm}
With the notation of Theorem \ref{t:Fredholm}, we have
\begin{equation}\label{e:fr-r}
u_s \left(\frac{D_{s+3}}{D_{s+2}}-\frac{D_{s+2}}{D_{s+1}}\right) =
\frac{\bigl(\mu^{22}_s+(p_s/q_s)\cdot\mu^{12}_s\bigr)^2}{\eta\cdot
d_1(\pi_{s+2})\cdot d_2(\pi_{s+2})} \cdot u_{s+1}
\left(\frac{D_{s+2}}{D_{s+1}}-\frac{D_{s+1}}{D_s}\right)
\end{equation}
for all $s\in\zpk$, $s\leq N-3$.
\end{cor}
\begin{proof}
Immediate from \eq{fr} and \eq{ms11}.
\end{proof}

\subsection{}\label{ss:fr-4} It turns out that it is possible to extend the
results of Theorem \ref{t:Fredholm} to some cases where the
orthogonality set $\fX$ is discrete but not locally finite, i.e.,
has an accumulation point. The motivation for
such an extension is the following. Even though the DRHP
$(\fX,w)$, as it is stated in \S\ref{s:DRHP}, is not well posed if
$\fX$ is not locally finite (for then we need to impose additional conditions near the accumulation point of the poles), the
definition \eq{eq:kernel} of the kernel $K$ still makes sense: the polynomials $\phi(\ze)=P_k(\ze)$ and
$\psi(\ze)=(P_{k-1},P_{k-1})_\om^{-1}\cdot P_{k-1}(\ze)$ are
well-defined. In fact, there exist
several classical families of basic hypergeometric orthogonal
polynomials for which the orthogonality set is discrete but not
locally finite (see e.g. \cite{KS}, Chapter 3). On the other hand, the
solutions $m_s(\ze)$ of ``restricted'' DRHPs, and hence all the
quantities derived from them, are also defined for a non-locally
finite orthogonality set $\fX$, because their definitions involve
only finite subsets of $\fX$. In particular, we can still consider
the corresponding Lax pair as in subsection \ref{ss:fr-3} and the
scalar sequences $p_s,q_s,m_s^{ij},\mu_s^{ij},\nu_s^{ij}$ defined
by \eq{lax1} and \eq{matrix-elements}. It is therefore natural to
ask whether the recurrence relation \eq{fr} remains valid in the
case where $\fX$ is not locally finite. We will see in subsection
\ref{ss:fr-6} that it does.

\subsection{}\label{ss:fr-5} Let us recall the following probability-theoretic
interpretation of the Fredholm determinants $D_s$, see e.g. \cite{TW1}.
In general, if $\fX$ is a discrete, not necessarily locally finite
subset of $\bR$ of cardinality $N+1$ ($N\in\zp\cup\{\infty\}$),
$\om:\fX\rightarrow\bR$ is a strictly positive weight function
whose moments are finite, $\{P_n(\ze)\}_{n=0}^N$ is the
corresponding family of orthogonal polynomials and $k$ is a
natural number, $k\leq N$, we can consider a probability
distribution $\fP$ on the set of all subsets of $\fX$ of
cardinality $k$, defined by
\begin{equation}\label{e:distribution}
\fP\bigl(\{x_1,\dotsc,x_k\}\bigr)=\frac{1}{Z}\cdot \prod_{1\leq
i<j\leq k} (x_i-x_j)^2\cdot \prod_{i=1}^k \om(x_i).
\end{equation}
Here $Z$ is the unique constant for which the measure of the set
of all subsets of $\fX$ of cardinality $k$ is equal to $1$:
\begin{equation}\label{e:ZZ}
Z=\sum_{\{x_1,\dotsc,x_k\}\subseteq\fX} \prod_{1\leq i<j\leq k}
(x_i-x_j)^2\cdot \prod_{i=1}^k \om(x_i).
\end{equation}
Now if $K$ denotes the kernel \eq{eq:kernel} defined in subsection
\ref{ss:fr-1}, then for any subset $\fY\subseteq\fX$, we have
\begin{equation}\label{e:fr-prob}
\det\bigl(1-K\big\vert_{\fY\times\fY}\bigr)=\sum_{x_i\not\in\fY}
\fP\bigl(\{x_1,\dotsc,x_k\}\bigr)
\end{equation}
(the sum on the RHS is taken over all subsets
$\{x_1,\dotsc,x_k\}\subseteq\fX$ of cardinality $k$ which are
disjoint from $\fY$).

\subsection{}\label{ss:fr-6}
We now assume that $\fX=\{\pi_x\}_{x=0}^\infty\subset\bR$ is
discrete but not necessarily locally finite. We consider, as
before, the subsets $\fY_s=\{\pi_x\}_{x=s}^\infty$ of $\fX$, and
we are interested in the sequence $\{D_s\}_{s=k}^\infty$ defined
by
\[
D_s=\det (1-K_s),\quad K_s=K\big\vert_{\fY_s\times\fY_s},
\]
where $K$ is the kernel \eq{eq:kernel}. The proof of Lemma
\ref{l:invertible} is still valid and gives $D_s\neq 0$ for all
$s\in\zpk$. Thus we see from the discussion of subsection
\ref{ss:fr-4} that both sides of \eq{fr} are at least well defined
in this situation.
\begin{prop}\label{p:Fredholm}
All formulas of Theorem \ref{t:Fredholm} and Corollary
\ref{c:Fredholm} remain valid in the present situation.
\end{prop}
\begin{proof}
It is obvious from the proof of Theorem \ref{t:Fredholm} that
\eq{ms11} remains valid. Now let us fix $s\in\zpk$ and prove that
formula \eq{fr} also holds in the present situation (then
\eq{e:fr-r} follows automatically). Let $L\in\bZ$, $L\geq s+2$. We
write $\fX^{(L)}=\{\pi_x\}_{x=0}^L\subset\fX$. Since $\fX^{(L)}$
is finite, all of the discussion of subsections
\ref{ss:fr-1}--\ref{ss:fr-3} is valid for $\fX^{(L)}$ in place of
$\fX$. Since $L\geq s+2$, it is clear that replacing $\fX$ by
$\fX^{(L)}$ (and keeping the same weight function) has no effect
on $m_s(\ze)$, $p_s,q_s,m_s^{ij},\mu_s^{ij},\nu_s^{ij}$. But of
course, the quantities $D_s$ do change. Let $D_s^{(L)}$ denote the
Fredholm determinants defined as in subsection \ref{ss:fr-2} for
$\fX^{(L)}$ in place of $\fX$. Then Theorem \ref{t:Fredholm}(b)
gives
\[
\frac{D_{s+2}^{(L)}}{D_{s+1}^{(L)}}-\frac{D_{s+1}^{(L)}}{D_s^{(L)}}=
\frac{\om(s)\cdot u_s\cdot (m_s^{11})^2}{\eta\cdot d_1(\pi_{s+1})\cdot
d_2(\pi_{s+1})}.
\]
It remains to observe that
$D_{s+1}^{(L)}\big/D_s^{(L)}=D_{s+1}\big/D_s$ for all $s$. Indeed,
the discussion of subsection \ref{ss:fr-5} gives the formula
\begin{equation}\label{e:Ds}
D_s=\sum_{0\leq i_1<i_2<\dotsb<i_k\leq s-1}
\fP\bigl(\{\pi_{i_1},\dotsc,\pi_{i_k}\}\bigr),
\end{equation}
where $\fP$ is given by \eq{e:distribution}. If $\fP^{(L)}$
denotes the probability distribution on the set of all subsets of
$\fX^{(L)}$ of cardinality $k$ defined similarly to $\fP$, then we
also have
\begin{equation}\label{e:DsN}
D_s^{(L)}=\sum_{0\leq i_1<i_2<\dotsb<i_k\leq s-1}
\fP^{(L)}\bigl(\{\pi_{i_1},\dotsc,\pi_{i_k}\}\bigr)
\end{equation}
We note that the summations in \eq{e:Ds} and \eq{e:DsN} are over
the same index set, and the only difference between the
definitions of $\fP$ and $\fP^{(L)}$ is in the normalization
constant $Z$. Of course, when we take the ratios
$D_{s+1}^{(L)}\big/D_s^{(L)}$ and $D_{s+1}\big/D_s$, the
normalization constants cancel each other, completing the proof.
\end{proof}

\section{Compatibility conditions for Lax
pairs}\label{s:compatibility}

\subsection{} In this section we study the compatibility conditions for the Lax
pairs of the form considered in \S\ref{s:Lax}, \S\ref{s:Fredholm}:
\begin{equation}\label{l1}
m_{s+1}(\ze)=\left(I+\frac{A_s}{\ze-\pi_s}\right) m_s(\ze),
\end{equation}
\begin{equation}\label{l2}
m_s(\sg\ze)=M_s(\ze) m_{s+1}(\ze) D(\ze)^{-1}.
\end{equation}
The general notation and conventions are those of \S\ref{s:Lax},
\S\ref{s:Fredholm}. As in the second part of \S\ref{s:Fredholm},
we do not assume that the orthogonality set $\fX$ is locally
finite: we have already remarked in subsection \ref{ss:fr-4} that
all of our arguments related to Lax pairs only involve finite
subsets of $\fX$.
\begin{lem}\label{l:res}
Fix $s\in\zp$, $s\leq N$, let $A\in Mat(2,\bC)$, and define
$m(\ze)=\bigl(I+(\ze-\pi_s)^{-1}A\bigr)m_s(\ze)$. Then
$m(\ze)=m_{s+1}(\ze)$ if and only if the matrix $A$ satisfies the
following two conditions
\begin{gather}
\label{A-cond1} A\cdot m_s(\pi_s) w(\pi_s) = 0, \\
\label{A-cond2} m_s(\pi_s) w(\pi_s) + A\cdot m_s'(\pi_s) w(\pi_s)
= A\cdot m_s(\pi_s).
\end{gather}
In particular, for a fixed $s$, there is a unique matrix $A$
satisfying \eq{A-cond1} and \eq{A-cond2}, namely, $A=A_s$.
\end{lem}
\begin{proof}
By the uniqueness of $m_{s+1}(\ze)$, we only have to verify that
$m(\ze)$ satisfies the same residue conditions as $m_{s+1}(\ze)$
if and only if \eq{A-cond1} and \eq{A-cond2} hold (note that the
asymptotics of $m(\ze)$ and $m_{s+1}(\ze)$ as $\ze\to\infty$ are
clearly the same). Now if $0\leq x\leq s-1$, then since
$\bigl(I+(\ze-\pi_s)^{-1}A\bigr)$ is analytic near $\pi_x$, it is
clear that the residue condition at $\pi_x$ for $m_s(\ze)$ implies
one for $m(\ze)$. Thus, we only need to consider the residue
condition at the pole $\ze=\pi_s$. Since $m_s(\ze)$ is analytic
near $\pi_s$, we have $\mathrm{Res}_{\ze=\pi_s} m(\ze) = A\cdot
m_s(\pi_s)$. On the other hand, the limit
\[
\lim_{\ze\to\pi_s} m(\ze) w(\pi_s) =
\lim_{\ze\to\pi_s}\bigl(I+(\ze-\pi_s)^{-1}A\bigr)m_s(\ze)w(\pi_s)
\]
exists if and only if \eq{A-cond1} holds. Moreover, if
\eq{A-cond1} holds, this limit equals $m_s(\pi_s)w(\pi_s)+A\cdot
m_s'(\pi_s)w(\pi_s)$, so $m(\ze)$ satisfies the required residue
condition at $\ze=\pi_s$ if and only if \eq{A-cond2} holds.
\end{proof}
\begin{thm}[Compatibility conditions for Lax pairs]\label{t:compatibility}
Fix $s\in\zpk$, $s\leq N-1$.
\begin{enumerate}[(a)]
\item We have
\begin{equation}\label{e:compatibility}
\left(I+\frac{A_s}{\sg\ze-\pi_s}\right) M_s(\ze)=M_{s+1}(\ze)
\left(I+\frac{A_{s+1}}{\ze-\pi_{s+1}}\right).
\end{equation}
\item Conversely, assume that $M:\bC\rightarrow Mat(2,\bC)$ is an
analytic function, $A\in Mat(2,\bC)$ is a nilpotent matrix, and
\begin{equation}\label{e:comp}
\left(I+\frac{A_s}{\sg\ze-\pi_s}\right) M_s(\ze)=M(\ze)
\left(I+\frac{A}{\ze-\pi_{s+1}}\right),
\end{equation}
where $M_s(\ze)$ and $A_s$ are defined by \eq{l1}, \eq{l2}. Then
$M(\ze)=M_{s+1}(\ze)$ and $A=A_{s+1}$.
\end{enumerate}
\end{thm}
Equation \eq{e:compatibility} is the {\em compatibility condition}
for the Lax pair \eq{l1}, \eq{l2}.
\begin{rem}\label{r:recipe}
This theorem provides a recipe for computing $A_{s+1}$ and
$M_{s+1}(\ze)$ if we know $A_s$ and $M_s(\ze)$. Indeed, one simply
needs to find the unique solution of the compatibility condition
which satisfies $A_{s+1}^2=0$.
\end{rem}
\begin{proof}
(a) If we replace $\ze$ by $\sg\ze$ in \eq{l1} and then substitute
\eq{l2} into the result, we obtain
\begin{equation}\label{s1}
m_{s+1}(\sg\ze)=\left(I+\frac{A_s}{\sg\ze-\pi_s}\right) M_s(\ze)
m_{s+1}(\ze) D(\ze)^{-1}.
\end{equation}
On the other hand, if we substitute \eq{l1} into \eq{l2} and then
replace $s$ by $s+1$, we get
\begin{equation}\label{s2}
m_{s+1}(\sg\ze)=M_{s+1}(\ze)
\left(I+\frac{A_{s+1}}{\ze-\pi_{s+1}}\right) m_{s+1}(\ze)
D(\ze)^{-1}.
\end{equation}
Comparing \eq{s1} and \eq{s2} yields \eq{e:compatibility}.

\smallbreak

\noindent
(b) From \eq{e:compatibility} and \eq{e:comp}, we find that
\begin{equation}\label{e:a}
M(\ze)\left(I+\frac{A}{\ze-\pi_{s+1}}\right)=M_{s+1}(\ze)
\left(I+\frac{A_{s+1}}{\ze-\pi_{s+1}}\right).
\end{equation}
Using Lemma \ref{determinants} and \eq{l2} with $s$ replaced by
$s+1$, we see that $\det M_{s+1}(\ze)\equiv \det D(\ze)$, and
hence $M_{s+1}(\ze)$ is invertible near $\pi_{s+1}$. Now since
$A^2=0$, we can rewrite \eq{e:a} as
\begin{equation}\label{e:b}
M_{s+1}^{-1}(\ze)\cdot M(\ze) =
\left(I+\frac{A_{s+1}}{\ze-\pi_{s+1}}\right) \cdot
\left(I-\frac{A}{\ze-\pi_{s+1}}\right).
\end{equation}
The RHS is analytic for $\ze\neq\pi_{s+1}$, and the LHS is
analytic near $\pi_{s+1}$. Thus, both sides of \eq{e:b} are entire
functions. But the RHS tends to $I$ as $\ze\to\infty$, so by
Liouville's theorem, both sides are equal to $I$ for all $\ze$.
This proves that $M(\ze)=M_{s+1}(\ze)$ and $A=A_{s+1}$.
\end{proof}

\subsection{} Recall that the formulas of Theorem \ref{t:Fredholm} have been
derived under the assumption that the parameter $p_s$ does not
vanish for all $s\geq k$. Let us now establish the non-vanishing
of $p_s$ for the weight functions $\om$ and orthogonality sets
$\fX$ satisfying the assumptions of subsection \ref{assumps}. We need the following well-known
fact. For the reader's convenience, we also provide a proof.
\begin{lem}[Zeroes of discrete orthogonal polynomials]\label{l:zeroes}
Let $\fZ\subset\bC$ be a {\rm finite} subset, and assume that
$\fZ$ is contained in a closed interval
$[a,b]\subseteq\bR\subseteq\bC$ such that $a$ is the minimal
element of $\fZ$ and $b$ is the maximal element of $\fZ$. Let
$\om:\fZ\rightarrow\bR$ be a strictly positive weight function.
Then there exists a unique family $\{P_n(\ze)\}_{n=0}^L$ of
orthogonal polynomials corresponding to $\om$, where
$L=\mathrm{card}(\fZ)-1$. The coefficients of each $P_n(\ze)$ are
real. Moreover, for any $0\leq n\leq L$, all zeroes of $P_n(\ze)$
are real and are contained in the open interval $(a,b)$.
\end{lem}
\begin{proof}
Since $\om$ is strictly positive, the restriction of the
corresponding inner product $(\cdot,\cdot)_\om$ to the space
$\bR[\ze]^{\leq d}$ of {\it real} polynomials of degree at most
$d$ is nondegenerate for each $0\leq d\leq L$. Thus, there exists
a unique family of real orthogonal polynomials
$\{P_n(\ze)\}_{n=0}^L$ corresponding to $\om$, and
$(P_n,P_n)_\om\neq 0$ for $0\leq n\leq L$. Now fix $1\leq n\leq
L$, and assume that $P_n(\ze)$ has fewer than $n$ zeroes in the
open interval $(a,b)$. Let $z_1,\dotsc,z_m$ be the zeroes of
$P_n(\ze)$ in $(a,b)$, listed with their multiplicities, and let
$Q(\ze)=(\ze-z_1)\dotsm(\ze-z_m)$. Then, since $P_n(\ze)$ and
$Q(\ze)$ are real polynomials, we have either $P_n(\ze) Q(\ze)\geq
0$ for all $\ze\in[a,b]$, or $P_n(\ze) Q(\ze)\leq 0$ for all
$\ze\in[a,b]$. In addition, since the degree of $P_n(\ze)$ is less
than the cardinality of $\fZ$, there exists $z\in\fZ$ such that
$P_n(z) Q(z)\neq 0$. This implies that $P_n(\ze)$ and $Q(\ze)$ are
not orthogonal with respect to $\om$. Since the degree of $Q(\ze)$
is less than that of $P_n(\ze)$, we have a contradiction with the
definition of orthogonal polynomials.
\end{proof}
Now we can prove
\begin{prop}\label{p:ps-nonzero}
With the notation and conventions of \S\ref{s:Lax} and
\S\ref{s:Fredholm}, assume that either $\pi_0<\pi_1<\pi_2\dotsb$,
or $\pi_0>\pi_1>\pi_2>\dotsb$. Then $p_s\neq 0$ for all $s>k$,
$s\leq N$.
\end{prop}
\begin{proof}
Fix $s>k$, $s\leq N$. By Lemma \ref{l:zeroes}, there exists a
family $\{P_n(\ze)\}_{n=0}^{s-1}$ of polynomials orthogonal on
$\{\pi_0,\dotsc,\pi_{s-1}\}$ with the weight function given by the
restriction of $\om$ to $\{0,1,\dotsc,s-1\}$. Moreover, as $\pi_s$
lies outside the interval between $\pi_0$ and $\pi_{s-1}$, we have
$P_n(\pi_s)\neq 0$ for all $0\leq n\leq s-1$. Now since $k\leq
s-1$, we know from Theorem \ref{t:DRHP} that the first column of
the matrix $m_s(\ze)$ has the form $\bigl(P_k(\ze),c\cdot
P_{k-1}(\ze)\bigr)^t$, where $c$ is a nonzero constant. On the
other hand, by Lemma \ref{l:res}, we have
\begin{equation}\label{e:aux}
m_s(\pi_s) w(\pi_s) = A_s\cdot(m_s(\pi_s)-m_s'(\pi_s)w(\pi_s)).
\end{equation}
If $p_s=0$, then because the matrix $A_s$ is nilpotent by Theorem
\ref{t:Lax}(a), we have either $q_s=0$ or $r_s=0$, i.e., either
the first or the second row of $A_s$ is zero. By \eq{e:aux}, this
implies that either the $(1,2)$ or the $(2,2)$ element of the
matrix $m_s(\pi_s) w(\pi_s)$ is zero. This contradicts
$P_k(\pi_s),P_{k-1}(\pi_s)\neq 0$.
\end{proof}

\section{Initial conditions for recurrence
relations}\label{s:initial}

\subsection{} In this section we derive the initial conditions for the
recurrence relations \eq{fr} and \eq{e:compatibility}. We keep the
general notation and conventions of \S\ref{s:Fredholm} and
\S\ref{s:compatibility}. Recall in particular that $k$ is a
natural number that controls the asymptotics at infinity of the
solutions of all DRHPs that we consider. As in
\S\ref{s:compatibility}, we do not assume that the orthogonality
set $\fX$ is locally finite.
\begin{prop}\label{mk-init}
The solution $m_k(\ze)$ of the DRHP
$\bigl(\{\pi_0,\dotsc,\pi_{k-1}\},w\big\vert_{\{\pi_0,\dotsc,\pi_{k-1}\}}\bigr)$
with the asymptotics

\noindent
$m_k(\ze)\sim\Bigl(\begin{smallmatrix} \ze^k & 0 \\ 0 & \ze^{-k}
\end{smallmatrix} \Bigr)$ as $\ze\to\infty$ is given by
\begin{equation}\label{e:mk}
m_k(\ze)=\matr{(\ze-\pi_0)(\ze-\pi_1)\dotsm(\ze-\pi_{k-1})}{0}
{(\ze-\pi_0)(\ze-\pi_1)\dotsm(\ze-\pi_{k-1})\sum_{m=0}^{k-1}
\frac{\rho_m}{\ze-\pi_m}}{(\ze-\pi_0)^{-1}(\ze-\pi_1)^{-1}\dotsm(\ze-\pi_{k-1})^{-1}},
\end{equation}
where
\begin{equation}\label{e:rhom}
\rho_m=\om(m)^{-1}\cdot\prod_{\substack{0\leq j\leq k-1 \\ j\neq
m}} (\pi_m-\pi_j)^{-2}
\end{equation}
for all $0\leq m\leq k-1$.
\end{prop}
\begin{proof}
Let $m(\ze)$ be the matrix defined by the RHS of \eq{e:mk}. It is
clear that $m(\ze)$ has the required asymptotics at infinity.
Hence we only have to show that \eq{e:rhom} is the (unique) choice
of constants $\rho_m$ which makes $m(\ze)$ satisfy the required
residue conditions. Now if $0\leq x\leq k-1$, then the $(2,2)$
element of the matrix $\mathrm{Res}_{\ze=\pi_x} m(\ze)$ equals
\[
\prod_{\substack{0\leq l\leq k-1 \\
l\neq x}} (\pi_x-\pi_l)^{-1},
\]
and the $(2,2)$ element of the matrix $\lim_{\ze\to\pi_x} m(\ze)
w(\pi_x)$ equals
\[
\om(x)\cdot\rho_x\cdot\prod_{\substack{0\leq l\leq k-1 \\
l\neq x}} (\pi_x-\pi_l).
\]
The other elements of both matrices are zero. Equating the last
two expressions yields \eq{e:rhom}.
\end{proof}

\subsection{} Now we use Lemma \ref{l:res} to find the matrix $A_k$.
\begin{prop}\label{Ak-init}
The elements of the matrix
\[
A_k=\matr{p_k}{q_k}{r_k}{-p_k}
\]
are given by the following formulas:
\begin{equation}\label{e:qk}
q_k=\left\{\rho_k+\sum_{m=0}^{k-1}\frac{\rho_m}{(\pi_k-\pi_m)^2}\right\}^{-1}=\left\{\sum_{m=0}^k\left[
\om(m)^{-1} \prod_{\substack{0\leq j\leq k \\ j\neq m}}
(\pi_m-\pi_j)^{-2} \right]\right\}^{-1},
\end{equation}
where
\[
\rho_k=\om(k)^{-1}\cdot\prod_{j=0}^{k-1} (\pi_k-\pi_j)^{-2};
\]
\begin{equation}\label{e:pk}
p_k=-q_k\cdot\sum_{m=0}^{k-1} \frac{\rho_m}{\pi_k-\pi_m},
\end{equation}
and
\begin{equation}\label{e:rk}
r_k=-q_k\cdot\left\{\sum_{m=0}^{k-1}
\frac{\rho_m}{\pi_k-\pi_m}\right\}^2.
\end{equation}
\end{prop}
\begin{rem}\label{r:pk-nonzero}
It follows from \eq{e:rhom}, \eq{e:qk} and \eq{e:pk} that if the
orthogonality set $\fX$ is contained in $\bR$ and either
$\pi_0>\pi_1>\pi_2>\dotsb$ or $\pi_0<\pi_1<\pi_2<\dotsb$ (and the
weight function $\om$ is strictly positive), then $\rho_m>0$ for
$0\leq m\leq k-1$, $q_k>0$, and hence $p_k\neq 0$.
\end{rem}
\begin{proof}
It follows from Lemma \ref{l:res} that $A_k$ is the unique matrix
satisfying the following system of equations:
\begin{gather}
\label{e:Ak1} A_k\cdot m_k(\pi_k) w(\pi_k) = 0, \\
\label{e:Ak2} m_k(\pi_k) w(\pi_k) + A_k\cdot m_k'(\pi_k) w(\pi_k)
= A_k\cdot m_k(\pi_k).
\end{gather}
Substituting \eq{e:mk} into \eq{e:Ak1} yields \eq{e:pk}. It
remains to prove \eq{e:qk}, for \eq{e:rk} then follows from
$p_k^2=-q_k r_k$. To this end, we consider the $(1,2)$ elements of
both sides of \eq{e:Ak2}. The first summand on the LHS contributes
$\om(k)\cdot\prod_{j=0}^{k-1}(\pi_k-\pi_j)$ to the $(1,2)$
element. Now we consider the second summand. We can rewrite it as
\[
\frac{d}{d\ze}\Big\vert_{\ze=\pi_k} \bigl(A_k\cdot m_k(\ze)
w(\pi_k)\bigr)=\frac{d}{d\ze}\Big\vert_{\ze=\pi_k} \left\{
(\ze-\pi_0)\dotsm(\ze-\pi_{k-1})\cdot\om(k)\cdot
A_k\cdot\matr{0}{1}{0}{\sum_{m=0}^{k-1}\frac{\rho_m}{\ze-\pi_m}}\right\}.
\]
If the derivative falls onto the factor
$(\ze-\pi_0)\dotsm(\ze-\pi_{k-1})$, the corresponding term is zero
because of \eq{e:Ak1}. Hence the whole expression equals
\[
(\pi_k-\pi_0)\dotsm(\pi_k-\pi_{k-1})\cdot A_k\cdot\om(k)\cdot
\matr{0}{0}{0}{-\sum_{m=0}^{k-1}\frac{\rho_m}{(\pi_k-\pi_m)^2}}=\om(k)
\cdot \prod_{j=0}^{k-1}
(\pi_k-\pi_j)\cdot\matr{0}{-q_k\cdot\sum_{m=0}^{k-1}\frac{\rho_m}{(\pi_k-\pi_m)^2}}{0}{\ast}.
\]
Finally, the $(1,2)$ element of $A_k\cdot m_k(\pi_k)$ equals
$q_k\cdot\prod_{j=0}^{k-1}(\pi_k-\pi_j)^{-1}$. Thus, comparing the
$(1,2)$ elements of both sides of \eq{e:Ak2} yields
\[
\left\{\left[\prod_{j=0}^{k-1}(\pi_k-\pi_j)\right]\cdot\om(k)\cdot
\sum_{m=0}^{k-1}\frac{\rho_m}{(\pi_k-\pi_m)^2}+\prod_{j=0}^{k-1}
(\pi_k-\pi_j)^{-1}\right\}\cdot q_k =
\om(k)\cdot\prod_{j=0}^{k-1}(\pi_k-\pi_j),
\]
which gives \eq{e:qk}.
\end{proof}
\begin{rem}\label{r:Mk}
The two propositions we have just proved give explicit formulas
for the matrices $m_k(\ze)$ and $A_k$. Using these formulas, we
can also find the functions $m_{k+1}(\ze)$ and $M_k(\ze)$. Indeed,
from the Lax pair \eq{e:Lax1} and \eq{e:Lax2}, we have
\begin{equation}\label{e:mk1}
m_{k+1}(\ze)=\left(I+\frac{A_k}{\ze-\pi_k}\right) m_k(\ze),
\end{equation}
\begin{equation}\label{e:Mk}
M_k(\ze)=m_k(\sg\ze)D(\ze)m_{k+1}^{-1}(\ze)=m_k(\sg\ze)D(\ze)m_k^{-1}(\ze)\left(I-\frac{A_k}{\ze-\pi_k}\right).
\end{equation}
\end{rem}

\noindent
Even though this gives an explicit formula for $M_k(\ze)$, it is
cumbersome to use it in practice. In the case where the matrix
$D(\ze)$ is linear in $\ze$, a more explicit version is available:
\begin{prop}\label{p:Mk}
Assume that $d_1(\ze)=\la_1\ze+\mu_1$, $d_2(\ze)=\la_2\ze+\mu_2$,
where $\la_1,\la_2,\mu_1,\mu_2\in\bC$ are constants (some of which
could be zero). Then
\begin{equation}\label{e:Mk-lin}
M_k(\ze)=\matr{\eta^k(\la_1\ze+\mu_1)+\eta^k\la_1(\pi_0-\pi_k-p_k)}{-\eta^k\la_1
q_k}{-\eta^{-k}\la_2
r_k+(\eta^{k-1}\la_1-\eta^{-k}\la_2)\sum_{m=0}^{k-1}\rho_m}{\eta^{-k}(\la_2\ze+
\mu_2)+\eta^{-k}\la_2(p_k+\pi_k-\pi_0)},
\end{equation}
where $p_k,q_k,r_k,\rho_m$ are given by \eq{e:pk}, \eq{e:qk},
\eq{e:rk} and \eq{e:rhom}.
\end{prop}
\begin{proof}
Since $M_k(\ze)$ is an entire function, it suffices by Liouville's
theorem to show that \eq{e:Mk-lin} holds up to terms of order
$\ze^{-1}$ as $\ze\to\infty$. To that end, note that by \eq{e:mk}
and \eq{e:eta}, we have
\begin{equation}\label{e:mksg}
m_k(\sg\ze)=\matr{\eta^k(\ze-\pi_1)\dotsm(\ze-\pi_k)}{0}{\eta^{k-1}
(\ze-\pi_1)\dotsm(\ze-\pi_k)\sum_{m=0}^{k-1}
\frac{\rho_m}{\ze-\pi_{m+1}}}{\eta^{-k}(\ze-\pi_1)^{-1}\dotsm(\ze-\pi_k)^{-1}}.
\end{equation}
But it follows from \eq{e:Mk} that
\[
M_k(\ze)=\left\{m_k(\sg\ze)\diag{\ze^{-k}}{\ze^k}\right\}\cdot
D(\ze)\cdot
\left\{m_k(\ze)\diag{\ze^{-k}}{\ze^k}\right\}^{-1}\cdot\left(I-
\frac{A_k}{\ze-\pi_k}\right).
\]
Substituting \eq{e:mk} and \eq{e:mksg} into the last formula, we
arrive at \eq{e:Mk-lin}.
\end{proof}

\subsection{}\label{ss:fr-init} The final result of this section is the computation of the
Fredholm determinants $D_k$ and $D_{k+1}$. We use the
probability-theoretic interpretation of the Fredholm determinants
$D_s$ given in subsection \ref{ss:fr-5}. One can show (this is a standard random matrix theory argument, see e.g. \cite{Me}) that the
constant $Z$ given by \eq{e:ZZ} is equal to the product of the
norms squared of the first $k$ monic orthogonal polynomials:
\begin{equation}\label{e:Z}
Z=\prod_{i=0}^{k-1} (P_i,P_i)_\om.
\end{equation}
Now we prove
\begin{prop}\label{fr-init}
With the notation of \S\ref{s:Fredholm}, let $\fX\subset\bR$ be a
discrete set, let
$\{P_n(\ze)\}$ be the family of orthogonal polynomials
corresponding to a strictly positive weight function
$\om:\fX\rightarrow\bR$, and let $Z$ be given by \eq{e:Z}. Then
\begin{gather}
\label{e:Dk} D_k=\frac{1}{Z}\cdot\prod_{0\leq i<j\leq
k-1}(\pi_i-\pi_j)^2\cdot\prod_{l=0}^{k-1} \om(l), \\
\label{e:Dk1} D_{k+1}=\om(k)\cdot q_k^{-1}\cdot D_k\cdot
\prod_{l=0}^{k-1} (\pi_k-\pi_l)^2,
\end{gather}
where $q_k$ is given by \eq{e:qk}.
\end{prop}
\begin{proof}
Recall that for all $s\in\zpk$, $s\leq N$, we have defined
$\fZ_s=\{\pi_x\}_{x=0}^{s-1}$, $\fY_s=\fX\setminus\fZ_s$, and
$D_s=\det(1-K\big\vert_{\fY_s\times\fY_s})$. Hence a subset of
$\fX$ is disjoin from $\fY_s$ if and only if it is contained in
$\fZ_s$. There exists only one subset of $\fZ_k$ of cardinality
$k$, namely, $\fZ_k=\{\pi_0,\dotsc,\pi_{k-1}\}$ itself. Applying
\eq{e:fr-prob} yields \eq{e:Dk}. Next, there are $k+1$ subsets of
$\fZ_{k+1}$ of cardinality $k$, namely, those of the form
$\fZ_{k+1}\setminus\{\pi_m\}$ for $0\leq m\leq k$. Applying
\eq{e:fr-prob} gives
\[
D_{k+1}=\frac{1}{Z}\cdot \prod_{l=0}^k \om(l)\cdot \left\{
\sum_{m=0}^k \left[\frac{1}{\om(m)} \prod_{\substack{0\leq i<j\leq
k \\ i,j\neq m}} (\pi_i-\pi_j)^2 \right]\right\}.
\]
Using \eq{e:rhom}, \eq{e:qk} and \eq{e:Dk}, we see that the last
equation is equivalent to \eq{e:Dk1}.
\end{proof}

\section{Lax pairs for discrete orthogonal polynomials of the
Askey scheme}\label{s:Lax-Askey}

\subsection{} In this section we specialize to weight functions
appearing in the orthogonality relations for the hypergeometric
orthogonal polynomials and the basic hypergeometric orthogonal
polynomials of the Askey scheme. We use \cite{KS} as our main
reference for the orthogonal polynomials of the Askey scheme. We
are only interested in those families for which the orthogonality
set is discrete. Since our ultimate goal is to derive a recurrence
relation for the associated Fredholm determinants, we do not
impose the local finiteness condition on $\fX$. However, the basic
assumptions of subsection \ref{assumps} have to be satisfied in order to use our
approach (in its present form). These assumptions are not satisfied for the
following families of discrete orthogonal polynomials: the Racah
polynomials (\cite{KS}, \S1.2), the dual Hahn polynomials
(\cite{KS}, \S1.6), the $q$-Racah polynomials (\cite{KS}, \S3.2),
the big $q$-Jacobi polynomials (\cite{KS}, \S3.5), the big
$q$-Legendre polynomials (\cite{KS}, \S3.5.1), the dual $q$-Hahn
polynomials (\cite{KS}, \S3.7), the big $q$-Laguerre polynomials
(\cite{KS}, \S3.11), the dual $q$-Krawtchouk polynomials
(\cite{KS}, \S3.17), the Al-Salam-Carlitz I polynomials
(\cite{KS}, \S3.24), and the discrete $q$-Hermite I polynomials
(\cite{KS}, \S3.28).

\subsection{}\label{ss:list}
Now we list the families of hypergeometric and basic
hypergeometric orthogonal polynomials for which our results do apply. Instead of writing out the whole Lax pair
in each case, we only give the orthogonality set $\fX$, the weight
function $\om(x)$, the affine transformation
$\sg:\bC\rightarrow\bC$, and the corresponding entire functions
$d_1(\ze)$, $d_2(\ze)$ which satisfy the assumption of Theorem
\ref{t:Lax}(b). For the basic hypergeometric polynomials, we
assume from now on that $0<q<1$. This restriction ensures that
$\fX\subset\bR$ and the weight function is strictly positive and
has finite moments.

\begin{itemize}
\item Hahn polynomials (\cite{KS}, \S1.5): \quad $\fX=\{0,\dotsc,N\}$, $N\in\zp$;
\[
\om(x)={\al+x \choose x}{\be+N-x \choose N-x},\quad
\mathrm{where}\ \al,\be>-1\ \mathrm{or}\ \al,\be<-N;
\]
\[
\sg\ze=\ze-1,\quad d_1(\ze)=\ze(\ze-\be-N-1),\quad
d_2(\ze)=(\ze-N-1)(\ze+\al).
\]
\item Meixner polynomials (\cite{KS}, \S1.9): \quad $\fX=\zp$;
\[
\om(x)=\frac{(\be)_x}{x!}c^x,\quad \mathrm{where}\ \be>0\
\mathrm{and}\ 0<c<1;
\]
\[
\sg\ze=\ze-1,\quad d_1(\ze)=\ze,\quad d_2(\ze)=c\ze+c(\be-1).
\]
\item Krawtchouk polynomials (\cite{KS}, \S1.10): \quad $\fX=\{0,\dotsc,N\}$, $N\in\zp$;
\[
\om(x)={N\choose x} p^x (1-p)^{N-x},\quad \mathrm{where}\ 0<p<1;
\]
\[
\sg\ze=\ze-1,\quad d_1(\ze)=\ze,\quad
d_2(\ze)=\frac{p}{p-1}(\ze-N-1).
\]
\item Charlier polynomials (\cite{KS}, \S1.12): \quad $\fX=\zp$;
\[
\om(x)=\frac{a^x}{x!},\quad \mathrm{where}\ a>0;
\]
\[
\sg\ze=\ze-1,\quad d_1(\ze)=\ze,\quad d_2(\ze)=a.
\]
\item $q$-Hahn polynomials (\cite{KS}, \S3.6):\quad
$\fX=\{q^{-x}\big\vert x=0,\dotsc,N\}$, $N\in\zp$;
\[
\om(x)=\frac{(\al q;q)_x (q^{-N};q)_x} {(q;q)_x (\be^{-1}
q^{-N};q)_x} (\al\be q)^{-x},\quad \mathrm{where}\
0<\al,\be<q^{-1}\ \mathrm{or}\ \al,\be>q^{-N};
\]
\[
\sg\ze=q\ze,\quad d_1(\ze)=\al\be
(\ze-1)(\ze-\be^{-1}q^{-N-1}),\quad
d_2(\ze)=(\ze-\al)(\ze-q^{-N-1}).
\]
\item Little $q$-Jacobi polynomials (\cite{KS}, \S3.12):\quad
$\fX=\{q^x\big\vert x\in\zp\}$;
\[
\om(x)=\frac{(bq;q)_x}{(q;q)_x} (aq)^x,\quad \mathrm{where}\
0<a<q^{-1}\ \mathrm{and}\ b<q^{-1};
\]
\[
\sg\ze=q^{-1}\ze,\quad d_1(\ze)=\ze-1,\quad d_2(\ze)=a(b\ze-1).
\]
\item $q$-Meixner polynomials (\cite{KS}, \S3.13):\quad
$\fX=\{q^{-x}\big\vert x\in\zp\}$;
\[
\om(x)=\frac{(bq;q)_x} {(q;q)_x (-bcq;q)_x} c^x q^{x \choose
2},\quad \mathrm{where}\ 0<b<q^{-1}\ \mathrm{and}\ c>0;
\]
\[
\sg\ze=q\ze,\quad d_1(\ze)=(\ze-1)(\ze+bc),\quad
d_2(\ze)=c(\ze-b).
\]
\item Quantum $q$-Krawtchouk polynomials (\cite{KS}, \S3.14):\quad
$\fX=\{q^{-x}\big\vert x=0,\dotsc,N\}$, $N\in\zp$;
\[
\om(x)=\frac{(pq;q)_{N-x}}{(q;q)_x (q;q)_{N-x}} (-1)^{N-x} q^{x
\choose 2},\quad \mathrm{where}\ p>q^{-N};
\]
\[
\sg\ze=q\ze,\quad d_1(\ze)=(\ze-1)(pq^{N+1}\ze-1),\quad
d_2(\ze)=1-q^{N+1}\ze.
\]
\item $q$-Krawtchouk polynomials (\cite{KS}, \S3.15):\quad
$\fX=\{q^{-x}\big\vert x=0,\dotsc,N\}$, $N\in\zp$;
\[
\om(x)=\frac{(q^{-N};q)_x}{(q;q)_x} (-p)^{-x},\quad
\mathrm{where}\ p>0;
\]
\[
\sg\ze=q\ze,\quad d_1(\ze)=p(\ze-1),\quad d_2(\ze)=q^{-N}-q\ze.
\]
\item Affine $q$-Krawtchouk polynomials (\cite{KS}, \S3.16):\quad
$\fX=\{q^{-x}\big\vert x=0,\dotsc,N\}$, $N\in\zp$;
\[
\om(x)=\frac{(pq;q)_x (q;q)_N}{(q;q)_x (q;q)_{N-x}}(pq)^{-x},\quad
\mathrm{where}\ 0<p<q^{-1};
\]
\[
\sg\ze=q\ze,\quad d_1(\ze)=p(\ze-1),\quad
d_2(\ze)=(\ze-p)(q^{N+1}\ze-1).
\]
\item Little $q$-Laguerre/Wall polynomials (\cite{KS}, \S3.20):\quad
$\fX=\{q^x\big\vert x\in\zp\}$;
\[
\om(x)=\frac{(aq)^x}{(q;q)_x},\quad \mathrm{where}\ 0<a<q^{-1};
\]
\[
\sg\ze=q^{-1}\ze,\quad d_1(\ze)=\ze-1,\quad d_2(\ze)=-a.
\]
\item Alternative $q$-Charlier polynomials (\cite{KS}, \S3.22):\quad
$\fX=\{q^x\big\vert x\in\zp\}$;
\[
\om(x)=\frac{a^x}{(q;q)_x} q^{x+1\choose 2},\quad \mathrm{where}\
a>0;
\]
\[
\sg\ze=q^{-1}\ze,\quad d_1(\ze)=\ze-1,\quad
d_2(\ze)=-\frac{a}{q}\ze.
\]
\item $q$-Charlier polynomials (\cite{KS}, \S3.23):\quad
$\fX=\{q^{-x}\big\vert x\in\zp\}$;
\[
\om(x)=\frac{a^x}{(q;q)_x} q^{x\choose 2},\quad \mathrm{where}\
a>0;
\]
\[
\sg\ze=q\ze,\quad d_1(\ze)=\ze-1,\quad d_2(\ze)=a.
\]
\item Al-Salam-Carlitz II polynomials (\cite{KS}, \S3.25):\quad
$\fX=\{q^{-x}\big\vert x\in\zp\}$;
\[
\om(x)=\frac{q^{x^2} a^x}{(q;q)_x (aq;q)_x},\quad \mathrm{where}\
a>0;
\]
\[
\sg\ze=q\ze,\quad d_1(\ze)=(\ze-1)(\ze-a),\quad d_2(\ze)=a.
\]
\end{itemize}

\subsection{}\label{ss:solve}
The next three sections (\S\S\ref{s:general}--\ref{s:Jimbo-Sakai})
deal with the various possible ways of ``solving'' the
compatibility conditions for the Lax pairs listed above. By a
``solution'' of a compatibility condition of the form
\eq{e:compatibility} we mean a collection of formulas which allow
us to express the entries of the matrices $M_{s+1}^{(i)}$,
$A_{s+1}$ as rational functions of the entries of the matrices
$M_s^{(i)}$, $A_s$, where $M_s(\ze)=M_s^{(l)} \ze^l + \dotsb +
M_s^{(0)}$, $M_s^{(i)}\in Mat(2,\bC)$, for all $s$. The most
general case where we have been able to solve the compatibility
condition explicitly is the one where the functions $d_1(\ze)$ and
$d_2(\ze)$ are either linear or constant; this is described in
\S\ref{s:general}. The resulting formulas can be used for
practical computations, but they do not appear to be related to
any known systems of difference equations. In certain more
specialized cases we have been able to reduce the compatibility
condition to one of the equations of H.~Sakai's hierarchy in
\cite{Sak}.

\subsection{}
In \S\ref{s:dP} we solve the compatibility condition
\eq{e:compatibility} in the case where the orthogonality set has
the form $\fX=\{x\in\zp\big\vert x\leq N\}$
($N\in\zp\cup\{\infty\}$) and the functions $d_1(\ze)$ and
$d_2(\ze)$ are linear, by a method different from the one used in
\S\ref{s:general} . We show that if both $d_1$ and $d_2$ are
nonconstant, then the compatibility condition is (generically)
equivalent to the $d-P_V$ equation of H.~Sakai \cite{Sak}, and if
$d_2$ is constant, then the compatibility condition is
(generically) equivalent to the $d-P_{IV}$ equation ibid. (see
Theorem \ref{t:dP}). This result allows us to write down explicit
solutions for the recurrence relations corresponding to the
Meixner polynomials, the Krawtchouk polynomials, and the Charlier
polynomials --- see \S\ref{s:Charlier}.

\subsection{}\label{ss:JS}
As for the basic hypergeometric orthogonal polynomials, we show in
\S\ref{s:Jimbo-Sakai} (see Theorem \ref{qp6}(b)) that in the case
where the orthogonality set has the form $\fX=\{q^{-s}\}_{s=0}^N$
($N\in\zp\cup\{\infty\}$) and the functions $d_1(\ze)$ and
$d_2(\ze)$ are linear and nonconstant, the compatibility condition
for the corresponding Lax pair is equivalent to the $q-P_{VI}$
system of M.~Jimbo and H.~Sakai, for a certain choice of the
parameters. Since the case where $\fX=\{q^s\big\vert s\in\zp\}$ is
reduced to the former one after replacing $q$ by $q^{-1}$, this
situation suits the following families of basic hypergeometric
orthogonal polynomials: the little $q$-Jacobi polynomials and the
$q$-Krawtchouk polynomials (it does \textit{not} suit the
alternative $q$-Charlier polynomials because the constant terms of
the functions $d_1(\ze)$ and $d_2(\ze)$ must be nonzero,
cf.~Theorem \ref{t:JS}). If one of the functions $d_1(\ze)$ and
$d_2(\ze)$ is linear and the other one is constant, it is possible
to reduce the corresponding compatibility condition to a
degeneration of the $q-P_{VI}$ system. This process is described
in subsection \ref{ss:degeneration}. It allows us to solve the
compatibility conditions for the little $q$-Laguerre/Wall
polynomials and the $q$-Charlier polynomials.

It turns out, however, that the method of solving the
compatibility condition by reducing it to the $q-P_{VI}$ system
(or its degeneration) is rather difficult to carry out in
practice. So to find a recurrence relation for the Fredholm
determinants associated to classical families of basic hypergeometric
orthogonal polynomials, we prefer to use the more general formulas
of \S\ref{s:general} (see \S\ref{s:Charlier}). The disadvantage of the formulas of \S\ref{s:general}, as compared to $q-P_{VI}$, is the fact that the recurrence step substantially involves more than two sequences, while for $q-P_{VI}$ two sequences suffice, cf. Theorems \ref{t:general} and 
\ref{t:JS} below.

\section{Solution of the compatibility condition: the general
case}\label{s:general}

\subsection{}
In this section we ``solve'' (in the sense of subsection
\ref{ss:solve}) the compatibility condition
\begin{equation}\label{e:compat}
\left(I+\frac{\eta^{-1} A_s}{\ze-\pi_{s+1}}\right) M_s(\ze) =
M_{s+1}(\ze) \left(I+\frac{A_{s+1}}{\ze-\pi_{s+1}}\right)
\end{equation}
derived in \S\ref{s:compatibility} (cf. equation
\eq{e:compatibility}), in the case where the matrix $D(\ze)$
defined in Theorem \ref{t:Lax}(b) depends linearly on $\ze$. Our
method is based on the following simple observation.
\begin{prop}\label{p:general}
If $M_s(\ze)=\La\cdot\ze+C_s$ for all $s$, where $C_s$ does not
depend on $\ze$ and $\La$ is a fixed matrix independent both of
$\ze$ and of $s$, then under the assumption that $A_{s+1}^2=0$,
the compatibility condition \eq{e:compat} is equivalent to the
following system of linear equations:
\begin{gather}
\label{e:comp1} (\pi_{s+1}\La+C_s+\eta^{-1}A_s\La)\cdot A_{s+1} =
\eta^{-1} A_s\cdot(\pi_{s+1}\La+C_s), \\
\label{e:comp2} C_{s+1}=C_s+\eta^{-1}A_s\La-\La A_{s+1}.
\end{gather}
\end{prop}
\begin{proof}
Comparing the asymptotics of both sides of \eq{e:compat} as
$\ze\to\infty$ yields \eq{e:comp2}. If we take the residues of
both sides of \eq{e:compat} at $\ze=\pi_{s+1}$, we obtain
\[
\eta^{-1} A_s\cdot (\pi_{s+1}\La+C_s) =
(\pi_{s+1}\La+C_{s+1})\cdot A_{s+1}.
\]
Substituting \eq{e:comp2} into the last equation and using the
assumption that $A_{s+1}^2=0$ gives \eq{e:comp1}.
\end{proof}

\subsection{}
We now note that if the matrix $\pi_{s+1}\La+C_s+\eta^{-1}A_s\La$
is invertible, then the system \eq{e:comp1}, \eq{e:comp2} already
has a unique solution $(A_{s+1},C_{s+1})$. Since the compatibility
condition \eq{e:compat} always has a unique solution by Theorem
\ref{t:compatibility}(b), it follows from Proposition
\ref{p:general} that in this case, the solution of \eq{e:comp1},
\eq{e:comp2} is also the solution of \eq{e:compat}. Even though we
cannot prove that the matrix $\pi_{s+1}\La+C_s+\eta^{-1}A_s\La$ is
invertible in general, the explicit computations we have carried
out for five families of basic hypergeometric orthogonal
polynomials (see \S\ref{s:Charlier}) show that the result below
(Theorem \ref{t:general}) has practical significance.

\subsection{}
We introduce the following notation. Suppose that
$d_1(\ze)=\la_1\ze+\mu_1$, $d_2(\ze)=\la_2\ze+\mu_2$, where
$\la_1,\mu_1,\la_2,\mu_2\in\bC$ are constants. Then it follows
from Theorem \ref{t:Lax}(c) that the assumption of Proposition
\ref{p:general} is satisfied for $\La=\mathrm{diag}(\ka_1,\ka_2)$,
where $\ka_1=\eta^k\la_1$ and $\ka_2=\eta^{-k}\la_2$. Let us write
\[
A_s=\matr{p_s}{q_s}{r_s}{-p_s} \quad \text{and} \quad
C_s=\matr{\al_s}{\be_s}{\ga_s}{\de_s}.
\]
Finally, define $\eps_s=\det(\pi_{s+1}\La+C_s+\eta^{-1}A_s\La)$.
Now we can state
\begin{thm}\label{t:general}
We have
\begin{equation}\label{e:epss}
\eps_s = d_1(\pi_{s+1})d_2(\pi_{s+1}) +
\eta^{-1}\ka_1(p_s\de_s-r_s\be_s) -
\eta^{-1}\ka_2(p_s\al_s+q_s\ga_s).
\end{equation}
If $\eps_s\neq 0$, then the following formulas hold:
\begin{gather}
\label{e:ps1} p_{s+1}=-\eta^{-1} p_s^{-1} \eps_s^{-1}\cdot
(p_s\be_s+q_s\de_s+\ka_2\pi_{s+1}q_s)\cdot
(r_s\al_s-p_s\ga_s+\ka_1\pi_{s+1}r_s), \\
\label{e:qs1} q_{s+1}=\eta^{-1} q_s^{-1} \eps_s^{-1} \cdot
(p_s\be_s+q_s\de_s+\ka_2\pi_{s+1}q_s)^2, \\
\label{e:rs1} r_{s+1}=\eta^{-1} r_s^{-1} \eps_s^{-1} \cdot
(r_s\al_s-p_s\ga_s+\ka_1\pi_{s+1}r_s)^2, \\
\label{e:als1} \al_{s+1}=\al_s+\eta^{-1}\ka_1 p_s - \ka_1 p_{s+1}, \\
\label{e:bes1} \be_{s+1}=\be_s+\eta^{-1}\ka_2 q_s - \ka_1 q_{s+1}, \\
\label{e:gas1} \ga_{s+1}=\ga_s+\eta^{-1}\ka_1 r_s - \ka_2 r_{s+1}, \\
\label{e:des1} \de_{s+1}=\de_s-\eta^{-1}\ka_2 p_s + \ka_2 p_{s+1}, \\
\label{e:us}
u_s=-\ka_2\ga_s+\frac{p_s}{q_s}\cdot(\ka_1\de_s-\ka_2\al_s) +
\frac{p_s^2}{q_s^2}\cdot\ka_1\be_s,
\end{gather}
where $u_s$ is defined in Theorem \ref{t:Fredholm}(b).
\end{thm}
We omit the proof, as it consists entirely of straightforward
computations. We first derive \eq{e:epss} using the identity $\det
M_s(\ze)\equiv\det D(\ze)$ which follows from Lemma
\ref{determinants}. If $\eps_s\neq 0$, we rewrite \eq{e:comp1} as
\[
A_{s+1} = \eta^{-1}\cdot
(\pi_{s+1}\La+C_s+\eta^{-1}A_s\La)^{-1}\cdot
A_s\cdot(\pi_{s+1}\La+C_s).
\]
Writing out the matrix product on the RHS explicitly yields
\eq{e:ps1}--\eq{e:rs1}. Then \eq{e:comp2} gives
\eq{e:als1}--\eq{e:des1}. Finally, \eq{e:us} follows immediately
from the definition of $u_s$.

\section{The fifth and the fourth discrete Painlev\'e equations}\label{s:dP}

\subsection{} In this section we assume that the orthogonality set is of the
form $\fX=\{x\in\zp\big\vert x\leq N\}$ ($N\in\zp\cup\{\infty\}$),
so that, with the notation of \S\ref{s:Lax}, $\sg\ze=\ze-1$ and
$\eta=1$. Our goal is to prove that if the functions $d_1(\ze)$
and $d_2(\ze)$ are linear, then the compatibility condition
\eq{e:compatibility} is equivalent to either the fifth or the
fourth discrete Painlev\'e equation of H.~Sakai \cite{Sak}.

Recall from Theorem \ref{t:Lax} that since $\pi_0=0$, we must have
$d_1(0)=0$, and because only the ratio $d_1(\ze)/d_2(\ze)$
matters, we may assume without loss of generality that
$d_1(\ze)=\ze$, and write $d_2(\ze)=\xi\ze+\tau$ for some
$\xi,\tau\in\bC$ (unless otherwise explicitly stated, we do not
exclude the possibility $\xi=0$). By Theorem \ref{t:Lax}(c), we
can write
\[
M_s(\ze)=\La\ze+C_s,\quad \text{where } \La=\diag{1}{\xi} \text{
and } C_s=\matr{C_s^{11}}{C_s^{12}}{C_s^{21}}{C_s^{22}}.
\]
Then the compatibility condition \eq{e:compatibility} takes the
form
\begin{equation}\label{e:c-lin}
\left(I+\frac{A_s}{\ze-(s+1)}\right)\cdot\bigl(\La\ze+C_s\bigr) =
\bigl(\La\ze+C_{s+1}\bigr)\cdot\left(I+\frac{A_{s+1}}{\ze-(s+1)}\right),
\end{equation}
where $A_s=\Bigl(\begin{smallmatrix} p_s & q_s \\ r_s & -p_s
\end{smallmatrix} \Bigr)$. The following result allows us to find
a convenient reparameterization of the matrices $A_s$ and $C_s$
which leads to an explicit solution of \eq{e:c-lin}.
\begin{lem}\label{l:Cs}
We have
\begin{equation}\label{e:Cs}
C_s^{11}+p_s=-k,\quad C_s^{22}-\xi p_s=\xi k + \tau,\quad C_s^{12}
C_s^{21} = C_s^{11} C_s^{22}
\end{equation}
for all $s\in\zpk$, $s\leq N$.
\end{lem}
\begin{proof}
Taking the asymptotics of both sides of \eq{e:c-lin} as
$\ze\to\infty$ gives
\begin{equation}\label{cla}
C_s+A_s\La=C_{s+1}+\La A_{s+1},
\end{equation}
which implies that
\[
C_s^{11}+p_s=C_{s+1}^{11}+p_{s+1}\quad \text{and}\quad
C_s^{22}-\xi p_s=C_{s+1}^{22}-\xi p_{s+1}
\]
for all $s\in\zpk$, $s<N$. This means that the expressions
$C_s^{11}+p_s$ and $C_s^{22}-\xi p_s$ do not depend on $s$. But we
know from Proposition \ref{p:Mk} that
\begin{equation}\label{e:Ck-lin}
C_k=\matr{-k-p_k}{-q_k}{-\xi r_k + (1-\xi)
\sum_{m=0}^{k-1}\rho_m}{\tau+\xi(p_k+k)},
\end{equation}
whence $C_k^{11}+p_k=-k$ and $C_k^{22}-\xi p_k=\xi k+\tau$,
proving the first two equalities in \eq{e:Cs}. To prove the third
equality, note that $\det M_s(\ze)=\det D(\ze)$ for all $\ze$, by
\eq{e:Lax2} and Lemma \ref{determinants}, and take $\ze=0$.
\end{proof}

\subsection{} It follows from Lemma \ref{l:Cs} that for all $s\in\zpk$, $s\leq
N$, the matrices $A_s$ and $C_s$ can be naturally parameterized as
follows:
\begin{equation}\label{e:AsCs-param}
A_s=(k+b_s)\matr{-1}{-\al_s\be_s}{1/(\al_s\be_s)}{1},\quad
C_s=\matr{b_s}{b_s\be_s}{(\tau-\xi b_s)/\be_s}{\tau-\xi b_s}.
\end{equation}
\begin{rem}\label{r:generic}
This parameterization is taken from \cite{B} (see equation (6.8)
ibid.). The proof of the theorem below is based on the same idea
as the proof of Proposition 6.3 in \cite{B}. In fact, the
situation considered in \S6 of \cite{B} corresponds, with minor
modifications, to the weight function for Meixner polynomials (see
\S\ref{s:Lax-Askey} above). The only essential
difference with the present paper is that we consider a slightly
more general situation by letting $d_2(\ze)$ be an arbitrary
linear function, which allows us to treat the cases of the
Krawtchouk and Charlier polynomials, as well as the case of the
Meixner polynomials.

Note also that the parameterization \eq{e:AsCs-param} is only
valid if the matrices $A_s$ and $C_s$ are sufficiently generic.
If, for instance, $C_s^{12}=0$, but $C_s^{11}\neq 0$, then
\eq{e:AsCs-param} does not make sense. When we try to solve
\eq{e:c-lin} in terms of the parameterization \eq{e:AsCs-param},
we will encounter a similar difficulty: the formulas will involve
rational functions of the parameters, and it is not clear a priori
that the denominators of all fractions do not vanish. We refer the
reader to \cite{B}, \S6, where this problem is discussed in
detail. The argument of \cite{B} can be easily adapted to our
situation.
\end{rem}
\begin{thm}\label{t:dP}
\begin{enumerate}[(a)]
\item Assume that $\xi\neq 0$. Introduce new variables $f_s$, $g_s$ by
\begin{equation}\label{e:fsgs}
f_s=-k-b_s+\frac{s}{1-\al_s},\quad g_s=-\al_s.
\end{equation}
Then with the parameterization \eq{e:AsCs-param}, the recurrence
relation \eq{e:c-lin} has the following solution:
\begin{equation}\label{dPV-1}
f_{s+1}+f_s=-(k+\tau/\xi)+\frac{s}{1+g_s}+\frac{\tau/\xi+s+1}{1+\xi
g_s},
\end{equation}
\begin{equation}\label{dPV-2}
g_{s+1} g_s = \frac{(f_{s+1}-1-s)(f_{s+1}-1-s+k)}{\xi f_{s+1}
(f_{s+1}+k+\tau/\xi)},
\end{equation}
\begin{equation}\label{dPV-3}
\frac{\be_{s+1}}{\be_s}=-\frac{\xi g_s}{g_{s+1}}\cdot
\frac{(1+g_{s+1})f_{s+1}+(k+\tau/\xi)g_{s+1}-s-1}{(1+\xi
g_s)f_{s+1}+k-s-1}.
\end{equation}
\item Now let $\xi=0$, and introduce new variables $f_s$, $g_s$ by
\begin{equation}\label{e:fsgs-d}
f_s=\al_s^{-1},\quad g_s=\tau\al_s+b_s+s+1.
\end{equation}
Then with the parameterization \eq{e:AsCs-param}, the recurrence
relation \eq{e:c-lin} has the following solution:
\begin{equation}\label{dPIV-1}
f_s f_{s+1} = \frac{\tau g_s}{(g_s-s-1)(g_s+k-s-1)},
\end{equation}
\begin{equation}\label{dPIV-2}
g_s+g_{s+1} = \frac{\tau}{f_{s+1}}-\frac{s+1}{1-f_{s+1}}-k+2s+3.
\end{equation}
\begin{equation}\label{dPIV-3}
\frac{\be_{s+1}}{\be_s}= \frac{\tau}{f_s(g_s+k-s-1)}.
\end{equation}
\end{enumerate}
\end{thm}
\begin{rem}\label{r:dP-Sak}
\begin{enumerate}[(a)]
\item If we set $f=f_s$, $\fb=f_{s+1}$, $g=g_s$, $\gb=g_{s+1}$, then the
relations \eq{dPV-1}, \eq{dPV-2} form a special case of the difference
Painlev\'e V equation ($d-P_V$) of \cite{Sak}, \S7. The parameters
$\la,a_0,a_1,a_2,a_3,a_4$ in our case are as follows:
\[
a_0=\tau/\xi+s+1,\quad a_1=s,\quad a_2=-s,
\]
\[
a_3=-(k+\tau/\xi),\quad a_4=k,\quad \la=a_1+2a_2+a_3+a_4+a_0=1.
\]
\item If we set $f=f_{s+1}$, $\fb=f_s$, $g=g_{s+1}$, $\gb=g_s$, then the
relations \eq{dPIV-1}, \eq{dPIV-2} form a special case of the difference
Painlev\'e IV equation ($d-P_{IV}$) of \cite{Sak}, \S7. The parameters
$\la,a_0,a_1,a_2,a_3$ in our case are as follows:
\[
a_0=-s-2,\quad a_1=1,\quad a_2=k,
\]
\[
a_3=s+2-k,\quad \la=a_1+a_2+a_3+a_0=1.
\]
\end{enumerate}
\end{rem}

\medbreak

\noindent
\textit{Proof of Theorem \ref{t:dP}.} For the first part of the
proof we do not need to distinguish between the cases where
$\xi\neq 0$ and $\xi=0$. Taking the residues of both sides of
\eq{e:c-lin} at $\ze=s+1$ yields
\[
A_s\cdot\bigl((s+1)\La+C_s\bigr)=\bigl((s+1)\La+C_{s+1}\bigr)\cdot
A_{s+1},
\]
i.e.,
\begin{multline*}
(k+b_s)\cdot\matr{-1}{-\al_s\be_s}{1/(\al_s\be_s)}{1}\cdot
\matr{s+1+b_s}{b_s\be_s}{(\tau-\xi b_s)/\be_s}{\xi(s+1)+\tau-\xi b_s} \\
=(k+b_{s+1})\cdot \matr{s+1+b_{s+1}}{b_{s+1}\be_{s+1}} {(\tau-\xi
b_{s+1})/\be_{s+1}}{\xi(s+1)+\tau-\xi b_{s+1}}
\cdot\matr{-1}{-\al_{s+1}\be_{s+1}}{1/(\al_{s+1}\be_{s+1})}{1}.
\end{multline*}
Comparing the diagonal terms on both sides, we get a system of two
equations:
\begin{equation}\label{e:1}
(k+b_s)\cdot\bigl[-(\tau-\xi
b_s)\al_s-(s+1)-b_s\bigr]=(k+b_{s+1})\cdot\bigl[b_{s+1}/\al_{s+1}-(s+1)-b_{s+1}
\bigr],
\end{equation}
\begin{equation}\label{e:2}
(k+b_s)\cdot\bigl[b_s/\al_s+\xi(s+1)+\tau-\xi
b_s\bigr]=(k+b_{s+1})\cdot\bigl[-(\tau-\xi
b_{s+1})\al_{s+1}+\xi(s+1)+\tau-\xi b_{s+1}\bigr].
\end{equation}
If we multiply \eq{e:1} by $\al_{s+1}$, \eq{e:2} by $\al_s$, and
add the results, we obtain an equation which can be written as
follows:
\begin{equation}\label{e:3}
\begin{split}
(k+b_s)\cdot\bigl[(1-\xi\al_s)(1-\al_{s+1})b_s + \tau\al_s(1-\al_{s+1})+\xi(s+1)\al_s-(s+1)\al_s\bigr] \\
=(k+b_{s+1})\cdot\bigl[(1-\xi\al_s)(1-\al_{s+1})b_{s+1} +
\tau\al_s(1-\al_{s+1})+\xi(s+1)\al_s-(s+1)\al_s\bigr].
\end{split}
\end{equation}
Now we subtract the LHS of the last equation from its RHS and
divide the result by $(1-\xi\al_s)(1-\al_{s+1})(b_{s+1}-b_s)$;
noting that
$(k+b_{s+1})b_{s+1}-(k+b_s)b_s=(k+b_{s+1}+b_s)(b_{s+1}-b_s)$, we
get
\begin{equation}\label{e:4}
k+b_s+b_{s+1}+\frac{\tau\al_s+s+1}{1-\xi\al_s}-\frac{s+1}{1-\al_{s+1}}=0.
\end{equation}

\medbreak

Let us now assume that $\xi\neq 0$. In this case, it is easy to
see that with the notation \eq{e:fsgs}, the last equation is
equivalent to \eq{dPV-1}. To obtain \eq{dPV-2}, we divide \eq{e:2}
by \eq{e:1}, which yields
\begin{equation}\label{e:5}
\frac{b_s/\al_s+\xi(s+1)+\tau-\xi b_s}{-(\tau-\xi
b_s)\al_s-(s+1)-b_s} =\frac{-(\tau-\xi
b_{s+1})\al_{s+1}+\xi(s+1)+\tau-\xi b_{s+1}}
{b_{s+1}/\al_{s+1}-(s+1)-b_{s+1}}.
\end{equation}
From \eq{e:fsgs} and \eq{e:4}, we have
\[
b_s/\al_s+\xi(s+1)+\tau-\xi
b_s=\frac{1}{\al_s}(1-\xi\al_s)b_s+\xi(s+1)+\tau
=\frac{1-\xi\al_s}{\al_s}(f_{s+1}-1-s),
\]
\[
-(\tau-\xi b_s)\al_s-(s+1)-b_s=
-\bigl[(1-\xi\al_s)b_s+\tau\al_s+s+1\bigr]= -(1-\xi\al_s)f_{s+1},
\]
\[
-(\tau-\xi b_{s+1})\al_{s+1}+\xi(s+1)+\tau-\xi b_{s+1}=
\tau(1-\al_{s+1})-\xi\cdot\bigl[(1-\al_{s+1})b_{s+1}-(s+1)\bigr]=
(1-\al_{s+1})(\tau+\xi f_{s+1}+\xi k),
\]
\[
b_{s+1}/\al_{s+1}-(s+1)-b_{s+1}=
\frac{1}{\al_{s+1}}\cdot\bigl[(1-\al_{s+1})b_{s+1}-(s+1)\al_{s+1}\bigr]=
\frac{1-\al_{s+1}}{\al_{s+1}}(s+1-f_{s+1}-k).
\]
This computation immediately implies that \eq{e:5} is equivalent
to \eq{dPV-2}. To complete the proof of part (a), we equate the
$(2,1)$ elements of both sides of \eq{cla}, which gives
\[
(\tau-\xi b_s)/\be_s+(k+b_s)/(\al_s\be_s)= (\tau-\xi
b_{s+1})/\be_{s+1}+\xi(k+b_{s+1})/(\al_{s+1}\be_{s+1}).
\]
We can rewrite the last equation as
\[
\frac{1}{\be_s}\bigl[ (\tau-\xi b_s)+(k+b_s)/\al_s \bigr] =
\frac{1}{\be_{s+1}} \bigl[(\tau-\xi
b_{s+1})+\xi(k+b_{s+1})/\al_{s+1}\bigr].
\]
It is easily seen to be equivalent to \eq{dPV-3}, by \eq{e:fsgs}.

\medbreak

Now we assume that $\xi=0$. Then \eq{e:4} becomes
\begin{equation}\label{e:6}
k+b_s+b_{s+1}+\tau\al_s+s+1-\frac{s+1}{1-\al_{s+1}}=0.
\end{equation}
It is clear that with the notation \eq{e:fsgs-d}, the last
equation is equivalent to \eq{dPIV-1}. To obtain \eq{dPIV-2}, we
divide \eq{e:2} by \eq{e:1}, which gives
\begin{equation}\label{e:7}
\frac{b_s/\al_s+\tau}{-\tau\al_s-(s+1)-b_s} =
\frac{-\tau\al_{s+1}+\tau} {b_{s+1}/\al_{s+1}-(s+1)-b_{s+1}}.
\end{equation}
From \eq{e:fsgs-d} and \eq{e:6}, we have
\[
b_s/\al_s+\tau=\frac{1}{\al_s}(\tau\al_s+b_s)=\frac{1}{\al_s}(g_s-s-1),
\]
\[
-\tau\al_s-b_s-(s+1)=-g_s,
\]
\[
-\tau\al_{s+1}+\tau=(1-\al_{s+1})\tau,
\]
\[
b_{s+1}/\al_{s+1}-(s+1)-b_{s+1}=
\frac{1}{\al_{s+1}}\cdot\bigl[(1-\al_{s+1})b_{s+1}-(s+1)\al_{s+1}\bigr]=
\frac{1-\al_{s+1}}{\al_{s+1}}(s+1-k-g_s).
\]
This computation immediately implies that \eq{e:7} is equivalent
to \eq{dPIV-2}. Finally, we compare the $(2,1)$ elements of both
sides of \eq{cla}. This yields
\[
\frac{\tau}{\be_s}+\frac{k+b_s}{\al_s\be_s}=\frac{\tau}{\be_{s+1}},\quad
\text{i.e.,}\quad
\frac{\be_{s+1}}{\be_s}=\frac{\tau}{\al_s^{-1}(\tau\al_s+k+b_s)},
\]
which gives \eq{dPIV-3}, completing the proof of part (b). \qed

\section{A connection with the $q-P_{VI}$ equation of M.~Jimbo and
H.~Sakai}\label{s:Jimbo-Sakai}

\subsection{Reduction to the $q-P_{VI}$ system}
In this section we show that the compatibility
conditions for the Lax pairs corresponding to some of the families
of polynomials orthogonal on $q$-lattices are equivalent to the
$q-P_{VI}$ system of M.~Jimbo and H.~Sakai (equations (19)-(20) in
\cite{JS}) for an appropriate choice of the parameters, or to a
certain degeneration of this system. Thus, we now assume that the
orthogonality set is of the form $\fX=\{q^{-s}\}$, where $s$ runs
either over $\zp$ or over $\{0,\dotsc,N\}$ ($N\in\zp$), and
 $\abs{q}\neq 0,1$. Hence we have
$\sg\ze=q\ze$ and $\eta=q$. Then the corresponding Lax pair has
the following form:
\begin{equation}\label{q-lax}
m_{s+1}(\ze)=\left(I+\frac{A_s}{\ze-q^{-s}}\right) m_s(\ze),\quad
m_s(q\ze)=M_s(\ze) m_{s+1}(\ze) D(\ze)^{-1},
\end{equation}
and the compatibility condition is (cf. equation
\eq{e:compatibility}):
\begin{equation}\label{q-comp}
\left(I+\frac{A_s}{q\ze-q^{-s}}\right) M_s(\ze)=M_{s+1}(\ze)
\left(I+\frac{A_{s+1}}{\ze-q^{-s-1}}\right).
\end{equation}
To relate our situation to the one considered in \cite{JS}, we
make the following change of notation: $x:=\ze$, $t:=q^{-s}$. Then
we define
\begin{gather}
\label{A} A(x,t)=M_s(x)\bigl((x-t)I+A_s\bigr), \\
\label{B0} B_0(t)=-qtI-A_{s-1}, \\
\label{B} B(x,t)=\frac{x\bigl(xI+B_0(t)\bigr)}{(x-qt)^2}.
\end{gather}
\begin{thm}\label{qp6}
\begin{enumerate}[(a)]
\item The compatibility condition \eq{q-comp} with $s$ replaced by $s-1$
is equivalent to the following equation:
\begin{equation}\label{q-compatibility}
A(x,qt)B(x,t)=B(qx,t)A(x,t).
\end{equation}
\item If the matrix $D(\ze)=\mathrm{diag}(d_1(\ze),d_2(\ze))$ is
linear in $\ze$, so that $d_1(\ze)=\la_1(\ze-a_3)$,
$d_2(\ze)=\la_2(\ze-a_4)$ with $\la_1,\la_2\neq 0$, then the
matrix $A(x,t)$ is quadratic in $x$, and if we write
\begin{equation}\label{e:Axt}
A(x,t)=A_0(t)+A_1(t)x+A_2 x^2,
\end{equation}
we have:
\begin{gather}
\label{e:Adiag} A_2=\diag{\ka_1}{\ka_2},\quad \ka_1=q^k\la_1,\
\ka_2=q^{-k}\la_2, \\
\label{e:A0} A_0(t)\ \mathrm{has\ eigenvalues}\ t\te_1,\,t\te_2, \\
\label{e:detA} \det A(x,t)=\ka_1 \ka_2 (x-t)^2 (x-a_3) (x-a_4),
\end{gather}
and the parameters $\ka_j$, $a_j$, $\te_j$ are independent of $t$.
\end{enumerate}
\end{thm}
Recall that the natural number $k$ in \eq{e:Adiag} defines the
asymptotics of the solutions $m_s(\ze)$ as $\ze\to\infty$ (see
Theorem \ref{t:Lax}).
\begin{rem}\label{formal}
The equation \eq{q-compatibility} can be viewed as the
compatibility condition for the following pair of $q$-difference
matrix equations, cf \cite{JS}:
\begin{equation}\label{nxt}
n(x,qt)=B(x,t)n(x,t),\quad n(qx,t)=A(x,t)n(x,t).
\end{equation}
One way to construct a matrix $n(x,t)$ which solves the system
\eq{nxt} is as follows. Let
$V(\ze)=\mathrm{diag}(v_1(\ze),v_2(\ze))$ be a diagonal matrix
such that $V(q\ze)=D(\ze)V(\ze)$ for all $\ze$. Then for all $s$,
we define a new matrix $n_s(\ze)$ by
\begin{equation}\label{ns}
m_s(\ze)=\left\{
\begin{array}{ll}
\ze^s q^{s\choose 2} \prod_{i=-\infty}^{s-1}
(q^i\ze-1)^{-1}\cdot n_s(\ze)\cdot V(\ze)^{-1} & \mathrm{if}\ \abs{q}>1, \\
\ze^s q^{s\choose 2} \prod_{i=s}^{+\infty} (q^i\ze-1)\cdot
n_s(\ze)\cdot V(\ze)^{-1} & \mathrm{if}\ 0<\abs{q}<1.
\end{array} \right.
\end{equation}
Substituting this into the first equation of the Lax pair
\eq{q-lax}, replacing $s$ by $s-1$ and simplifying yields
\[
\ze\cdot n_s(\ze) = \bigl((\ze-q^{-s+1})I+A_{s-1}\bigr)\cdot
n_{s-1}(\ze).
\]
Since $A_{s-1}^2=0$, this is equivalent to
\begin{equation}\label{s-shift}
n_{s-1}(\ze)=\frac{\ze\bigl(\ze I +
(-q^{-s+1}I-A_{s-1})\bigr)}{(\ze-q^{-s+1})^2} n_s(\ze).
\end{equation}
On the other hand, if we substitute the first equation of the Lax
pair \eq{q-lax} into the second one, use \eq{ns} and simplify the
result, we obtain
\begin{equation}\label{zeta-shift}
n_s(q\ze)=M_s(\ze)\cdot \bigl((\ze-q^{-s})+A_s\bigr) n_s(\ze).
\end{equation}
Now if we let $x=\ze$, $t=q^{-s}$ and $n(x,t)=n_s(\ze)$, then with
the notation \eq{A}, \eq{B0}, \eq{B} of Theorem \ref{qp6}, the
system \eq{s-shift}, \eq{zeta-shift} leads to \eq{nxt}.
\end{rem}

\noindent
{\it Proof of Theorem \ref{qp6}.} (a) Since $A_{s-1}^2=0$, we have
\[
A(x,qt)B(x,t)=M_{s-1}(\ze)\cdot\bigl((\ze-q^{-s+1})I+A_{s-1}\bigr)\cdot
\frac{\ze\bigl((\ze-q^{-s+1})I-A_{s-1}\bigr)}{(\ze-q^{-s+1})^2}
=\ze M_{s-1}(\ze)
\]
and
\begin{eqnarray*}
B(qx,t)A(x,t)&=&\frac{q\ze\bigl(q(\ze-q^{-s})I-A_{s-1}\bigr)}{q^2(\ze-q^{-s})^2}\cdot
M_s(\ze)\cdot\bigl((\ze-q^{-s})I+A_s\bigr) \\
&=& \ze\cdot\left(I-\frac{q^{-1}A_{s-1}}{\ze-q^{-s}}\right)\cdot
M_s(\ze) \cdot\left(I+\frac{A_s}{\ze-q^{-s}}\right).
\end{eqnarray*}
Hence, if we multiply both sides of \eq{q-compatibility} by
$\bigl(I-(\ze-q^{-s})^{-1} q^{-1}
A_{s-1}\bigr)^{-1}=I+(\ze-q^{-s})^{-1} q^{-1} A_{s-1}$ and divide
by $\ze$, we obtain
\[
\left(I+\frac{q^{-1}A_{s-1}}{\ze-q^{-s}}\right)\cdot M_{s-1}(\ze)
=M_s(\ze) \cdot\left(I+\frac{A_s}{\ze-q^{-s}}\right).
\]
Replacing $s$ by $s+1$ yields \eq{q-comp}.

\smallbreak

\noindent
(b) Since the matrix
$D(\ze)=\mathrm{diag}(\la_1(\ze-a_3),\la_2(\ze-a_4))$ is linear in
$\ze$, it follows from Theorem \ref{t:Lax}(c) that we can write
\[
M_s(\ze)=\diag{\ka_1\ze}{\ka_2\ze}+C_s
\]
for a constant matrix $C_s$, where $\ka_1,\ka_2$ are given by
\eq{e:Adiag}. This immediately implies that the matrix $A(x,t)$ is
quadratic in $x$ with the leading coefficient
$A_2=\mathrm{diag}(\ka_1,\ka_2)$. Since $\det M_s(\ze)=\det
D(\ze)$ for all $\ze$ by Lemma \ref{determinants}, and the matrix
$A_s$ is nilpotent, we see from \eq{A} that $\det
A(x,t)=\ka_1\ka_2(x-t)^2(x-a_3)(x-a_4)$. In particular, taking
$x=0$, we see that $\det A_0(t)=t^2\de$, where $\de$ is a constant
independent of $t$. To complete the proof, it therefore suffices
to show that $\tr A_0(t)=t\tau$, where $\tau$ is a constant
independent of $t$. To this end, we study the compatibility
condition \eq{q-comp}. Taking residues of both sides of
\eq{q-comp} at $\ze=q^{-s-1}$ yields
\begin{equation}\label{resi}
A_s\cdot\left\{\diag{\ka_1 q^{-s-1}}{\ka_2 q^{-s-1}}+C_s\right\}=
q\cdot\left\{\diag{\ka_1 q^{-s-1}}{\ka_2
q^{-s-1}}+C_{s+1}\right\}\cdot A_{s+1}.
\end{equation}
If we compare the asymptotics of both sides of \eq{q-comp} as
$\ze\to\infty$, we get
\begin{equation}\label{asy}
C_s+A_s\cdot\diag{\ka_1 q^{-1}}{\ka_2 q^{-1}} =
C_{s+1}+\diag{\ka_1}{\ka_2}\cdot A_{s+1}.
\end{equation}
Multiplying \eq{asy} by $q^{-s}$ and subtracting it from \eq{resi}
gives
\[
(A_s-q^{-s}I)C_s=qC_{s+1}(A_{s+1}-q^{-s-1}I).
\]
Since $A_0(t)=C_s(A_s-q^{-s}I)$, this shows that $\tr
A_0(qt)=q\cdot\tr A_0(t)$ for all $t$. \qed

\subsection{Solution of the $q-P_{VI}$ system} We are now in a
position to quote a result of M.~Jimbo and H.~Sakai \cite{JS}. The
situation considered in their work is more general: it is assumed
that $B(x,t)$ has the form
\begin{equation}\label{e:B-JS}
B(x,t)=\frac{x\bigl(xI+B_0(t)\bigr)}{(x-qta_1)(x-qta_2)},
\end{equation}
and instead of \eq{e:detA}, it is assumed that
\begin{equation}\label{e:detA-JS}
\det A(x,t) = \ka_1 \ka_2 (x-ta_1)(x-ta_2)(x-a_3)(x-a_4).
\end{equation}
Thus, the situation of Theorem \ref{qp6} corresponds to the
specialization $a_1=a_2=1$. Now we have
\begin{thm}[Jimbo--Sakai, \cite{JS}]\label{t:JS}
Assume \eq{e:B-JS}, \eq{e:Axt}, \eq{e:Adiag}, \eq{e:A0} and
\eq{e:detA-JS}. Also, suppose that $\ka_j,\te_j\neq 0$ {\rm
(}$j=1,2${\rm )}, $a_j\neq 0$ {\rm (}$j=1,2,3,4${\rm )} and
$\ka_1\neq\ka_2$.
\begin{enumerate}[(a)]
\item Define $y=y(t)$ and $z_i=z_i(t)$ ($i=1,2$) by
\[
A_{12}(y,t)=0,\qquad A_{11}(y,t)=\ka_1 z_1,\qquad
A_{22}(y,t)=\ka_2 z_2,
\]
where $A_{ij}(x,t)$ are the elements of the matrix $A(x,t)$, so
that $z_1 z_2 = (y-ta_1)(y-ta_2)(y-a_3)(y-a_4)$. In terms of
$y,z_1,z_2$, the matrix $A(x,t)$ can be parameterized as follows:
\[
A(x,t)=\matr{\ka_1\bigl((x-y)(x-\al)+z_1\bigr)}{\ka_2
w(x-y)}{\ka_1 w^{-1}(\ga
x+\de)}{\ka_2\bigl((x-y)(x-\be)+z_2\bigr)},
\]
where
\[
\al=\frac{1}{\ka_1-\ka_2}\Bigl[ y^{-1}\bigl((\te_1+\te_2)t-\ka_1
z_1-\ka_2 z_2\bigr)-\ka_2\bigl((a_1+a_2)t+a_3+a_4-2y\bigr)\Bigr],
\]
\[
\be=\frac{1}{\ka_1-\ka_2}\Bigl[-y^{-1}\bigl((\te_1+\te_2)t-\ka_1
z_1-\ka_2 z_2\bigr)+\ka_1\bigl((a_1+a_2)t+a_3+a_4-2y\bigr)\Bigr],
\]
\[
\ga=z_1+z_2+(y+\al)(y+\be)+(\al+\be)y-a_1 a_2
t^2-(a_1+a_2)(a_3+a_4)t-a_3 a_4,
\]
\[
\de=y^{-1}\bigl(a_1 a_2 a_3 a_4 t^2 - (\al y+z_1)(\be
y+z_2)\bigr).
\]
\item Define $z=z(t)$ by
\[
z_1=\frac{(y-ta_1)(y-ta_2)}{q\ka_1 z},\qquad
z_2=q\ka_1(y-a_3)(y-a_4)z.
\]
Introduce the notation $\yb=y(qt)$, $\zb=z(qt)$, $\wb=w(qt)$, and
set
\[
b_1=\frac{a_1 a_2}{\te_1},\quad b_2=\frac{a_1 a_2}{\te_2},\quad
b_3=\frac{1}{q\ka_1},\quad b_4=\frac{1}{\ka_2}.
\]
Then the compatibility condition \eq{q-compatibility} is
equivalent to the following system of equations:
\begin{gather}
\label{JS1} \frac{y\yb}{a_3
a_4}=\frac{(\zb-tb_1)(\zb-tb_2)}{(\zb-b_3)(\zb-b_4)}, \\
\label{JS2} \frac{z\zb}{b_3
b_4}=\frac{(y-ta_1)(y-ta_2)}{(y-a_3)(y-a_4)}, \\
\label{JS3}
\frac{\wb}{w}=\frac{b_4}{b_3}\cdot\frac{\zb-b_3}{\zb-b_4}.
\end{gather}
\end{enumerate}
\end{thm}
\begin{rem}
Equations \eq{JS1}--\eq{JS3} allow us to compute $(\yb,\zb,\wb)$
if we know $(y,z,w)$, and vice versa.
\end{rem}
\begin{rem}
Unfortunately, it was beyond our technical abilities to follow the
proof of Theorem \ref{t:JS} in \cite{JS}. However, we were able to
verify the statement of the theorem using computer simulations
with random values of the parameters $\ka_j, a_j, \te_j$.
\end{rem}

\subsection{Degeneration of $q-P_{VI}$}\label{ss:degeneration} We have mentioned
in subsection \ref{ss:JS} that the compatibility conditions for
Lax pairs corresponding to certain families of orthogonal
polynomials are equivalent not to the $q-P_{VI}$ system but to a degeneration of it. We
now describe this degeneration.
\begin{thm}\label{t:degeneration}
With the notation of Theorem \ref{qp6}, suppose that the matrix
$D(\ze)=\mathrm{diag}(d_1(\ze),d_2(\ze))$ is such that
$d_1(\ze)=\la_1(\ze-a^{\circ}_3)$, $d_2(\ze)=\la_2$, where
$\la_j,a^{\circ}_3\neq 0$. Then
\begin{enumerate}[(a)]
\item The matrix $A(x,t)$ is quadratic in $x$, and if we write
\begin{equation}\label{e:Axt-d}
A(x,t)=A_0(t)+A_1(t)x+A_2 x^2,
\end{equation}
we have:
\begin{gather}
\label{e:Adiag-d} A_2=\diag{\ka^{\circ}_1}{0},\quad \ka^{\circ}_1=q^k\la_1, \\
\label{e:A0-d} A_0(t)\ \mathrm{has\ eigenvalues}\ t\te^{\circ}_1,\,t\te^{\circ}_2, \\
\label{e:detA-d} \det A(x,t)=\ka^{\circ}_1 \ka^{\circ}_2 (x-t)^2
(x-a^{\circ}_3),\quad \ka^\circ_2=q^{-k}\la_2,
\end{gather}
and the parameters $\ka^{\circ}_j$, $\te^{\circ}_j$, $a^{\circ}_3$
are independent of $t$.
\item We can parameterize the matrix $A(x,t)$ as follows:
\begin{equation}\label{e:param}
A(x,t)=\matr{\ka^{\circ}_1\bigl((x-y^{\circ})(x-\al^{\circ})+z^{\circ}_1\bigr)}
{\ka^{\circ}_2w^{\circ}(x-y^{\circ})}{\ka^{\circ}_1
(w^{\circ})^{-1}(\ga^{\circ}
x+\de^{\circ})}{\ka^{\circ}_2(x-y^{\circ}+z^{\circ}_2)},
\end{equation}
where
\[
\al^\circ=\frac{1}{\ka^\circ_1}\Bigl[
(y^\circ)^{-1}\bigl((\te^\circ_1+\te^\circ_2)t-\ka^\circ_1
z^\circ_1-\ka^\circ_2 z^\circ_2\bigr)+\ka^\circ_2\Bigr],
\]
\[
\ga^\circ=z^\circ_2-(y^\circ+\al^\circ)-y^\circ+2t+a^\circ_3,
\]
\[
\de^\circ=(y^\circ)^{-1}\bigl(-a^\circ_3 t^2 + (\al^\circ
y^\circ+z^\circ_1)(y^\circ-z^\circ_2)\bigr).
\]
Define $z^\circ=z^\circ(t)$ by
\[
z^\circ_1=\frac{(y^\circ-t)^2}{q\ka^\circ_1 z^\circ},\qquad
z^\circ_2=q\ka^\circ_1 (y-a^\circ_3) z^\circ.
\]
Introduce the notation $\yb^\circ=y^\circ(qt)$,
$\zb^\circ=z^\circ(qt)$, $\wb^\circ=w^\circ(qt)$, and set
\[
b^\circ_1=\frac{1}{\te^\circ_1},\quad
b^\circ_2=\frac{1}{\te^\circ_2},\quad
b^\circ_3=\frac{1}{q\ka^\circ_1}.
\]
Then the compatibility condition \eq{q-compatibility} is
equivalent to the following system of equations:
\begin{gather}
\label{JS1-d} \frac{y^\circ\yb^\circ}{\ka^\circ_2 a^\circ_3}=
\frac{(\zb^\circ-tb^\circ_1)(\zb^\circ-tb^\circ_2)}{\zb^\circ-b^\circ_3}, \\
\label{JS2-d} \frac{z^\circ\zb^\circ}{b^\circ_3}=\frac{(y^\circ-t)^2}{\ka^\circ_2(y^\circ-a^\circ_3)}, \\
\label{JS3-d}
\frac{\wb^\circ}{w^\circ}=\frac{b^\circ_3-\zb^\circ}{b^\circ_3}.
\end{gather}
\end{enumerate}
\end{thm}
\begin{proof}
(a) The proof of this part is almost identical to that of Theorem
\ref{qp6}(b) and will therefore be omitted.

\noindent
(b) One checks directly that all formulas of \cite{JS} are
compatible with the following limit transition:
\[
\ka_1\to\ka^\circ_1,\quad a_1,a_2\to 1,\quad
\te_1\to\te^\circ_1,\quad \te_2\to\te^\circ_2,\quad a_3\to
a^\circ_3,\quad \al\to\al^\circ,\quad y\to y^\circ,\quad z_1\to
z^\circ_1,\quad z\to z^\circ,
\]
\[
\ka_2\to 0,\quad \ka_2 a_4\to -\ka^\circ_2,\quad \ka_2\be\to
-\ka^\circ_2,\quad \ka_2\ga\to\ka^\circ_2\ga^\circ,\quad
\ka_2\de\to\ka^\circ_2\de^\circ,\quad \ka_2 z_2\to\ka^\circ_2
z^\circ_2,\quad \ka_2 w\to\ka^\circ_2 w^\circ.
\]
This limit transition takes the parameterization of $A(x,t)$ given
in Theorem \ref{t:JS}(a) to the parameterization \eq{e:param}, and
it takes the system \eq{JS1}--\eq{JS3} to the system
\eq{JS1-d}--\eq{JS3-d}.
\end{proof}

\section{Applications: recurrence relations for some polynomials of the Askey scheme}\label{s:Charlier}

\subsection{Notation}
In this section we illustrate the results of
\S\S\ref{s:Lax}--\ref{s:Jimbo-Sakai} by considering several
specific examples: Charlier polynomials, Meixner polynomials,
Krawtchouk polynomials, $q$-Charlier polynomials, little
$q$-Laguerre/Wall polynomials, alternative $q$-Charlier
polynomials, little $q$-Jacobi polynomials and $q$-Krawtchouk
polynomials. In the first four cases we solve the compatibility
condition explicitly, we write down a recurrence relation for the
corresponding Fredholm determinants in terms of the solution, and
we provide the initial conditions for all of our recurrence
relations. This is done in subsections \ref{ss:charlier},
\ref{ss:meixner}, \ref{ss:krawtchouk} and \ref{ss:q-charlier}. For
the other four families, we contend ourselves with making some
general remarks in subsection \ref{ss:remarks}.

The general notation of this section is that of \S\ref{s:Fredholm}
and \S\ref{s:compatibility}. Recall that we are considering a
family $\{P_n(\ze)\}_{n=0}^N$ of monic polynomials orthogonal on a
discrete, but not necessarily locally finite subset
$\fX=\{\pi_x\}_{x=0}^N$ of $\bR$ (where $N\in\zp\cup\{\infty\}$),
with respect to a strictly positive weight function
$\om:\fX\rightarrow\bR$. If $k$ is a natural number, $k\leq N$, we
can consider a kernel $K$ on $\fX\times\fX$ defined by the formula
\begin{equation}\label{e:kernel}
K(\pi_x,\pi_y)=\left\{
\begin{array}{ll}
\norm{P_{k-1}}_\om^{-2}\sqrt{\om(x)\om(y)}
\dfrac{P_k(\pi_x)P_{k-1}(\pi_y)-P_{k-1}(\pi_x)P_k(\pi_y)}
{(\pi_x-\pi_y)},
& x\neq y, \\
\norm{P_{k-1}}_\om^{-2}\om(x)
\bigl(P_k'(\pi_x)P_{k-1}(\pi_x)-P_{k-1}'(\pi_x)P_k(\pi_x)\bigr), &
x=y,
\end{array}\right.
\end{equation}
where $\norm{P_{k-1}}_\om=(P_{k-1},P_{k-1})_\om^{1/2}$ denotes the
norm of $P_{k-1}(\ze)$ with respect to the inner product defined
by $\om$. Up to conjugation, this coincides with the kernel
introduced in the beginning of \S\ref{s:Fredholm} (see equation
\eq{eq:kernel}). For all $k\leq s\leq N$, we define a subset
$\fY_s=\{\pi_x\}_{x=s}^N\subseteq\fX$, and we are interested in
the Fredholm determinants
\begin{equation}\label{e:det}
D_s=\det\bigl(1-K\big\vert_{\fY_s\times\fY_s}\bigr).
\end{equation}

\subsection{Charlier polynomials (\cite{KS}, \S1.12)}\label{ss:charlier}
The \textit{$n$-th Charlier polynomial} is defined by
\begin{equation}\label{e:ch}
C_n(x;a)=\hypergeom{2}{0}{-n,-x}{-}{-\frac{1}{a}}.
\end{equation}
These polynomials satisfy the orthogonality relation 
\begin{equation}\label{e:ch-o}
\sum_{x=0}^\infty \frac{a^x}{x!} C_m(x;a) C_n(x;a) = a^{-n} e^a n!
\de_{mn},
\end{equation}
where $a>0$. Thus the orthogonality set for Charlier polynomials
is $\fX=\zp$, and the weight function is $\om(x)=\frac{a^x}{x!}$.
The polynomial $C_n(x;a)$ is not monic in general; in fact, its
leading coefficient is $(-a)^{-n}$. Hence the corresponding family
of orthogonal polynomials (recall that the orthogonal polynomials that we use are monic, see \S\ref{s:DRHP}) is $\{P_n(\ze)=(-a)^n
C_n(\ze;a)\}_{n=0}^\infty$. We call $P_n(\ze)$ the \textit{$n$-th
normalized Charlier polynomial}. Now from \eq{e:ch-o}, we find
that $(P_n,P_n)_\om=a^n e^a n!$ for all $n\geq 0$. After these
preliminaries, we can state our main result for Charlier
polynomials.
\begin{thm}\label{t:Charlier}
If $K$ is the kernel \eq{e:kernel} corresponding to the family
$\{P_n(\ze)\}_{n=0}^\infty$ of normalized Charlier polynomials,
then the Fredholm determinants $D_s$ defined by \eq{e:det} can be
computed from the following recurrence relation:
\begin{equation}\label{e:ch-fr}
\frac{D_{s+2}}{D_{s+1}}-\frac{D_{s+1}}{D_s}=
\frac{a^{s-1}}{(s+1)!}\cdot\frac{f_s^2}{e_s} \cdot (g_s-s-1) \cdot
h_s^2.
\end{equation}
Here, the scalar sequences $\{e_s\}_{s\geq k}$, $\{f_s\}_{s\geq
k}$, $\{g_s\}_{s\geq k}$ and $\{h_s\}_{s\geq k}$ satisfy the
following recurrence relations:
\begin{gather}
\label{e:ch-e}
e_{s+1}= \frac{a e_s}{f_s(g_s+k-s-1)}, \\
\label{e:ch-f}
f_{s+1} = \frac{a g_s}{f_s(g_s-s-1)(g_s+k-s-1)}, \\
\label{e:ch-g}
g_{s+1} = \frac{a}{f_{s+1}}-\frac{s+1}{1-f_{s+1}}-g_s-k+2s+3, \\
\label{e:ch-h} h_{s+1}=a^{-1}\cdot f_s\cdot(g_s-s-1)\cdot h_s.
\end{gather}
The initial conditions for the recurrence relations
\eq{e:ch-fr}--\eq{e:ch-h} are given by
\begin{gather}
\label{ch-fr-init}
D_k=e^{-ak},\quad D_{k+1}=e^{-ak}\cdot \Phi(-k;1;-a), \\
\label{ch-e-init}
e_k=\frac{a^k\cdot (k-1)!}{\Phi(1-k;1;-a)}, \\
\label{ch-f-init}
f_k=-\frac{a\cdot \Phi(1-k;2;-a)}{\Phi(1-k;1;-a)}, \\
\label{ch-g-init} g_k=k+1-\frac{(k+1)\cdot \Phi(1-k;1;-a)\cdot
\Phi(-k;2;-a)}
{\Phi(-k;1;-a)\cdot \Phi(1-k;2;-a)}, \\
\label{ch-h-init} h_k=k!,
\end{gather}
where
\[
\Phi(u;w;z)=\hypergeom{1}{1}{u}{w}{z}.
\]
\end{thm}
\begin{rem}\label{r:ch-dp4}
As we have have seen in \S\ref{s:dP} (cf. Remark
\ref{r:dP-Sak}(b)), the recurrence relations
\eq{e:ch-f},~\eq{e:ch-g} form a special case of the $d-P_{IV}$
equation of \cite{Sak}.
\end{rem}
\begin{proof}
The proof is a straightforward application of our previous results. For the reader's convenience we provide some
remarks; the same ones apply to Theorems \ref{t:Meixner} and
\ref{t:Krawtchouk}, and hence the proofs of those results will be
omitted.

To make the notation of the present section more uniform, we have
been writing $e_s$ for $\be_s$ and $h_s$ for $m_s^{11}$, where
$\{\be_s\}$ and $\{m_s^{11}\}$ are the scalar sequences defined in
\S\ref{s:dP} and \S\ref{s:Fredholm}, respectively. The symbols
$f_s$ and $g_s$ have the same meaning as in \S\ref{s:dP}. Then the
recurrence relations \eq{e:ch-fr}--\eq{e:ch-h} are obtained
directly from Theorem \ref{t:Fredholm}(b) and Theorem
\ref{t:dP}(b). To find the initial conditions, one uses the
definitions of $b_k,\al_k,\be_k,f_k,g_k$, together with
Propositions \ref{mk-init}, \ref{Ak-init}, \ref{fr-init}, and the
obvious identities $h_k=m_k^{11}=k!$, $b_k\be_k=-q_k$, which
follow from \eq{e:mk}, \eq{e:Ck-lin}.
\end{proof}

\subsection{}\label{ss:charlier-diff}
We now illustrate the concluding remark of \S\ref{s:Lax} by
showing how one can use Proposition \ref{p:Lax} to obtain
difference equations satisfied by orthogonal polynomials.
\begin{prop}[cf. \cite{KS}, \S1.12, equation (1.12.5)]
\label{p:ch-diff} The $k$-th normalized Charlier polynomial
$P_k(\ze)$ solves the following difference equation:
\begin{equation}\label{e:ch-diff}
-k P_k(\ze) = a P_k(\ze+1) - (\ze+a) P_k(\ze) + \ze P_k(\ze-1).
\end{equation}
\end{prop}
\begin{proof}
We use the notation of Proposition \ref{p:Lax}. Recall that in the
case of Charlier polynomials, we have
$D(\ze)=\mathrm{diag}(\ze,a)$. We also observe that from the proof
of Theorem \ref{t:DRHP} it follows that the matrix $\mx(\ze)$ has
a full asymptotic expansion in $\ze$ as $\ze\to\infty$; in
particular, we can write, by \eq{asymp},
\[
\mx(\ze)\cdot\diag{\ze^{-k}}{\ze^k}=
I+\matr{\al}{\be}{\ga}{\de}\cdot\ze^{-1} + O(\ze^{-2}).
\]
Therefore, as $\ze\to\infty$, we have
\begin{eqnarray*}
M(\ze)&=&\mx(\ze-1)\cdot D(\ze)\cdot \mx^{-1}(\ze) \\
&=& \mx(\ze-1)\cdot \diag{\ze^{-k}}{\ze^k}\cdot \diag{\ze}{a}\cdot
\diag{\ze^k}{\ze^{-k}}\cdot \mx^{-1}(\ze) \\
&=& \left\{ \mx(\ze-1)\cdot\diag{(\ze-1)^{-k}}{(\ze-1)^k} \right\}
\cdot \diag{(1-1/\ze)^k}{(1-1/\ze)^{-k}}\cdot\diag{\ze}{a} \\
&&\times\left\{\mx(\ze)\cdot\diag{\ze^{-k}}{\ze^k}\right\}^{-1}
\end{eqnarray*}
\begin{eqnarray*}
&=& \matr{1+\al\ze^{-1}}{\be\ze^{-1}}{\ga\ze^{-1}}{1+\de\ze^{-1}}
\cdot \diag{\ze-k}{a} \cdot
\matr{1-\al\ze^{-1}}{-\be\ze^{-1}}{-\ga\ze^{-1}}{1-\de\ze^{-1}}+
O(\ze^{-1}) \\
&=& \matr{\ze-k}{-\be}{\ga}{a} + O(\ze^{-1}).
\end{eqnarray*}
Since $M(\ze)$ is entire by Proposition \ref{p:Lax}, the last term
$O(\ze^{-1})$ is identically zero by Liouville's theorem. Hence
the system of equations \eq{eq1}, \eq{eq2} takes the following
form:
\begin{eqnarray}
\label{eq3}
\ze\cdot P_k(\ze-1)&=&(\ze-k)\cdot P_k(\ze) - \be\cdot c P_{k-1}(\ze), \\
\label{eq4} \ze\cdot c P_{k-1}(\ze-1)&=&\ga\cdot P_k(\ze) + a\cdot
c P_{k-1}(\ze).
\end{eqnarray}
Note that since $\det M(\ze)=\det D(\ze)$ for all $\ze$ by Lemma
\ref{determinants}, we have $\be\ga=ak$; in particular, $\be\neq
0$. Now from \eq{eq3}, we find that
\[
c P_{k-1}(\ze) = \frac{1}{\be}\cdot \bigl[(\ze-k)P_k(\ze)-\ze
P_k(\ze-1)\bigr].
\]
Substituting this into \eq{eq4}, multiplying the result by $\be$
and using $\be\ga=ak$, we obtain
\[
\ze\cdot\bigl[(\ze-1-k)P_k(\ze-1)-(\ze-1)P_k(\ze-2)\bigr] =
ak\cdot P_k(\ze) + a\cdot\bigl[(\ze-k)P_k(\ze)-\ze
P_k(\ze-1)\bigr].
\]
Dividing the last equation by $\ze$ and replacing $\ze$ by
$\ze+1$, we find that it is equivalent to \eq{e:ch-diff}.
\end{proof}

\subsection{Meixner polynomials (\cite{KS}, \S1.9)}\label{ss:meixner}
The
\textit{$n$-th Meixner polynomial} is defined by
\begin{equation}\label{e:me}
M_n(x;\be,c)=\hypergeom{2}{1}{-n,-x}{\be}{1-\frac{1}{c}}.
\end{equation}
These polynomials satisfy the orthogonality relation 
\begin{equation}\label{e:me-o}
\sum_{x=0}^\infty \frac{(\be)_x}{x!} c^x M_m(x;\be,c) M_n(x;\be,c)
= \frac{c^{-n} n!}{(\be)_n (1-c)^\be} \de_{mn},
\end{equation}
where $\be>0$ and $0<c<1$. Thus the orthogonality set for Meixner
polynomials is $\fX=\zp$, and the weight function is
$\om(x)=\frac{(\be)_x}{x!}c^x$. The leading coefficient of the
polynomial $M_n(x;\be,c)$ is $\frac{(1-1/c)^n}{(\be)_n}$, so the
corresponding family of monic orthogonal polynomials is
$\{P_n(\ze)=(\be)_n (1-1/c)^{-n} M_n(\ze;\be,c)\}_{n=0}^\infty$.
We call $P_n(\ze)$ the \textit{$n$-th normalized Meixner
polynomial}. Now from \eq{e:me-o}, we find that
$(P_n,P_n)_\om=\frac{(\be)_n c^n n!}{(1-c)^{\be+2n}}$ for all
$n\geq 0$. Then we have
\begin{thm}\label{t:Meixner}
If $K$ is the kernel \eq{e:kernel} corresponding to the family
$\{P_n(\ze)\}_{n=0}^\infty$ of normalized Meixner polynomials,
then the Fredholm determinants $D_s$ defined by \eq{e:det} can be
computed from the following recurrence relation:
\begin{equation}\label{e:me-fr}
\frac{D_{s+2}}{D_{s+1}}-\frac{D_{s+1}}{D_s}=
\frac{(\be)_s}{\be+s}\cdot \frac{c^{s-1}}{(s+1)!}\cdot \frac{1+c
g_s}{e_s g_s^2}\cdot \bigl[(1+c g_s)f_{s+1}-s-1\bigr] \cdot h_s^2.
\end{equation}
Here, the scalar sequences $\{e_s\}_{s\geq k}$, $\{f_s\}_{s\geq
k}$, $\{g_s\}_{s\geq k}$ and $\{h_s\}_{s\geq k}$ satisfy the
following recurrence relations:
\begin{gather}
\label{e:me-e}
e_{s+1}= -\frac{c e_s g_s}{g_{s+1}}\cdot \frac{(1+g_{s+1})f_{s+1}+(\be+k-1)g_{s+1}-s-1}{(1+c g_s)f_{s+1}+k-s-1}, \\
\label{e:me-f}
f_{s+1} = 1-\be-k-f_s+\frac{s}{1+g_s}+\frac{\be+s}{1+c g_s}, \\
\label{e:me-g}
g_{s+1} = \frac{(f_{s+1}-1-s)(f_{s+1}-1-s+k)}{cg_sf_{s+1}(f_{s+1}+\be+k-1)}, \\
\label{e:me-h} h_{s+1}=\frac{(1+c
g_s)(s+1-f_{s+1})}{c(\be+s)g_s}\cdot h_s.
\end{gather}
The initial conditions for the recurrence relations
\eq{e:me-fr}--\eq{e:me-h} are given by
\begin{gather}
\label{me-fr-init}
D_k=(1-c)^{k(\be+k-1)},\quad D_{k+1}=\frac{(\be)_k}{k!}\cdot c^k\cdot (1-c)^{k(\be+k-1)} \cdot F(-k,-k;\be;1/c), \\
\label{me-e-init}
e_k=\frac{\be c\cdot (k-1)!^2}{F(1-k,1-k;1+\be;1/c)}, \\
\label{me-f-init}
f_k=0, \\
\label{me-g-init}
g_k=\frac{k}{\be c}\cdot \frac{F(1-k,1-k;1+\be;1/c)}{F(-k,1-k;\be;1/c)}, \\
\label{me-h-init} h_k=k!,
\end{gather}
where
\[
F(u,v;w;z)=\hypergeom{2}{1}{u,v}{w}{z}.
\]
\end{thm}
\begin{proof}
See the proof of Theorem \ref{t:Charlier}.
\end{proof}
\begin{rem}\label{r:me-dp5}
As we have have seen in \S\ref{s:dP} (cf. Remark
\ref{r:dP-Sak}(a)), the recurrence relations
\eq{e:me-f},~\eq{e:me-g} form a special case of the $d-P_{V}$
equation of \cite{Sak}.
\end{rem}

\subsection{Krawtchouk polynomials (\cite{KS}, \S1.10)}
\label{ss:krawtchouk}
The
\textit{$n$-th Krawtchouk polynomial} is defined by
\begin{equation}\label{e:kr}
K_n(x;p,N)=\hypergeom{2}{1}{-n,-x}{-N}{\frac{1}{p}}.
\end{equation}
These polynomials satisfy the orthogonality relation 
\begin{equation}\label{e:kr-o}
\sum_{x=0}^N {N \choose x} p^x (1-p)^{N-x} K_m(x;p,N) K_n(x;p,N) =
\frac{(-1)^n n!}{(-N)_n} \left(\frac{1-p}{p}\right)^n \de_{mn},
\end{equation}
where $N\in\zp$ and $0<p<1$. Thus the orthogonality set for
Krawtchouk polynomials is $\fX=\{0,\dotsc,N\}$, and the weight
function is $\om(x)={N \choose x} p^x (1-p)^{N-x}$. The leading
coefficient of the polynomial $K_n(x;p,N)$ is $(-N)_n^{-1}
p^{-n}$, so the corresponding family of monic orthogonal polynomials is
$\{P_n(\ze)=(-N)_n p^n K_n(\ze;p,N)\}_{n=0}^N$. We call $P_n(\ze)$
the \textit{$n$-th normalized Krawtchouk polynomial}. Now from
\eq{e:kr-o}, we find that $(P_n,P_n)_\om=(-1)^n n! (-N)_n p^n
(1-p)^n$ for all $0\leq n\leq N$. Then we have
\begin{thm}\label{t:Krawtchouk}
If $K$ is the kernel \eq{e:kernel} corresponding to the family
$\{P_n(\ze)\}_{n=0}^N$ of normalized Krawtchouk polynomials, then
the Fredholm determinants $D_s$ defined by \eq{e:det} can be
computed from the following recurrence relation:
\begin{equation}\label{e:kr-fr}
\frac{D_{s+2}}{D_{s+1}}-\frac{D_{s+1}}{D_s}= {N\choose s+1}\cdot
\frac{p^{s-1}(1-p)^{N-s+1}}{(N-s)^2}\cdot \frac{1+p g_s/(p-1)}{e_s
g_s^2}\cdot \Bigl[\bigl(1+p g_s/(p-1)\bigr)f_{s+1}-s-1\Bigr] \cdot
h_s^2.
\end{equation}
Here, the scalar sequences $\{e_s\}_{s\geq k}$, $\{f_s\}_{s\geq
k}$, $\{g_s\}_{s\geq k}$ and $\{h_s\}_{s\geq k}$ satisfy the
following recurrence relations:
\begin{gather}
\label{e:kr-e} e_{s+1}= \frac{p e_s g_s}{(1-p)g_{s+1}}\cdot
\frac{(1+g_{s+1})f_{s+1}+(k-N-1)g_{s+1}-s-1}
{\bigl(1+p g_s/(p-1)\bigr)f_{s+1}+k-s-1}, \\
\label{e:kr-f}
f_{s+1} = N+1-k-f_s+\frac{s}{1+g_s}+\frac{(1-p)(N-s)}{p-1+p g_s}, \\
\label{e:kr-g} g_{s+1} =
\frac{(1-p)(f_{s+1}-1-s)(f_{s+1}-1-s+k)}{pg_sf_{s+1}(N+1-k-f_{s+1})}, \\
\label{e:kr-h} h_{s+1}=\frac{(p-1+p
g_s)(f_{s+1}-s-1)}{p(N-s)g_s}\cdot h_s.
\end{gather}
The initial conditions for the recurrence relations
\eq{e:kr-fr}--\eq{e:kr-h} are given by
\begin{gather}
\label{kr-fr-init}
D_k=(1-p)^{k(N+1-k)},\quad D_{k+1}={N \choose k}\cdot p^k\cdot (1-p)^{k(N-k)} \cdot F(-k,-k;-N;1-1/p), \\
\label{kr-e-init}
e_k=\frac{Np(1-p)^{N-1}(k-1)!^2}{F(1-k,1-k;1-N;1-1/p)}, \\
\label{kr-f-init}
f_k=0, \\
\label{kr-g-init}
g_k=\frac{k(1-p)}{Np}\cdot \frac{F(1-k,1-k;1-N;1-1/p)}{F(-k,1-k;-N;1-1/p)}, \\
\label{kr-h-init} h_k=k!,
\end{gather}
where
\[
F(u,v;w;z)=\hypergeom{2}{1}{u,v}{w}{z}.
\]
\end{thm}
\begin{proof}
See the proof of Theorem \ref{t:Charlier}.
\end{proof}
\begin{rem}\label{r:kr-dp5}
As we have have seen in \S\ref{s:dP} (cf. Remark
\ref{r:dP-Sak}(a)), the recurrence relations
\eq{e:kr-f},~\eq{e:kr-g} form a special case of the $d-P_{V}$
equation of \cite{Sak}.
\end{rem}

\subsection{$q$-Charlier polynomials (\cite{KS}, \S3.23)}\label{ss:q-charlier}
In this subsection, we assume that $q$ is a fixed real number,
$0<q<1$.The \textit{$n$-th $q$-Charlier polynomial} is defined by
\begin{equation}\label{e:q-ch}
C_n(\ze;a;q)=\qhypergeom{2}{1}{q^{-n},\ze}{0}{-\frac{q^{n+1}}{a}}.
\end{equation}
These polynomials satisfy the orthogonality relation 
\begin{equation}\label{e:q-ch-o}
\sum_{x=0}^\infty \frac{a^x}{(q;q)_x} q^{x\choose 2}
C_m(q^{-x};a;q) C_n(q^{-x};a;q) = q^{-n} (-a;q)_\infty
(-a^{-1}q,q;q)_n \de_{mn},
\end{equation}
where $a>0$ and
\[
(-a;q)_\infty = \prod_{l=0}^\infty (1+a q^l).
\]
Thus the orthogonality
set for $q$-Charlier polynomials is $\fX=\{q^{-x}\}_{x=0}^\infty$,
and the weight function is $\om(x)=\frac{a^x}{(q;q)_x}q^{x\choose
2}$. The leading coefficient of the polynomial $C_n(\ze;a;q)$ is
$(-1)^n q^{n^2} a^{-n}$, so the corresponding family of orthogonal
polynomials is $\bigl\{P_n(\ze)=(-1)^n a^n q^{-n^2}
C_n(\ze;a;q)\bigr\}_{n=0}^\infty$. We call $P_n(\ze)$ the
\textit{$n$-th normalized $q$-Charlier polynomial}. Now from
\eq{e:q-ch-o}, we find that
\[
(P_n,P_n)_\om = a^{2n} q^{-2n^2-n} (-a;q)_\infty (-a^{-1}q,q;q)_n
\]
for all $n\geq 0$. Then we have
\begin{thm}\label{t:q-Charlier}
If $K$ is the kernel \eq{e:kernel} corresponding to the family
$\{P_n(\ze)\}_{n=0}^\infty$ of normalized $q$-Charlier
polynomials, then the Fredholm determinants $D_s$ defined by
\eq{e:det} satisfy the following recurrence relation:
\begin{equation}\label{e:q-ch-fr}
\frac{D_{s+2}}{D_{s+1}}-\frac{D_{s+1}}{D_s} =
\frac{a^{s-1}}{(q;q)_{s+1}}\cdot q^{s+1\choose 2}\cdot u_s\cdot
h_s^2,
\end{equation}
where
\begin{equation}\label{e:q-ch-us}
u_s=q^k\cdot\frac{p_s}{q_s^2}\cdot(p_s\be_s+aq^{-k}q_s)
\end{equation}
for all $s\geq 0$. The scalar sequences $\{p_s\}_{s\geq k}$,
$\{q_s\}_{s\geq k}$, $\{\be_s\}_{s\geq k}$ and $\{h_s\}_{s\geq k}$
can be computed from the following recurrence relations (which
involve additional sequences):
\begin{gather}
\label{e:q-ch-eps} \eps_s=a(q^{-s-1}-1)+q^{k-1}(aq^{-k}p_s-r_s\be_s), \\
\label{e:q-ch-p}
p_{s+1}=-q^{-1}p_s^{-1}\eps_s^{-1}\cdot(p_s\be_s+aq^{-k}q_s)\cdot(r_s\al_s-p_s\ga_s+q^{k-s-1}r_s), \\
\label{e:q-ch-q}
q_{s+1}=q^{-1}q_s^{-1}\eps_s^{-1}\cdot(p_s\be_s+aq^{-k}q_s)^2, \\
\label{e:q-ch-r}
r_{s+1} = q^{-1}r_s^{-1}\eps_s^{-1}\cdot (r_s\al_s-p_s\ga_s+q^{k-s-1}r_s)^2, \\
\label{e:q-ch-al}
\al_{s+1}=\al_s+q^{k-1}p_s-q^k p_{s+1}, \\
\label{e:q-ch-be}
\be_{s+1}=\be_s-q^k q_{s+1}, \\
\label{e:q-ch-ga}
\ga_{s+1}=\ga_s+q^{k-1} r_s, \\
\label{e:q-ch-h}
h_{s+1}=\frac{p_s\be_s+aq^{-k}q_s}{aq_s}\cdot h_s.
\end{gather}
The initial conditions for the recurrence relations \eq{e:q-ch-fr}
and \eq{e:q-ch-p}--\eq{e:q-ch-h} are provided by
\begin{gather}
\label{q-ch-dk}
D_k=(-a;q)_\infty^{-k}\cdot a^{-{k\choose 2}}\cdot
\prod_{n=0}^{k-1} \bigl[ (-a^{-1}q;q)_n^{-1} q^{n+1\choose 2}
\bigr], \\
\label{q-ch-dk1}
D_{k+1}=(-a;q)_\infty^{-k}\cdot\frac{a^{k-{k\choose
2}}}{(q;q)_k}\cdot q^{-{k\choose 2}}\cdot
G_q(q^{-k},q^{-k};-q^{2k}/a) \cdot \prod_{n=0}^{k-1} \bigl[
(-a^{-1}q;q)_n^{-1} q^{n+1\choose 2}
\bigr], \\
\label{q-ch-p-init}
p_k=(1-q^{-k})\cdot\frac{G_q(q^{-k},q^{1-k};-q^{2k-1}/a)}{G_q(q^{-k},q^{-k};-q^{2k}/a)}, \\
\label{q-ch-q-init}
q_k=(q;q)_k^2\cdot q^{-k(k+1)}\cdot G_q(q^{-k},q^{-k};-q^{2k}/a)^{-1}, \\
\label{q-ch-r-init}
r_k=\frac{q^{k^2}(1-q^{-k})}{(q;q)_k (q;q)_{k-1}}
\cdot\frac{G_q(q^{-k},q^{1-k};-q^{2k-1}/a)^2}{G_q(q^{-k},q^{-k};-q^{2k}/a)}, \\
\label{q-ch-al-init}
\al_k=-1-q^k p_k, \\
\label{q-ch-be-init}
\be_k=-q^k q_k, \\
\label{q-ch-ga-init}
\ga_k=\frac{q^{k^2-1}}{(q;q)_{k-1}^2}\cdot G_q(q^{1-k},q^{1-k};-q^{2k-2}/a), \\
\label{q-ch-h-init}
h_k=q^{-k^2}\cdot (q;q)_k,
\end{gather}
where
\[
G_q(u,v;z)=\qhypergeom{2}{0}{u,v}{-}{z}.
\]
\end{thm}
\begin{proof}
As before, the proof is quite straightforward. We have been
writing $h_s$ for $m_s^{11}$ (defined in \S\ref{s:Fredholm}). The
formulas \eq{e:q-ch-us} and \eq{e:q-ch-eps}--\eq{e:q-ch-ga} follow
immediately from Theorem \ref{t:general} (note that
$\de_s=aq^{-k}$ for all $s$, since we have $\de_{s+1}=\de_s$ from
\eq{e:des1}, and $\de_k=aq^{-k}$ from \eq{e:Mk-lin}). Then
\eq{e:q-ch-fr} and \eq{e:q-ch-h} are deduced from \eq{fr} and
\eq{ms11}, respectively. Finally, the initial conditions
\eq{q-ch-dk}--\eq{q-ch-h-init} are easily obtained from
Proposition \ref{fr-init}, \ref{Ak-init} and \ref{p:Mk}.
\end{proof}
\begin{rem}\label{r:q-ch}
As we have mentioned in subsection \ref{ss:JS}, the recurrence
relation \eq{e:q-ch-p}--\eq{e:q-ch-ga} for $q$-Charlier
polynomials is in fact equivalent to a certain degeneration of the
$q-P_{VI}$ equation of \cite{JS}. This is a special case of
Theorem \ref{t:degeneration}.
\end{rem}
\begin{rem}\label{r:q-charlier}
In the case of $q$-Charlier polynomials, it is possible to solve
the compatibility condition for the corresponding Lax pair by a
method similar to the one used in Theorem \ref{t:dP}(b). Namely,
it is easy to see that the Lax pair can be parameterized as
follows:
\[
m_{s+1}(\zeta)=\left\{I+(\zeta-q^{-s})^{-1}\left(
\begin{array}{cc}
p_s & p_s a_s c_s \\
-p_s\!/(a_s c_s) & -p_s
\end{array}\right)\right\}\cdot m_s(\ze),
\]
\[
m_s(q\zeta)=\left(
\begin{array}{cc}
q^k(\zeta-1)+b_s & b_s c_s \\
a q^{-k}\!/c_s & a q^{-k}
\end{array}\right) m_{s+1}(\zeta) \left(
\begin{array}{cc}
(\zeta-1)^{-1} & 0 \\
0 & a^{-1}
\end{array}\right).
\]
Then the compatibility condition gives the following recurrence
relations for the parameters $a_s,b_s,c_s,p_s$:
\[
p_{s+1}=\frac{p_s(b_s+aq^{-k}a_s)(q^{k-s-1}-q^k+b_s+aq^{-k}a_s)}{q^k
p_s(b_s+aq^{-k}a_s)+q\cdot (q^{k-s-1}-q^k)\cdot aq^{-k}a_s},
\]
\[
b_{s+1}=b_s + q^{k-1} p_s - q^k p_{s+1},
\]
\[
a_{s+1}=\frac{b_s-q^k p_{s+1}+aq^{-k}a_s} {q^{k-s-1}-q^k+b_s-q^k
p_{s+1}+aq^{-k}a_s},
\]
\[
c_{s+1}=\frac{aq^{-k}a_s c_s}{aq^{-k}a_s-q^{k-1}p_s}.
\]
With this notation, the recurrence relation for the Fredholm
determinants is \eq{e:q-ch-fr}, the same as before, but now we
have
\[
u_s=\frac{q^k\cdot(aq^{-k}+b_s/a_s)}{a_s c_s},
\]
and the recurrence relation for $h_s$ is given by
\[
h_{s+1}=a^{-1}\cdot(aq^{-k}+b_s/a_s)\cdot h_s.
\]
The initial values $a_k,b_k,c_k$ can be easily found from
\eq{q-ch-p-init}--\eq{q-ch-ga-init}.

The main difference between this situation and that of
\S\ref{s:dP} is that we cannot further reduce the recurrence
relations for $a_s,b_s,p_s$ to relations involving only two
sequences of parameters. So this method cannot be used to show
that our recurrence relations in the case of $q$-Charlier
polynomials are equivalent to one of H.~Sakai's $q$-difference
equations \cite{Sak}. From the computational point of view, this
method is slightly easier to use than the one presented in Theorem
\ref{t:q-Charlier}.
\end{rem}

\subsection{Concluding remarks}\label{ss:remarks}
It follows from Theorem \ref{qp6} that the recurrence relations
corresponding to the little $q$-Jacobi polynomials and the
$q$-Krawtchouk polynomials can be reduced to special cases of the
$q-P_{VI}$ system of \cite{JS}. Also, it follows from Theorem
\ref{t:degeneration} that the recurrence relations corresponding
to the $q$-Charlier polynomials and the little $q$-Laguerre/Wall
polynomials can be reduced to special cases of a certain
degeneration of the $q-P_{VI}$ system described in subsection
\ref{ss:degeneration}. However, it is more convenient to use the
formulas of \S\ref{s:general} for practical computations. In
addition, the method of \S\ref{s:general} covers the case of the
alternative $q$-Charlier polynomials, whereas we do not know if
the recurrence relation corresponding to these polynomials can be
reduced to one of the equations of H.~Sakai's hierarchy.

As far as using the formulas of \S\ref{s:general} is concerned,
there is no essential difference between the $q$-Charlier
polynomials and the other four families of basic hypergeometric
orthogonal polynomials that we consider here. So we have decided not to write
out explicitly the results we have obtained for these four
families. On the other hand, we have carried out all the
calculations for some specific values of parameters in Maple, and
in \S\ref{s:computations} we present a few plots of the ``density
function'' (difference or $q$-derivative of $D_s$) for the eight families of
orthogonal polynomials considered in this section.


\section{Numerical computations}\label{s:computations}

\subsection{} 
The plots in this section have been obtained in Maple by using the formulas of \S\ref{s:general} and subsections \ref{ss:charlier}, \ref{ss:meixner} and \ref{ss:krawtchouk} for the specific values of parameters indicated below.

\subsection{} 
The following are two plots of the density function $D_{s+1}-D_s$ for the family of Meixner polynomials. The parameters (cf. subsection \ref{ss:meixner} or subsection \ref{ss:list}) are $k=4$, $c=0.01$, $\be=3000$ for the first graph and $k=4$, $c=0.9$, $\be=0.5$ for the second graph. The $x$-coordinate in each case is $s$.

\medbreak

\begin{center}
\rotatebox[origin=c]{270}{\includegraphics[scale=0.24]{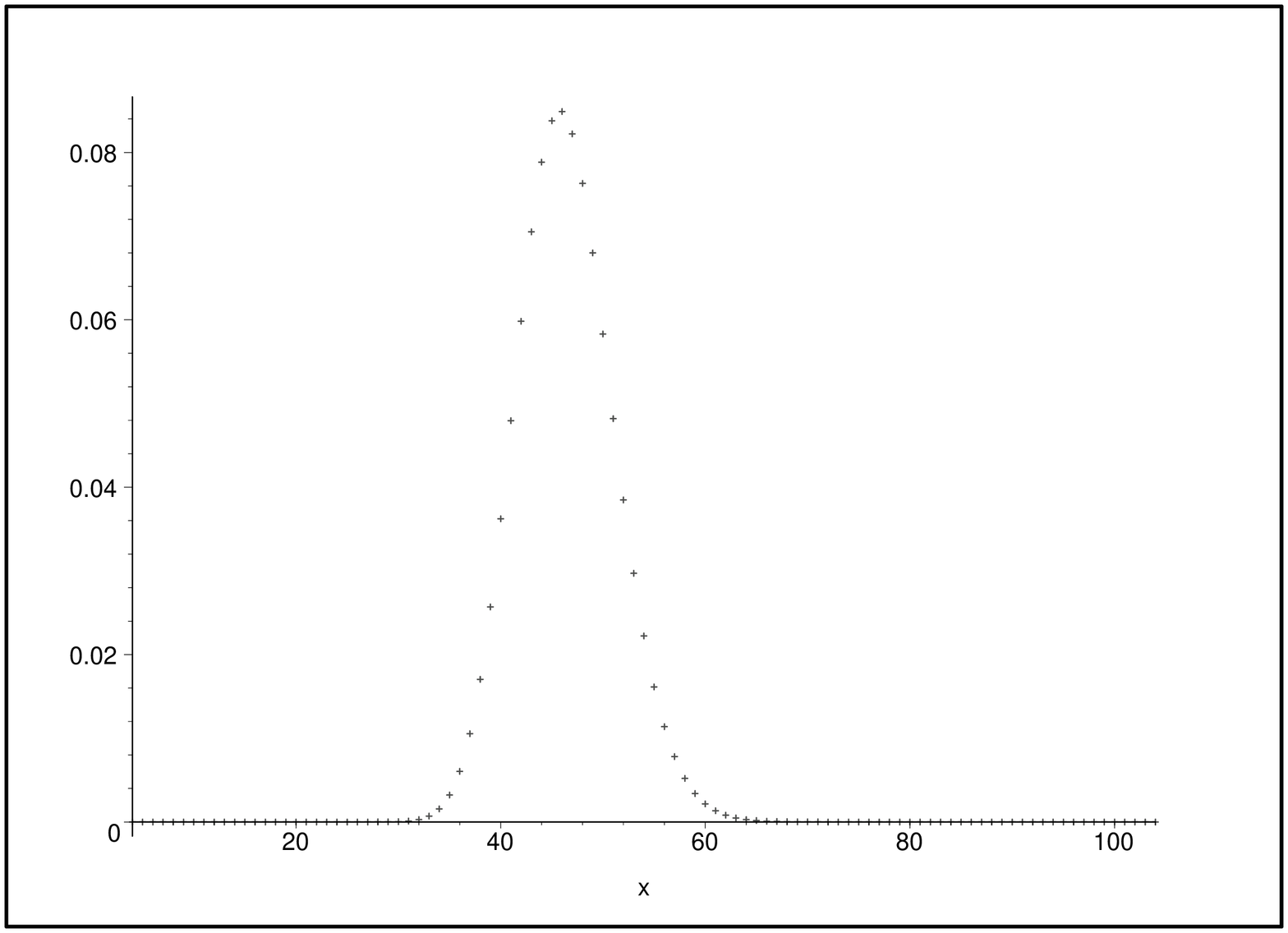}}
\rotatebox[origin=c]{270}{\includegraphics[scale=0.24]{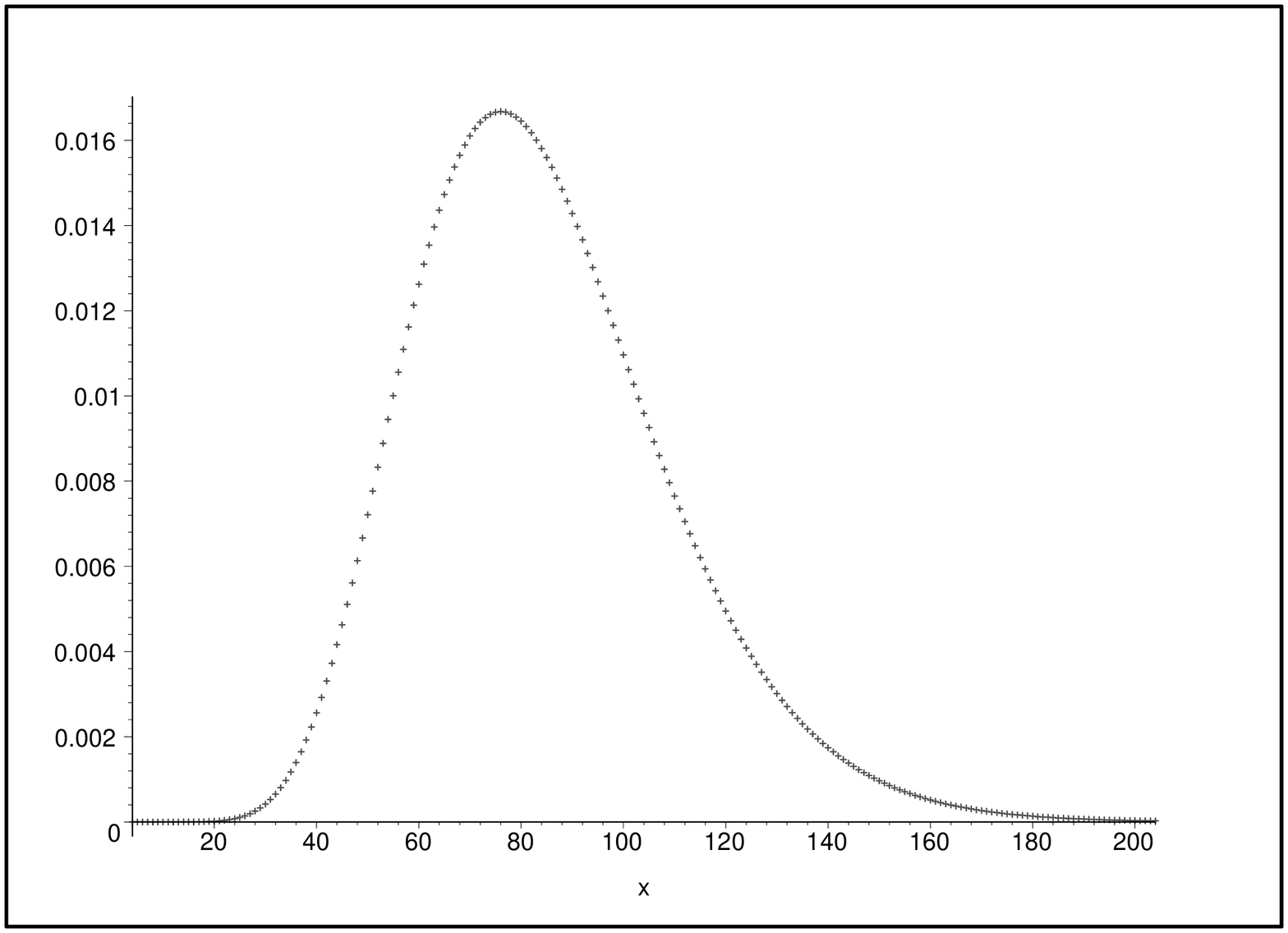}}
\end{center}

\subsection{} 
The following are the plots of the density function for the families of Charlier, $q$-Charlier and alternative $q$-Charlier polynomials (left to right). The parameters (cf. subsection \ref{ss:list}) are $k=6$, $a=20$ for the Charlier polynomials (first graph) and $k=6$, $a=20$, $q=0.96$ for the $q$-Charlier and the alternative $q$-Charlier polynomials (last two graphs). In the case of Charlier polynomials we plot the difference derivative, $D_{s+1}-D_s$, of $D_s$, and the $x$-coordinate is $s$, while in the other two cases we plot the $q$-derivative, $q^s\cdot (D_{s+1}-D_s)/(1-q)$, of $D_s$, and the $x$-coordinate is $q^{-s}$.

\medbreak

\begin{center}
\rotatebox[origin=c]{270}{\includegraphics[scale=0.24]{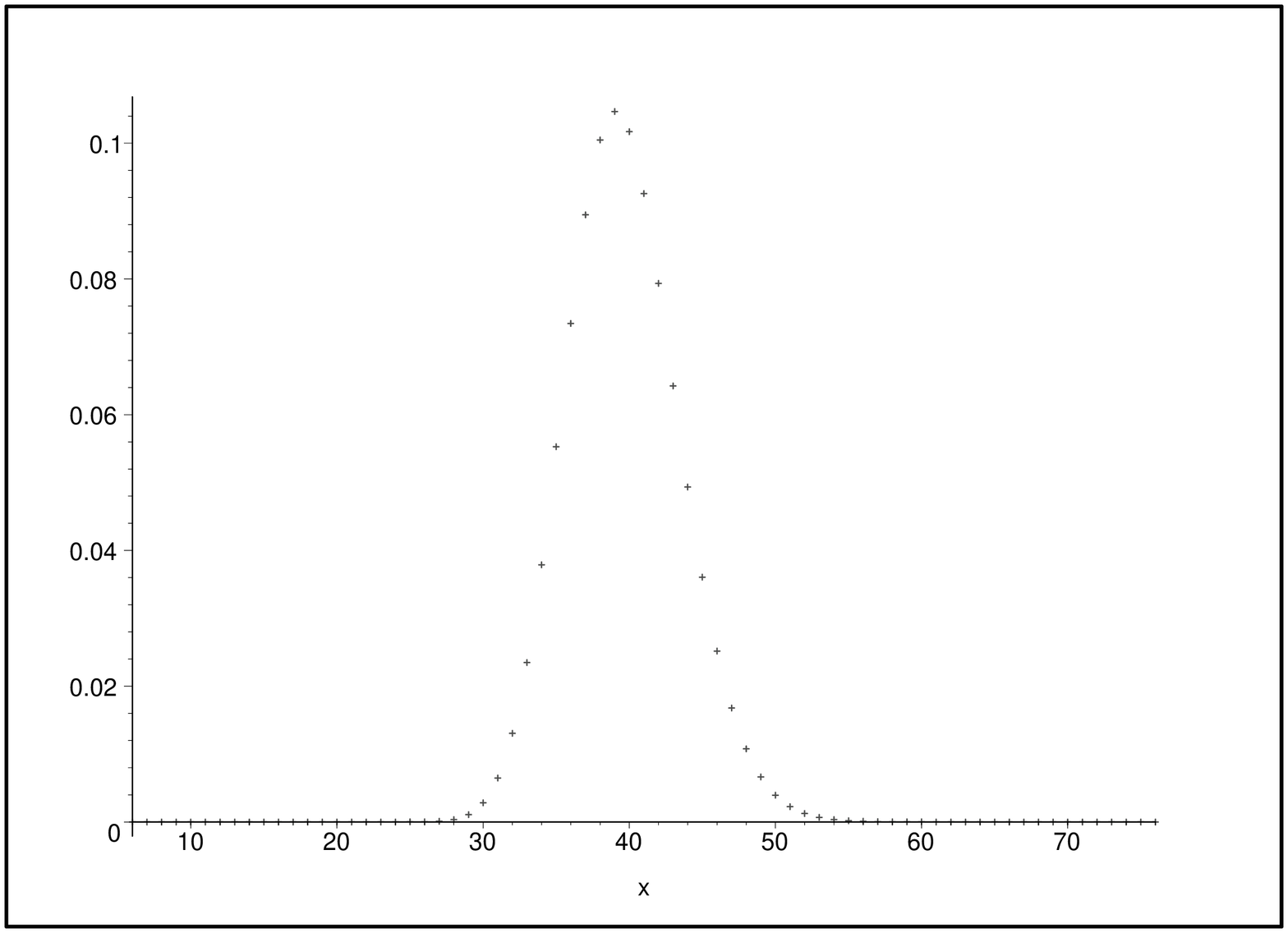}}
\rotatebox[origin=c]{270}{\includegraphics[scale=0.24]{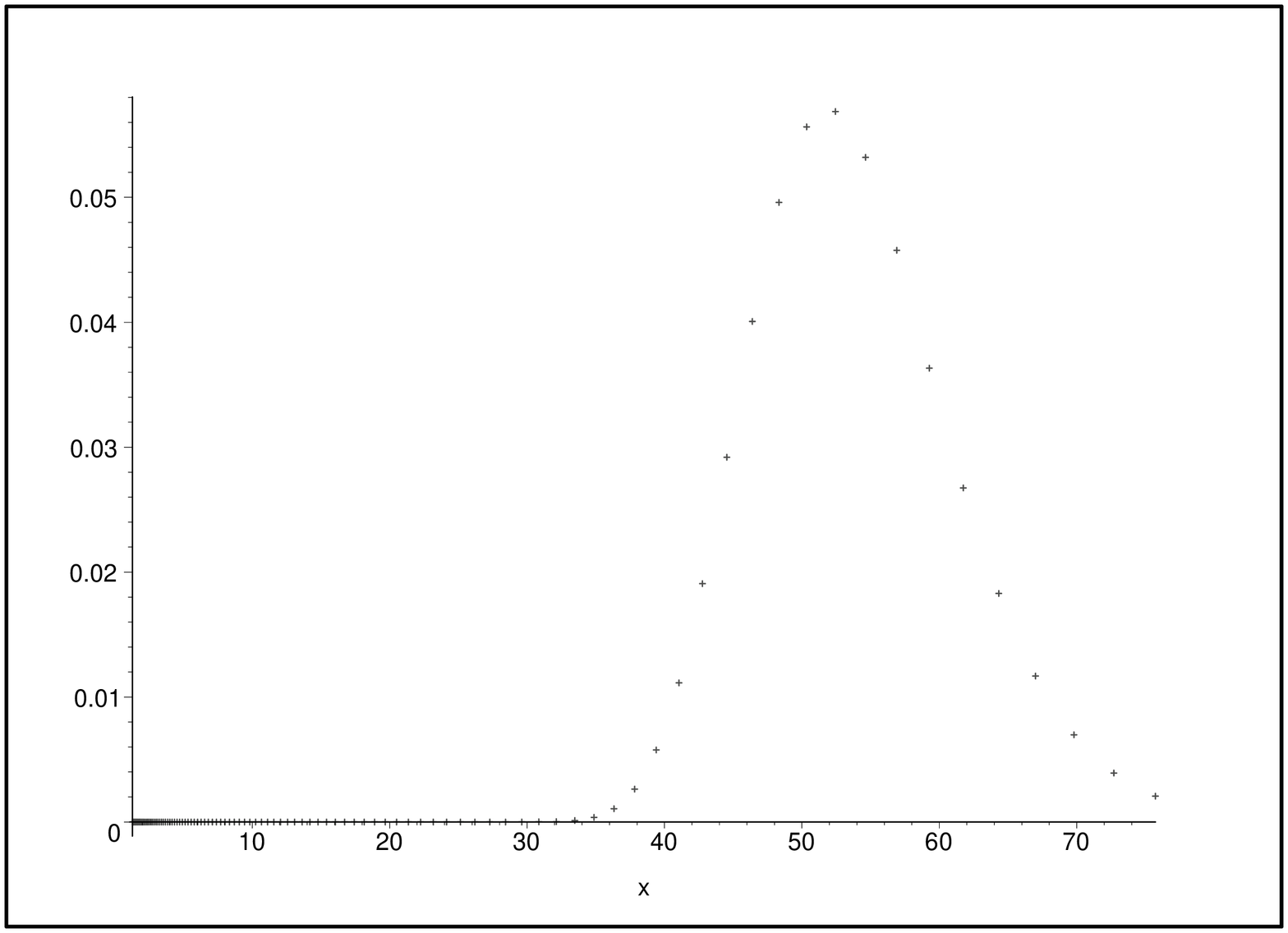}}
\rotatebox[origin=c]{270}{\includegraphics[scale=0.24]{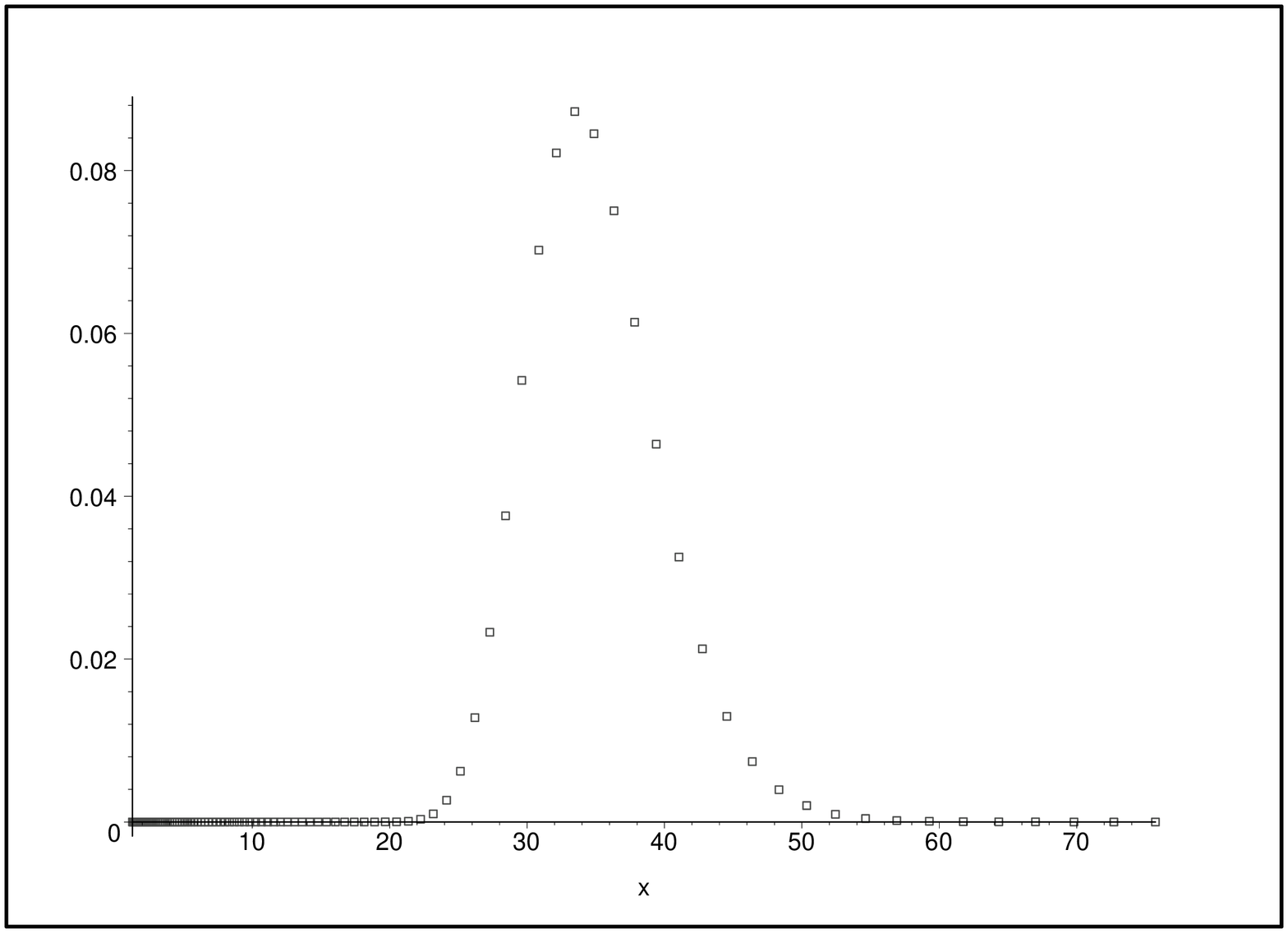}}
\end{center}

\subsection{} 
The following are the plots of the density function $q^{-s}\cdot (D_{s+1}-D_s)/(1-q)$ for the families of little $q$-Laguerre/Wall polynomials (first graph) and little $q$-Jacobi polynomials (second graph). The parameters (cf. subsection \ref{ss:list}) are $k=6$, $a=0.5$, $q=0.9$ for the little $q$-Laguerre polynomials and $k=6$, $a=0.5$, $b=1.5$, $q=0.9$ for the little $q$-Jacobi polynomials. The $x$-coordinate in each case is $q^s$.


$ $

\vskip0.5cm

\begin{center}
\rotatebox[origin=c]{270}{\includegraphics[scale=0.24]{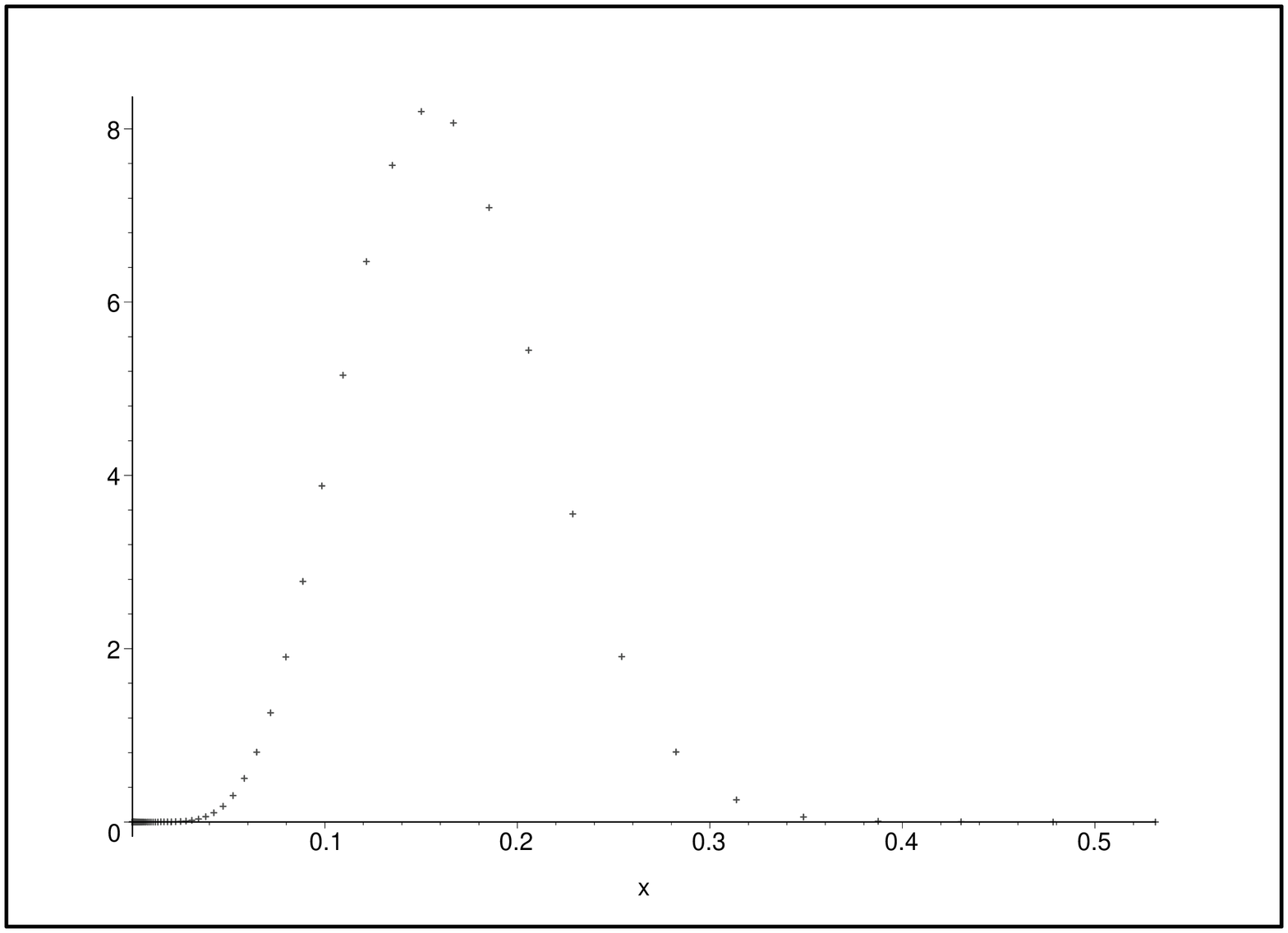}}
\rotatebox[origin=c]{270}{\includegraphics[scale=0.24]{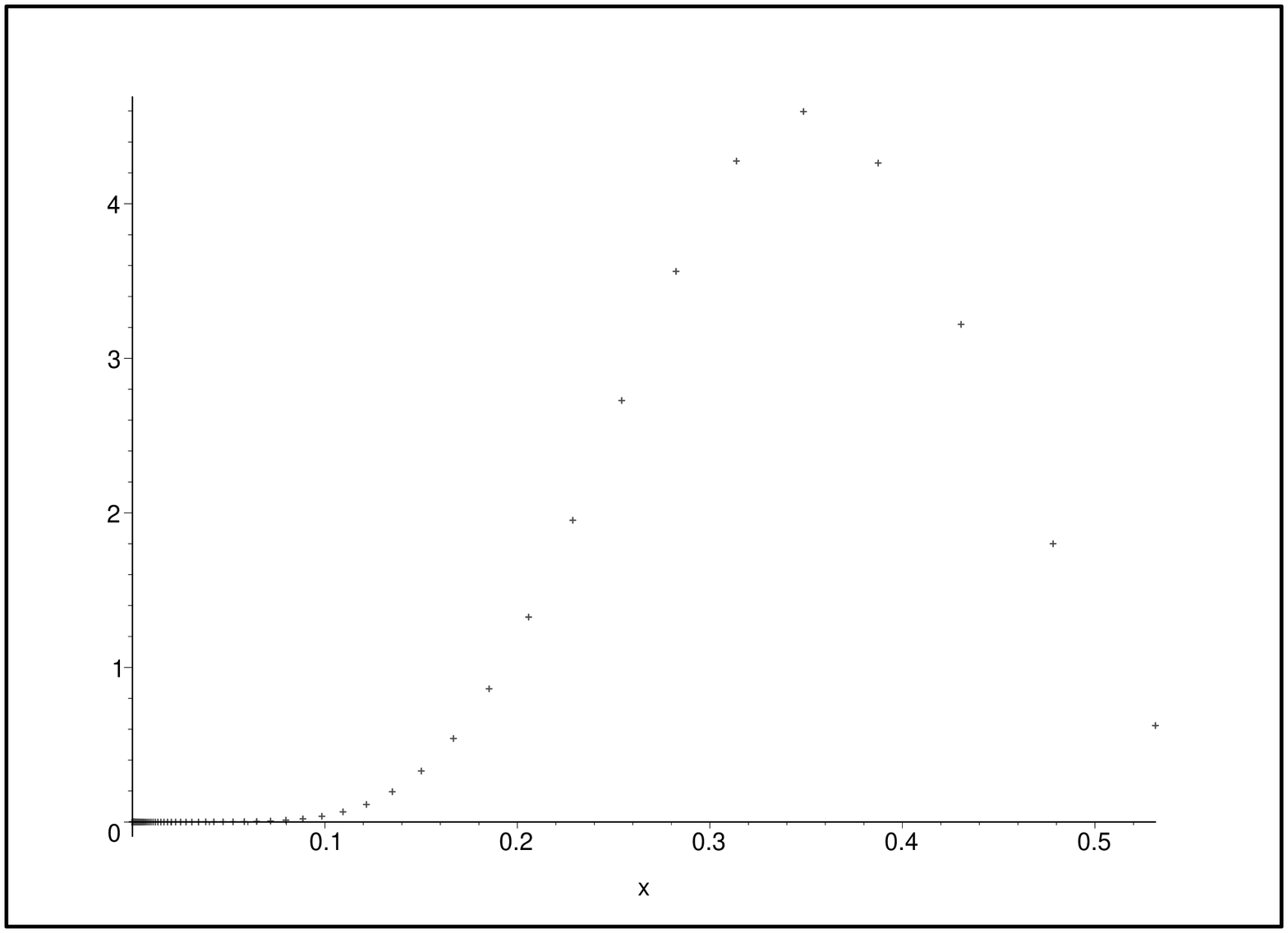}}
\end{center}

\subsection{} 
The following are the plots of the density function for the families of Krawtchouk polynomials (first graph) and $q$-Kratwchouk polynomials (second graph). The parameters (cf. subsection \ref{ss:list}) are $k=5$, $N=80$, $p=1/(0.7+1)$ for the Krawtchouk polynomials and $k=5$, $N=80$, $p=0.7$, $q=0.98$ for the $q$-Krawtchouk polynomials. In the first case we plot the difference derivative, $D_{s+1}-D_s$, of $D_s$, and the $x$-coordinate is $s$, while in the second case we plot the normalized $q$-derivative, $(q^{-N}-1)\cdot q^s\cdot (D_{s+1}-D_s)/(1-q)/N$, of $D_s$, and the $x$-coordinate is $N\cdot (q^{-s}-1)/(q^{-N}-1)$.

\medbreak

\begin{center}
\rotatebox[origin=c]{270}{\includegraphics[scale=0.24]{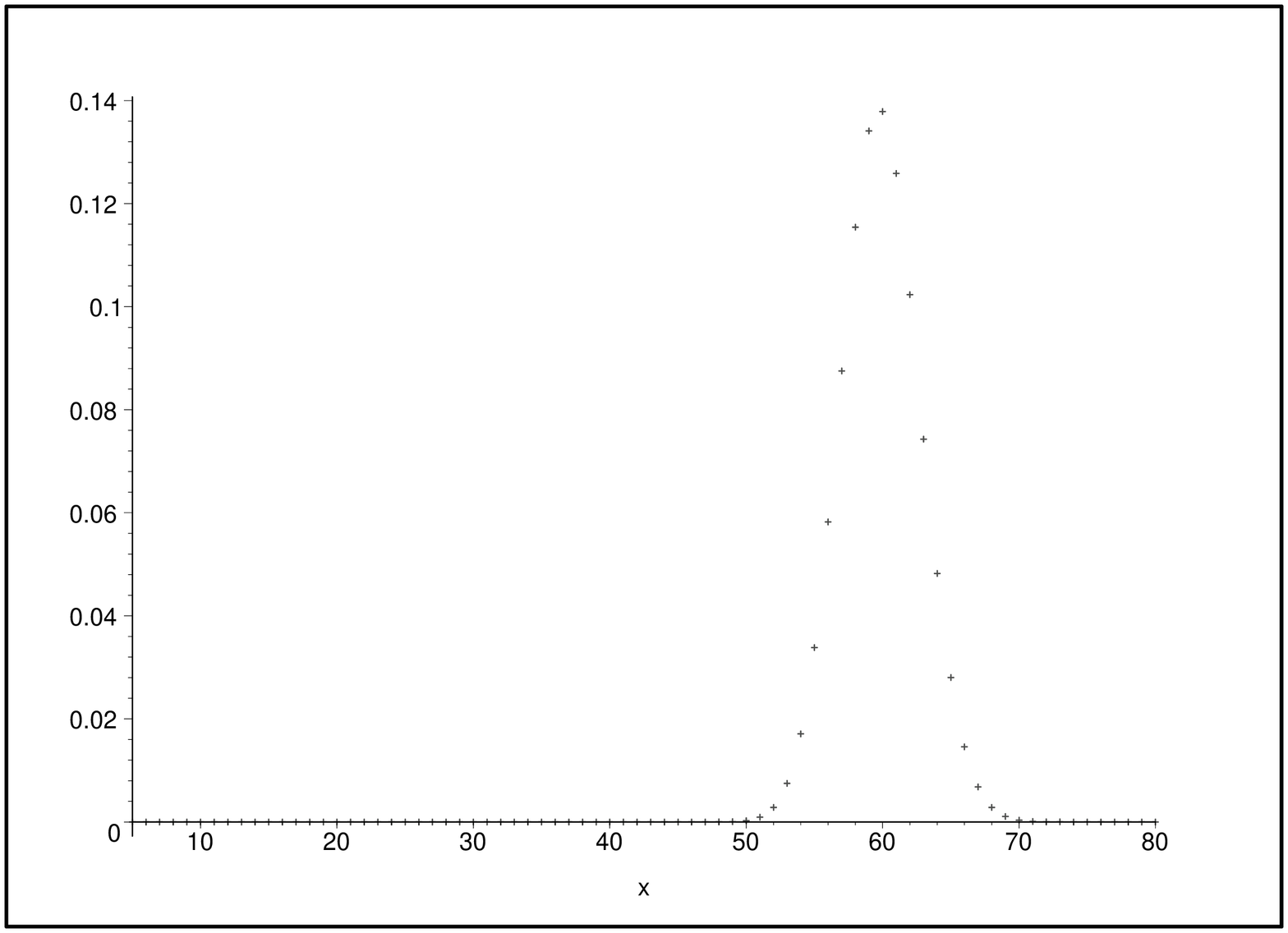}}
\rotatebox[origin=c]{270}{\includegraphics[scale=0.24]{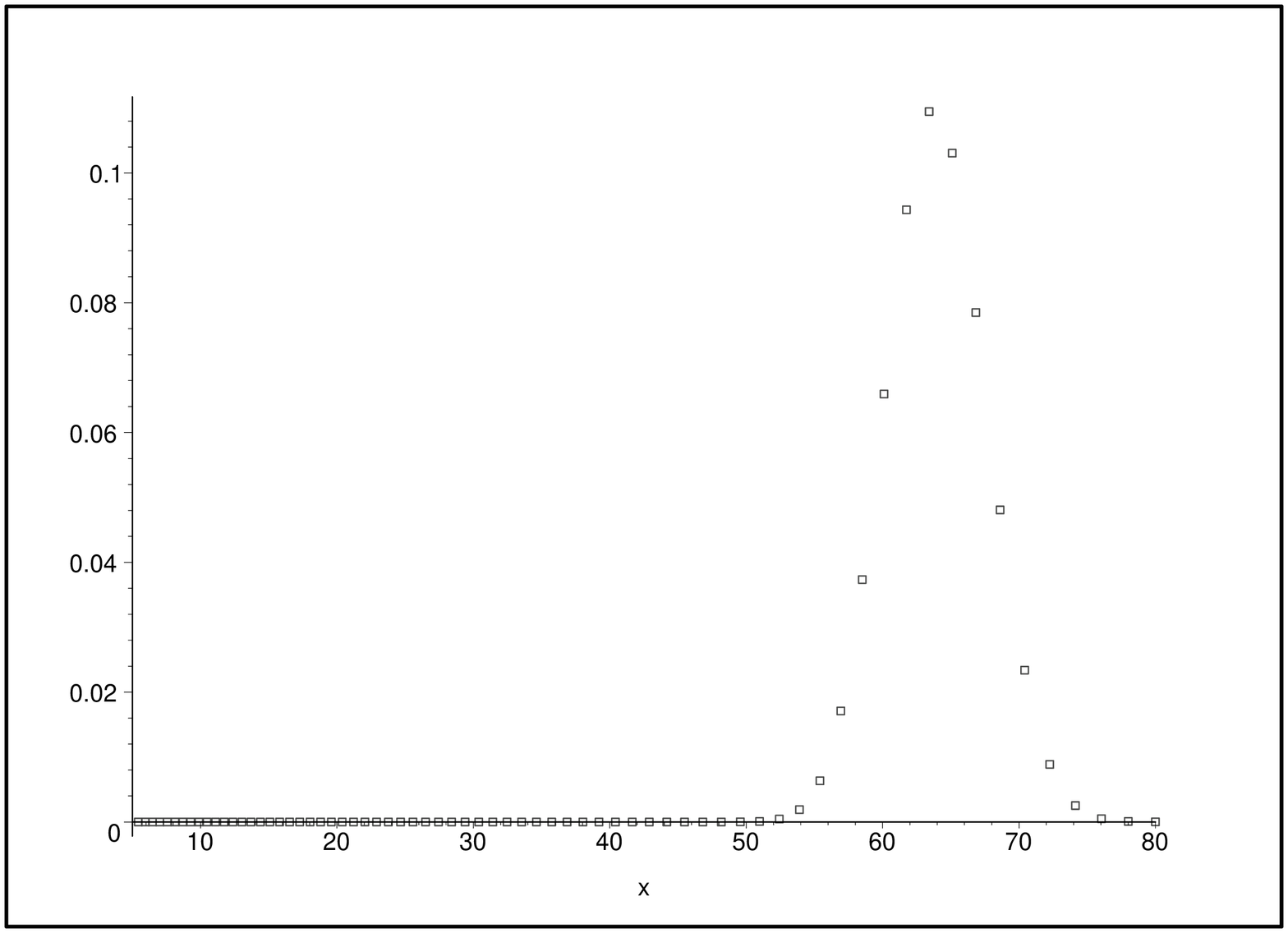}}
\end{center}

\end{document}